\documentclass[reprint,amsmath,amssymb,aip,showpacs,floatfix,superscriptaddress]{revtex4-1}

\usepackage{graphicx}
\usepackage{color}
\usepackage{verbatim}
\usepackage{nccmath}

\renewcommand{\figurename}{Fig.}
\renewcommand{\tablename}{Table}

\newcommand{\m}{\mathcal}
\newcommand{\ve}{\varepsilon}
\newcommand{\pd}{\partial}
\newcommand{\dif}{\mathrm{d}}

\newcommand{\Tr}{\mathrm{Tr}}

\newcommand{\bn}{\mathbf{n}}
\newcommand{\br}{\mathbf{r}}
\newcommand{\bu}{\mathbf{u}}

\newcommand{\mrL}{\mathrm{L}}
\newcommand{\mrR}{\mathrm{R}}
\newcommand{\mre}{\mathrm{e}}
\newcommand{\mrp}{\mathrm{p}}

\newcommand{\xx}{z}
\newcommand{\yy}{x}
\newcommand{\zz}{y}
\newcommand\cbl[1]{\color{blue}{#1}}
\newcommand\crd[1]{\color{red}{#1}}
\newcommand{\kappaone}{\kappa_{\mathrm{one}}}
\newcommand{\kappatwo}{\kappa_{\mathrm{two}}}

\begin{document}

\title{Designing $\pi$-stacked molecular structures to control heat transport through molecular junctions}

\author{Gediminas Kir\v{s}anskas}
\affiliation{Center for Quantum Devices, Niels Bohr Institute, University of Copenhagen, DK-2100 Copenhagen {\O}, Denmark}
\affiliation{Mathematical Physics and Nanometer Structure Consortium (nmC@LU), Lund University, Box 118, 221 00 Lund, Sweden}

\author{Qian Li}
\affiliation{Nano-Science Center and Department of Chemistry, University of Copenhagen, DK-2100 Copenhagen {\O}, Denmark}

\author{Karsten Flensberg}
\affiliation{Center for Quantum Devices, Niels Bohr Institute, University of Copenhagen, DK-2100 Copenhagen {\O}, Denmark}

\author{Gemma C. Solomon}
\affiliation{Nano-Science Center and Department of Chemistry, University of Copenhagen, DK-2100 Copenhagen {\O}, Denmark}

\author{Martin Leijnse}
\affiliation{Solid State Physics and Nanometer Structure Consortium (nmC@LU), Lund University, 221 00 Lund, Sweden}

\date{\today}

\begin{abstract}
We propose and analyze a new way of using $\pi$ stacking to design molecular junctions that either enhance or suppress a phononic heat current, but at the same time remain conductors for an electric current. Such functionality is highly desirable in thermoelectric energy converters, as well as in other electronic components where heat dissipation should be minimized or maximized. We suggest a molecular design consisting of two masses coupled to each other with one mass coupled to each lead. By having a small coupling (spring constant) between the masses, it is possible to either reduce, or perhaps more surprisingly enhance the phonon conductance. We investigate a simple model system to identify optimal parameter regimes and then use first principle calculations to extract model parameters for a number of specific molecular realizations, confirming that our proposal can indeed be realized using standard molecular building blocks.
\end{abstract}

\maketitle 

With decreasing device dimensions, managing heat flow in electronic circuits is becoming increasingly important, and recent research has proposed making thermal equivalents of functional electronic components, such as thermal transistors, diodes, and rectifiers.~\cite{Giazotto2006, Dubi2011} One application where controlling heat flow is essential is in thermoelectric devices,~\cite{Dubi2011, Rowe2005} in which a temperature gradient is used to generate an electric current or voltage (power conversion), or, conversely, an electric current is used to generate a temperature gradient (refrigeration).

A good thermoelectric material must have a large Seebeck coefficient, but also a reasonably good electric conductance combined with a poor heat conductance.~\cite{Goldsmid1964} In general, both electrons and phonons contribute to the heat conductance. In bulk materials the conductance and the electron contribution to the heat conductance are related by Wiedemann-Franz law,~\cite{Goldsmid1964} limiting the thermoelectric performance. Therefore, efforts have mainly focused on reducing the phonon contribution to the heat conductance, while trying to maintain electric conductivity. One way of phrasing it is to say that a good thermoelectric material should be an electron crystal, but a phonon glass.~\cite{Snyder2008}
Many recent theoretical works have investigated phonon transport in nanostructures~\cite{Segal2003,Mingo2003,Wang2007a,Galperin2007,Markussen2008} and experimental studies have shown reduced phonon conductance compared to bulk materials.~\cite{Boukai2008,Hochbaum2008,Blanc2013}


It has also been shown that the electronic properties can be much more suited for thermoelectrics in nanoscale systems,~\cite{Hicks1993a,Majumdar2004,Dresselhaus2007} where the Wiedemann-Franz law breaks down due to quantum confinement effects and strong electron-electron interactions.~\cite{Appleyard2000, Boese2001, Kubala2008} A promising candidate is molecular junctions,~\cite{Reddy2007} where sharp resonances originating from molecular orbitals~\cite{Murphy2008, Finch2009} or interference effects~\cite{Solomon2008, Karlstrom2011} could result in a large thermoelectric effect.~\cite{Bergfield2009} It has also been argued that molecular junctions should have a very small phonon contribution to the heat conductance because, in small molecules, the quantized molecular vibrational modes have frequencies above the Debye frequency of the electrodes, preventing phonon transport from the hot to the cold electrode via the molecule. However, this argument does not take into account the vibrational modes that are associated with center of mass motion of the entire molecule.~\cite{Seldenthuis2008} Since the mass of the entire molecule is very large compared with that of individual atoms, these modes have a low frequency, well inside the band of acoustic phonons in the electrodes, and might be detrimental for the thermoelectric performance.

\begin{figure*}[t!]
\begin{center}
\includegraphics[width=0.9\textwidth]{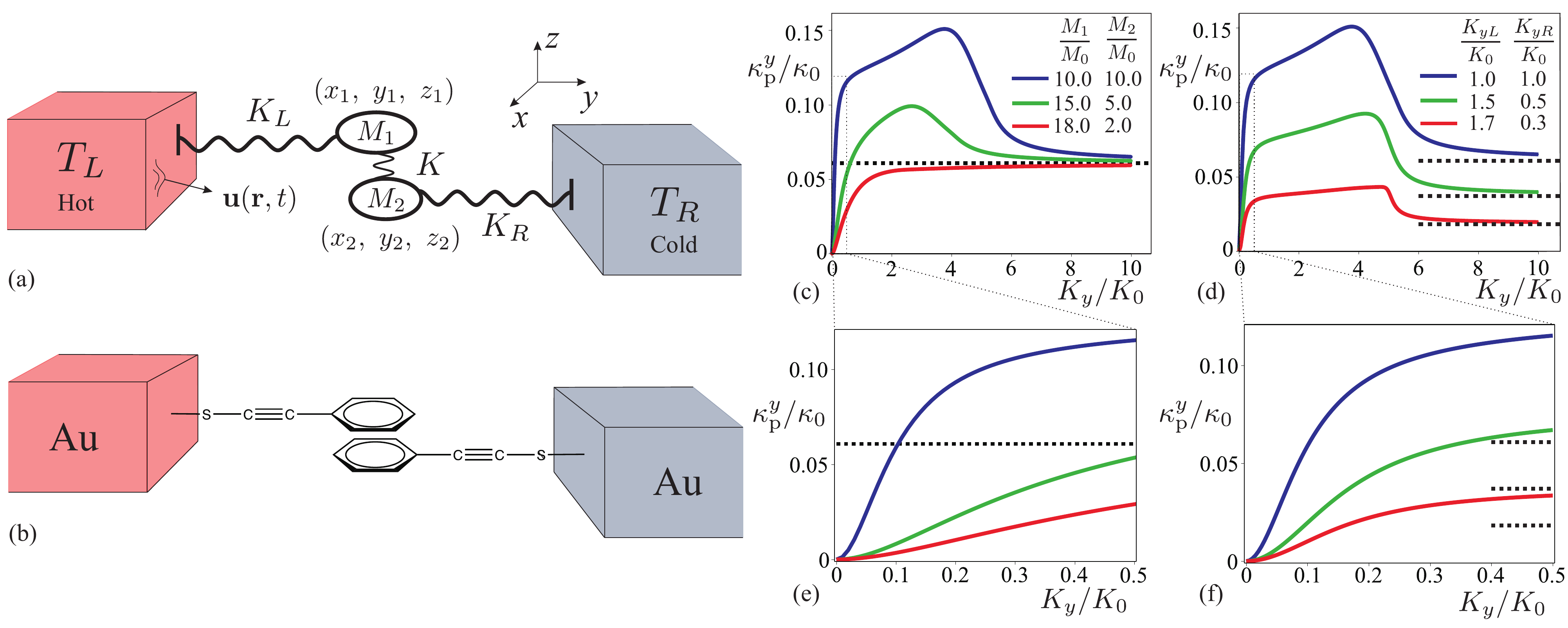}
\caption{\label{fig1}
(a) Effective description of the center of mass vibrational degrees of freedom in the system: Two masses $M_{1}$ and $M_{2}$, which have coordinates $(x_1,y_1,x_1)$ and $(x_2,y_2,z_2)$, are coupled to each other by a spring $K_{i}$ and to the leads by springs $K_{iL}$ and $K_{iR}$. The substrate is modeled as an elastic continuum described by the displacement vector $\bu(\br,t)$. $T_{L}$ and $T_{R}$ denote the temperature of the left and the right lead, respectively. (b) Sketch of a specific realization of the junction in (a) consisting of two $\pi$-stacked 2-phenylethyne-1-thiol molecules coupled to gold electrodes. (c)-(f) Phonon conductance, $\kappa_{\mrp}^{y}$, as a function of $K_y$, the middle spring constant in the $\zz$ direction.
In (c) and (e) the couplings to the leads are $K_{yL}=K_{yR}=K_0$ and in (d) and (f) the masses are $M_{1}=M_{2}=10 M_0$ ($K_0$, $M_0$, and $\kappa_0$ are defined in the main text). (e) and (f) Zoom of the small $K_y$ region of the phonon conductance in (c) and (d), respectively. The dotted black lines denote the asymptotic values for $\kappa_{\mrp}^{y}$ when $K_y \rightarrow \infty$. Thus, dividing the molecule into two sub units can be said to reduce/enhance the phonon heat conductance compared to the single-mass case for values of $K_y$ such that $\kappa_{\mrp}^{y}$ lies below/above this asymptotic value.}
\end{center}
\end{figure*}

In this work, we propose a solution to this problem by suggesting a simple mechanism to reduce the phonon conductance arising from the center of mass motion of a molecule in a transport junction. The idea is to have a molecular system that is divided into two sub units, where each sub unit binds strongly to one electrode, but where the two sub units are only weakly coupled to each other. In chemical terms, we break the bond between two parts of the molecule and consider two molecules that can only couple through-space. We show that the phonon conductance can be significantly reduced compared with the case of a single molecular unit in the junction. However, we also show that there is a regime where the heat current is enhanced due to the larger number of center-of-mass modes. We suggest and, using density functional theory (DFT), investigate concrete realizations of this proposal where the sub units are made by $\pi$-stacked aromatic rings, connected via $\pi$-$\pi$ electronic coupling, which provides a rather soft mechanical bridge while maintaining high electronic conductivity. $\pi$-$\pi$ stacking is efficiently used by nature to achieve directed long-range electron transport, such as in DNA base pairs~\cite{Giese2002,Watson2004} and amino acid residues. Break-junction experiments have found that the intermolecular $\pi$-$\pi$ stacking interaction can lead to molecular junctions with significant conductance.~\cite{Wu2008,Martin2010,Schneebeli2011} Thus, $\pi$-stacked systems could act as electric conductors, while the phonon conductance is tuned in order to meet the device requirements. We note that several recent experiments have investigated phonon transport in molecular monolayers.~\cite{Wang2007b,Losego2012,O'Brien2013,Meier2014}

To model the systems we have in mind, we consider two masses $M_{1}$ and $M_{2}$ coupled by the spring $K$ to each other and by springs $K_{L}, \ K_{R}$ to the leads as shown in \figurename~\ref{fig1}a. Figure~\ref{fig1}b shows an example of a specific realization of the junction in  \figurename~\ref{fig1}a with two $\pi$-stacked 2-phenylethyne-1-thiol molecules coupled to gold electrodes. When the spring constant between the masses is weak, the phonon conductance is reduced compared with the situation with the single mass $M=M_{1}+M_{2}$ in the junction.
We describe the vibrational degrees of freedom in a molecule and the coupling to the leads using the harmonic approximation and it is assumed that the molecule couples to a particular lead only at a single point $(x_{\alpha},y_{\alpha},z_{\alpha})$, with $\alpha=L,R$. In such a model, we need to specify the spring constants $K_{i}, \ K_{iL}, \ K_{iR}$ in all three directions $i=x,y,z$. The leads are modeled as a continuum governed by the equations of motion for an elastic medium~\cite{Landau1986,Ezawa1971,Patton2001} described by the displacement vector $\bu(\br,t)$.
The elastic description is reasonable because we are interested in the low energy behavior of the system, when phonons in the leads are long wavelength (acoustic) and have linear dispersion, and because the vibrational modes of the molecule contributing to the phonon conductance through the junction at room temperature have frequencies $< \omega_D$, the Debye frequency of the leads.

When there is a temperature difference $\Delta T=T_{L}-T_{R}$ between the leads, a heat current flows through the device. If the system has reached a steady state, the phonon conductance can be obtained from a Landauer-B\"{u}ttiker type expression~\cite{Buttiker1985}
\begin{align}
\kappa_{\mrp}&=\frac{\hbar}{\Delta{T}}\int_{0}^{+\infty}\frac{\dif\omega}{2\pi}\omega\m{T}_{\mrp}(\omega)(n_{L}-n_{R})\\
&\stackrel{\Delta{T}\rightarrow0}{=}\hbar\int_{0}^{+\infty}\frac{\dif\omega}{2\pi}\omega\m{T}_{\mrp}(\omega)\frac{\pd n(\omega,T)}{\pd T},
\end{align}
where $n_{\alpha}=1/\{\exp[\beta_{\alpha}\hbar\omega]-1\}$ is the Bose-Einstein distribution with $\beta_{\alpha}=1/(k_{\mathrm{B}}T_{\alpha})$ denoting the inverse temperature. The phonon transmission function $\m{T}_{\mrp}(\omega)$ is given by a Caroli type formula~\cite{Caroli1971}
$\m{T}_{\mrp}(\omega)=\Tr[\Gamma_{\mrL}(\omega)D^{R}(\omega)\Gamma_{\mrR}(\omega)D^{A}(\omega)],$
where $D^{R/A}$ is the retarded/advanced Green's function of the molecule and $\Gamma_{\mrL/\mrR}$ describes the coupling to the leads. In the Supplemental Material (SM)~\cite{SupplementalMaterial} we derive an exact analytical expression for the transmission.
We focus on operation of the device at room temperature, $T=300 \ \mathrm{K}$. For the molecules examined in this paper, the typical energies of the center of mass vibrational modes are smaller than $k_{\mathrm{B}} T$ at room temperature, so the phonon conductance corresponding to these modes saturates and becomes temperature independent, i.e., $\kappa_{\mrp} (T = 300 K) \approx \kappa_{\mrp} (T\rightarrow+\infty)=k_{\mathrm{B}}/2\pi\int_{0}^{+\infty}\dif\omega\m{T}_{\mrp}(\omega)$.
%

In our calculations, we use elastic parameters for gold, which has mass density $\rho=19.3 \ \mathrm{kg/cm^3}$, Young modulus $E_{\mathrm{Y}}=77.5 \ \mathrm{GPa}$, and Poisson ratio $\sigma=0.42$.~\cite{Samsonov1968} For frequency cut-off, we use the bulk Debye temperature $T_{D}=170 \ \mathrm{K}$, which corresponds to $\omega_{\mathrm{D}} \approx 22.2 \ \mathrm{THz}$.~\cite{Kittel2004} We normalize the spring constant, mass, and heat conductance, respectively, by: $K_{0}=1/A_{\perp}\omega_{\mathrm{D}}\approx 12.7 \ \mathrm{N/m}$, $M_{0}=1/(A_{\perp}\omega_{\mathrm{D}}^3)\approx 15.3 m_{\mathrm{H}}$, $\kappa_{0}=k_{\mathrm{B}}\omega_{\mathrm{D}}/(2\pi)\approx 49 \ \mathrm{pW/K}$, where $m_{\mathrm{H}}$ is the mass of the hydrogen atom and $A_{\perp}=1.445/(4\pi\rho c_{t}^3)$ with $c_{t}=\sqrt{E_{\mathrm{Y}}/\rho(1+\sigma)}$ denoting the velocity of the transverse wave.

Figures~\ref{fig1}c,d show the $\zz$-component, $\kappa_{\mrp}^{y}$, of the phonon conductance as a function of the middle spring constant $K_{\zz}$ in the $\zz$ direction. Figure~\ref{fig1}c depicts this dependence for different mass asymmetries with fixed total mass $M=M_{1}+M_{2}$ and symmetric coupling to the leads $K_{iL}=K_{iR}$. In the case of rather symmetric masses $M_{1} \approx M_{2}$ the conductance overshoots the value of conductance when there is a single mass in the junction. The reason for the overshooting is that in the case of two masses the number of modes doubles compared to the case of a single mass. Depending on the actual values of the parameters these modes can have a large transmission. The situation when the masses are symmetric $M_{1}=M_{2}$, but the coupling to the leads is changed, is shown in \figurename~\ref{fig1}d (the sum of the couplings to the left $K_{yL}$ and the right $K_{yR}$ leads is kept constant). Clearly, an asymmetric coupling reduces the phonon conductance.
We note that in the considered model the transmission separates for different directions $\m{T}_{\mrp}=\m{T}^{x}_{\mrp}+\m{T}^{y}_{\mrp}+\m{T}^{z}_{\mrp}$ and that for the $\xx$ and $\yy$ directions the dependence is analogous.
The dotted black lines in \figurename~\ref{fig1}c--f show the asymptotic values for $\kappa_{\mrp}^{y}$ when $K_{y} \rightarrow \infty$, which corresponds to having a molecule with a single mass $M=M_1+M_2$ in the junction. Our aim is to divide molecules into two sub units in such a way that $\kappa_{\mrp}^{i}$ is very different from this asymptotic value. A reduction of the phonon conductance is always possible if $K_{i} \ll K_{iL},K_{iR}$ (\figurename~\ref{fig1}e,f zooms in on the low $K_{y}$ region of decreased $\kappa_{\mrp}^{y}$).

\begin{figure}
\begin{center}
\includegraphics[width=0.9\columnwidth]{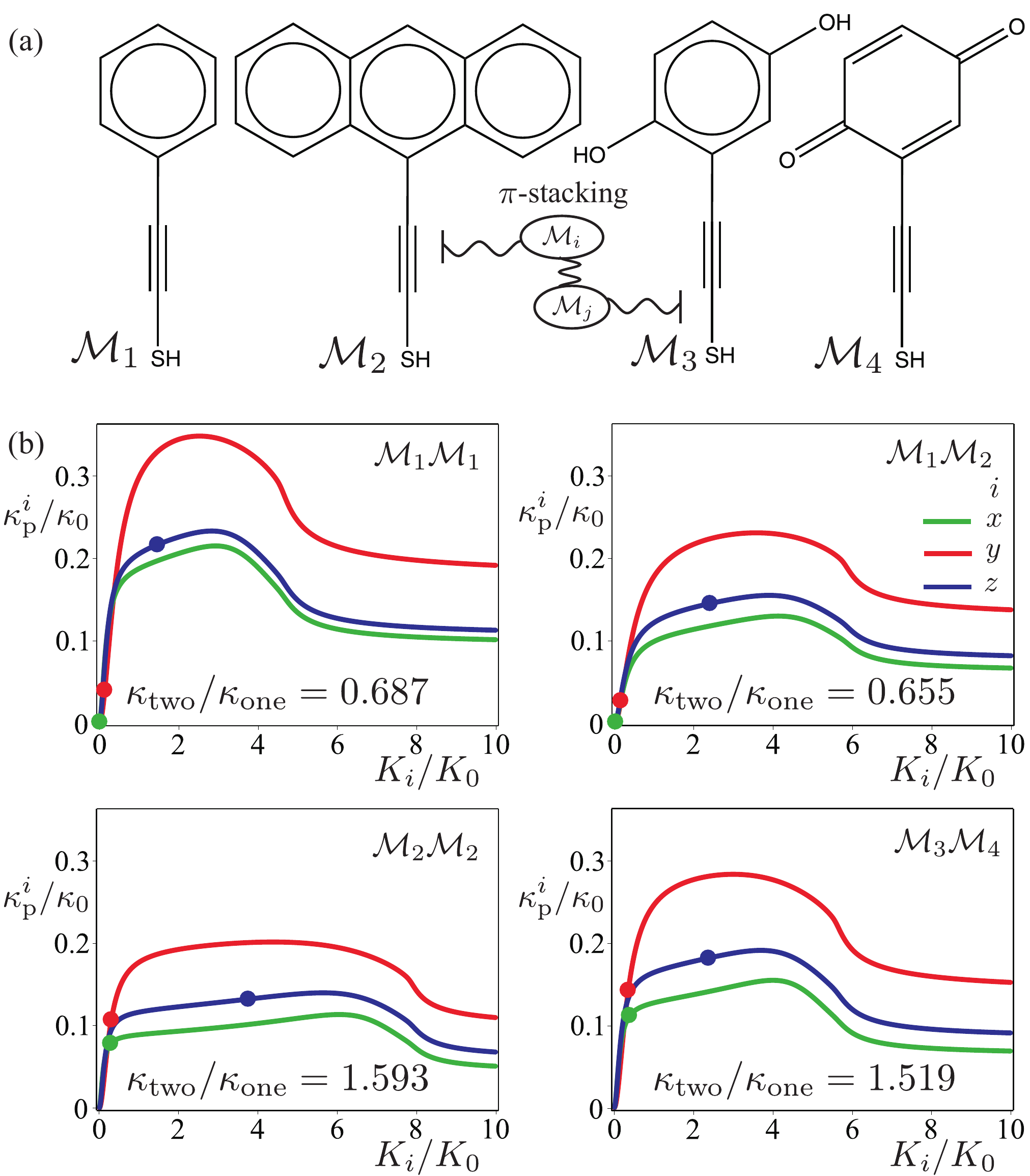}
\caption{\label{fig3} (a) The molecules considered. (b) $\kappa_{\mrp}^{i}$ as a function of $K_{i}$ in different directions $i = x,y,z$ for the considered stackings. Solid circles denote the conductance, $\kappatwo^{i}$, calculated with the middle spring constants obtained from DFT. }
\end{center}
\end{figure}

Having established exactly how molecules can be divided into sub units to perturb $\kappa_{\mrp}$, we now turn to the problem of finding molecular structures where these conditions are realized. The particular route we follow here is to use $\pi$-stacked aromatic rings to provide a soft mechanical connection that is not based on a chemical bond, while maintaining high electrical conductivity with $\pi$-$\pi$ coupling. The different molecular sub units we use for $\pi$-stacking are shown in \figurename~\ref{fig3}a.

We examine the following four stackings: $\m{M}_1\m{M}_1$, $\m{M}_1\m{M}_2$, $\m{M}_2\m{M}_2$, and $\m{M}_3\m{M}_4$.
The spring constants are calculated from GPAW~\cite{Enkovaara2010} (a grid-based real-space DFT code). The geometries of the isolated molecules are optimized using the finite difference model with the PBE exchange functional and the stacked structures are formed by combining two of these molecules. The molecules that couple to the lead are chemisorbed on a hollow site of the (111) surface of the gold electrodes with the terminal hydrogen atoms removed. The potential energy curves are calculated as a function of translation along the $x$, $y$ and $z$ axes of the molecules with van der Waals correction (TS09)~\cite{Tkatchenko2009} and the spring constants are obtained from the second derivative of the potential energy function. More details on the DFT calculations is given in the SM.~\cite{SupplementalMaterial}
Figure~\ref{fig3}b shows $\kappa_{\mrp}^{i}$ as a function of the middle spring constants $K_{i}$ for the considered stackings, where solid circles denote the conductance, $\kappatwo$, calculated with the middle spring constants obtained from DFT.
For comparison we define $\kappaone^{i}$ as the phonon conductance with a single mass $M = M_1 + M_2$ in the junction, i.e., $\kappaone^{i} = \mathrm{lim}_{K_i\rightarrow\infty}\kappa_{\mrp}^{i}$.
We see that for the stacking combinations $\m{M}_1\m{M}_1$ ($\kappatwo/\kappaone=0.687$) and $\m{M}_1\m{M}_2$ ($\kappatwo/\kappaone=0.655$), because of the small middle spring constant, the conductance is significantly reduced in the $\yy$ and $\zz$ directions compared with the case of a single mass in the junction.
Also, in the $\zz$ direction the coupling to the leads is the largest.
In all cases in the $\xx$ direction and in all directions for stackings $\m{M}_{2}\m{M}_{2}$ ($\kappatwo/\kappaone=1.593$), $\m{M}_{3}\m{M}_{4}$ ($\kappatwo/\kappaone=1.519$) the conductance is larger than in the single mass case.

These results can be understood by considering the chemical structure of the molecules. $\pi$-$\pi$ stacking results in appreciable mechanical coupling in the $\xx$ direction, holding the system together. However, for any pair of carbon atoms there is relatively weak interaction in the $\yy$ and $\zz$ directions. The larger coupling pushes the $\xx$ transport above the asymptotic limit in all these cases.

 For the stackings $\m{M}_1\m{M}_1$ and $\m{M}_1\m{M}_2$ the coupling in the $\yy$ and $\zz$ directions is relatively low and the overall phonon conductance remains well below the one molecule limit. Note that the largest reduction is for stacking $\m{M}_1\m{M}_2$, which has asymmetric masses. Extending the $\pi$-system, in the case of $\m{M}_{2}\m{M}_{2}$, or adding hydrogen bonds across the stack, in the case of ${\m{M}_3\m{M}_4}$, results in increased coupling in the $\yy$ and $\zz$ directions, allowing the system as a whole to exhibit increased conductance. Thus, for some, but not all, stacking combinations the suggested mechanism reduces the phonon conductance.

\begin{figure}
\begin{center}
\includegraphics[width=0.9\columnwidth]{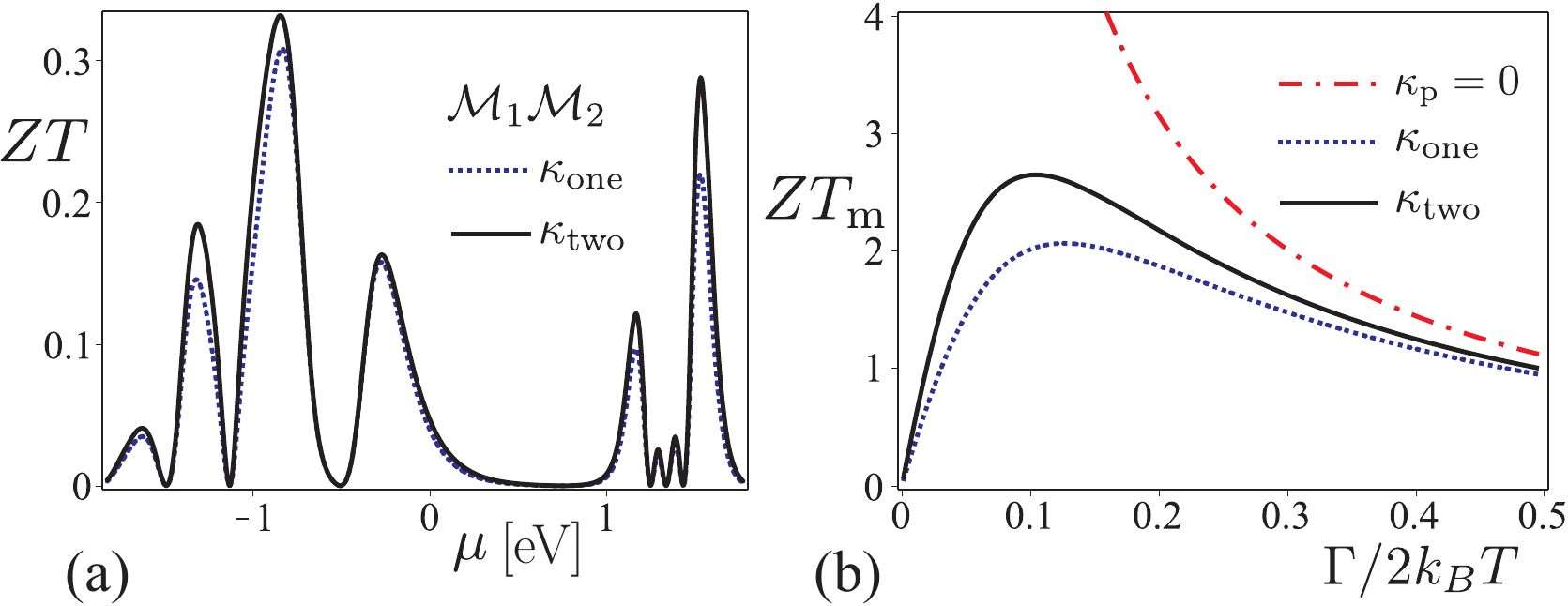}
\caption{\label{fig4} (a) $ZT$ as a function of $\mu$ at $T=300 \ \mathrm{K}$ for the stacking $\m{M}_1\m{M}_2$, which had the largest reduction in phonon conductance, with electronic transmission $\m{T}_{\mre}(E)$ obtained from DFT. The solid black curve shows $ZT$ when the system has a weak link in the middle (two masses model) and dashed blue curves show $ZT$ when the center of mass vibrational degrees of freedom are described by single mass in the junction. (b) $ZT_{\mathrm{m}}$ as a function of $\Gamma$ when the spinful resonant level electronic transmission $\m{T}_{\mre}(E)$ is used. The red dashed-dotted line shows the behavior of $ZT_{\mathrm{m}}$ when there is no phonon conductance, $\kappa_{\mrp}=0$.}
\end{center}
\end{figure}

To describe the thermoelectric efficiency of $\pi$-stacked molecules in the junction, we calculate the dimensionless figure of merit $ZT$, which is given by~\cite{Goldsmid1964}
%
$ZT=S^2\sigma_{\mre} T / (\kappa_{\mre}+\kappa_{\mrp}),$
%
where $S$ is the Seebeck coefficient, $\sigma_{\mre}$ is the electronic conductance, and $\kappa_{\mre}$ is the electron contribution to the thermal conductance.
We neglect inelastic scattering in our calculations and can therefore express the quantities $S$, $\sigma_{\mre}$, and $\kappa_{\mre}$ through the electron transmission $\m{T}_{\mre}(E)$ using Landauer-type expressions as described in Ref.~[\onlinecite{Sivan1986}].
The resulting $ZT$ at $T=300 \ \mathrm{K}$ for the stacking $\m{M}_1\m{M}_2$, which had the largest reduction in phonon conductance, is depicted in \figurename~\ref{fig4}a.  We see a slight increase in $ZT$ at some values of $\mu$, but the overall effect is rather small even though the phonon conductance is reduced. This clearly indicates that the main contribution to the thermal transport comes from electrons (see SM for the same calculations on the other three systems).~\cite{SupplementalMaterial} We also see that the values of $ZT$ for the examined molecules are not large, which is, however, mainly due to the electronic properties.
From this study, we have a clear idea of how we can tune the phonon contribution to the thermal conductance. However, it remains an interesting challenge for future studies to find $\pi$-stacked molecular compounds with electronic properties which are more favorable for thermoelectric devices.

One possibility of increasing $ZT$ is to reduce the electronic coupling to the leads, which can be done without significantly altering the mechanical coupling by, for example, changing the binding group from a thiol (chemisorbed) to some physisorbed system. It has been shown~\cite{Mahan1996,Humphrey2005,Murphy2008,Esposito2009} that in the limit of a small electronic coupling, $ZT$ becomes infinite if the molecular resonances can be correctly positioned relative to the lead Fermi level \emph{and if $\kappa_{\mrp}$ is neglected}. Therefore, the efficiency is limited only by the phonon heat current and our mechanism can lead to a significant improvement.
To illustrate this, we show in \figurename~\ref{fig4}b the maximal $ZT$ value as a function of the coupling strength, $ZT_m (\Gamma) = \mathrm{max}_{\ve_d} [ZT(\Gamma)]$, when the electron transmission has the form of a spinful resonant level, i.e. $\m{T}_{\mre}(E)=\Gamma^2/[(E-\ve_{d})^2+\Gamma^2]$, where $\ve_{d}$ is the position of the level and $\Gamma$ denotes the coupling strength to the leads. We note that including Coulomb interactions in the calculations would have a rather small effect on the result, as long as the associated energy scale is much smaller or larger than $k_{\mathrm{B}}T$ and $\Gamma$.~\cite{Leijnse2010}

In conclusion, we have proposed a simple mechanism to control the heat transport which occurs due to center of mass modes in molecular junctions. The idea is based on vibrationally decoupling the left lead from the right lead. As an example we examined $\pi$-stacked molecules, showing that the proposed mechanism can indeed significantly reduce the phonon conductance, but also increase it depending on the specific molecules and stackings used. For the molecules investigated here, we found a maximal decrease by 35\% and a maximal increase by 59\%, but the basic mechanism is general and it is an interesting challenge for future first principle calculations and experiments to search for molecular compounds where the phonon heat conductance is reduced or increased even further, preferably in combination with electronic properties which are favorable for applications in, for example, thermoelectric devices.

While we have focused our attention on the thermoelectric response of these systems, one can envisage situations where maximising the phononic heat current is also desirable, for example to maintain thermal equilibrium across a device, allow heat to dissipate more efficiently, etc. In these instances, using $\pi$-stacked systems to increase the heat current could be very useful. The increased heat current we observe is trivial to understand when we consider that we are increasing the number of center-of-mass modes; however, these systems present an interesting challenge to our intuition for heat transport. We have the clear, but somewhat paradoxical, result that we should break chemical bonds within a molecule in order to maximise the phononic heat transport in a molecular junction.

\begin{acknowledgments}
The research was carried out in the Danish-Chinese Centre for Molecular Nano-Electronics, and the Center for Quantum Devices, which is funded by the Danish National Research Foundation. M. L. acknowledges support from the Swedish Research Council (VR). G. C. S. and Q. L. were supported through funding from the European Research Council under the European Union's Seventh Framework Program (FP7/2007- 2013)/ERC Grant Agreement no. 258806 and The Danish Council for Independent Research Natural Sciences.
\end{acknowledgments}

\newpage\newpage

\renewcommand{\theequation}{S.\arabic{equation}}
\renewcommand{\thefigure}{S.\arabic{figure}}
\renewcommand{\thetable}{S.\arabic{table}}
\newcommand{\sectionname}{Section}

\newcommand{\e}{\mathrm{e}}
\renewcommand{\i}{\mathrm{i}}
\newcommand{\vphi}{\varphi}
\newcommand{\pdd}{\mathrm{d}}

\newcommand\abs[1]{\lvert#1\rvert}
\newcommand\absB[1]{\left\lvert#1\right\rvert}
\newcommand\avg[1]{\left\langle#1\right\rangle}
\newcommand\avgs[1]{\langle#1\rangle}
\newcommand{\Real}{\mathrm{Re}}
\newcommand{\Imag}{\mathrm{Im}}
\newcommand{\Realp}{\mathrm{Re}\hspace{0.025cm}}
\newcommand{\Imagp}{\mathrm{Im}\hspace{0.025cm}}

\newcommand{\bff}{\mathbf{f}}
\newcommand{\bk}{\mathbf{k}}
\newcommand{\bp}{\mathbf{p}}
\newcommand{\bv}{\mathbf{v}}
\newcommand{\bF}{\mathbf{F}}

\newcommand{\mrK}{\mathrm{K}}

\newcommand{\ad}{a^{\dag}}
\newcommand{\aan}{a^{\phantom{\dag}}}

\newcommand{\bpsi}{\boldsymbol{\psi}}
\newcommand{\bpi}{\boldsymbol{\pi}}
\newcommand\lbi{i}
\newcommand\lbl{l}
\newcommand\lbm{m}
\newcommand\lbn{n}
\newcommand\lbbeta{\beta}

\newcommand{\tr}{\tilde{a}}
\newcommand{\ti}{\tilde{b}}
\newcommand{\rr}{a}
\newcommand{\ii}{b}

\newcommand{\rFigOne}{1}
\newcommand\rmbr[1]{}

\onecolumngrid

\begin{center}
\textbf{\Large Supplemental Material \\ Designing $\pi$-stacked molecular structures to control heat transport through molecular junctions}
\end{center}

\setcounter{equation}{0}
\setcounter{figure}{0}
\setcounter{table}{0}
\setcounter{page}{1}
\renewcommand{\theequation}{S\arabic{equation}}
\renewcommand{\thefigure}{S\arabic{figure}}
\renewcommand{\bibnumfmt}[1]{[S#1]}
\renewcommand{\citenumfont}[1]{S#1}

\section{\label{sec:mdl}Model}

The system under consideration is a junction consisting of two leads described by vibrational modes of continuum and a molecule (middle region) described by harmonic oscillators. The Hamiltonian for the vibrational degrees of the system is
\begin{subequations}\label{ham}
\begin{equation}\label{hamP}
H=H_{0}+V,
\end{equation}
\begin{equation}\label{ham0}
H_{0}=\sum_{\lbi,\lbm}\frac{p_{\lbi\lbm}^2}{2M_{\lbm}}+H_{\mrL}+H_{\mrR}, \quad \lbi=x,y,z, \quad \lbm=1,2,
\end{equation}
\begin{equation}
H_{\alpha}=\sum_{\nu}\hbar\omega_{\alpha\nu}\ad_{\alpha\nu}\aan_{\alpha\nu}, \quad \alpha=\mrL,\mrR,
\end{equation}
\begin{equation}
V=\frac{1}{4}\sum_{\lbi\lbi'\lbm\lbm'}K_{\lbi\lbm,\lbi'\lbm'} (r_{\lbi\lbm}-r_{\lbi'\lbm'})^2+\frac{1}{2}\sum_{\lbi\lbi'\lbm\alpha \br}K_{\lbi\lbm,\lbi'\alpha}(\br)[r_{\lbi\lbm}-u_{\lbi'\alpha}(\br)]^2,
\end{equation}
\end{subequations}
where $\br_{m}=(x_m,y_m,z_m)$ are the coordinates of the mass $M_{m}$; $\bp_{m}$ is the corresponding momentum; $u_{\lbi'\alpha}(\br)\equiv u_{\lbi'\alpha}(\br,0)$ is the displacement vector in the lead $\alpha$ at time $t=0$; $K_{\lbi\lbm,\lbi'\lbm'}$ are the spring constants between masses and $K_{\lbi\lbm,\lbi'\alpha}(\br)$ are the couplings to the leads. The operator $\ad_{\alpha\nu}$ creates an excitation in the vibrational mode $\nu$ in the lead $\alpha$ with energy $\hbar\omega_{\alpha\nu}$, and it satisfies the canonical commutation relation $[\aan_{\alpha\nu},\ad_{\alpha'\nu'}]=\delta_{\alpha\alpha'}\delta_{\nu\nu'}$. The description of elastic continuum modes and quantization of displacement vector $\bu_{\alpha}(\br,t)$ is given in \sectionname~\ref{sec:emotl}. For the system depicted in \figurename~\rFigOne{a} we have the following non-zero couplings
\begin{equation}\label{twmsc}
K_{\lbi1,\lbi2}=K_{\lbi2,\lbi1}\equiv K_{\lbi}, \quad
K_{\lbi1,\lbi L}(\br)\equiv K_{\lbi L}\delta(\br-\br_{L}), \quad
K_{\lbi2,\lbi R}(\br)\equiv K_{\lbi R}\delta(\br-\br_{R}), \quad
\end{equation}
where $\br_{\alpha}\equiv (x_{\alpha},y_{\alpha},z_{\alpha})$ denotes points of attachment of the molecules to the leads.

We transform the coordinates into mass weighted coordinates, i.e.,
\begin{equation}\label{mwcoord}
\begin{aligned}
&r_{\lbn}\quad\rightarrow\quad \frac{r_{\lbn}}{\sqrt{M_{\lbn}}},
\quad p_{\lbn}\quad\rightarrow\quad p_{\lbn}\sqrt{M_{\lbn}},\\
&u_{\beta}\rmbr{(\br)}\quad\rightarrow\quad \frac{u_{\beta}\rmbr{(\br)}}{\sqrt{M_{\beta}}},
\quad \pi_{\beta}\rmbr{(\br)}\quad\rightarrow\quad \pi_{\beta}\rmbr{(\br)}\sqrt{M_{\beta}},
\end{aligned}
\end{equation}
where we have introduced the following shorthand notation $\lbn\equiv\lbi,\lbm$, $\lbbeta\equiv\lbi',\alpha,\br$ for the labels and $M_{\lbbeta}$ denotes the mass of the lead atoms. Then the Hamiltonian (\ref{ham}d) is rewritten in mass weighted coordinates as
\begin{equation}\label{hamV}
V=\frac{1}{2}\sum_{\lbn\lbn'}V_{\lbn\lbn'}r_{\lbn}r_{\lbn'}+\frac{1}{2}\sum_{\lbbeta}V_{\lbbeta}\rmbr{(\br)}u_{\lbbeta}^2\rmbr{(\br)}
+\sum_{\lbn\lbbeta}V_{\lbn\lbbeta}\rmbr{(\br)}r_{\lbn}u_{\lbbeta}\rmbr{(\br)},
\end{equation}
\begin{equation}
\begin{aligned}
&V_{\lbn\lbn}=\frac{1}{M_{\lbn}}\Big(\sum_{\lbn'}K_{\lbn\lbn'}+\sum_{\beta}K_{\lbn\lbbeta}\Big), \quad
V_{\lbn\lbn'}=-\frac{K_{\lbn\lbn'}}{\sqrt{M_{\lbn}M_{\lbn'}}}, \quad \lbn\neq \lbn',\\
&V_{\lbbeta}\rmbr{(\br)}=\frac{1}{M_{\lbbeta}}\sum_{\lbn'}K_{\lbn'\lbbeta}\rmbr{(\br)}, \quad
V_{\lbn\lbbeta}\rmbr{(\br)}=-\frac{K_{\lbn\lbbeta}\rmbr{(\br)}}{\sqrt{M_{\lbn}M_{\lbbeta}}},
\end{aligned}
\end{equation}
which is more convenient for the calculation of the current described in \sectionname~\ref{sec:hcpc}.

\section{\label{sec:emotl}Eigenmodes of the leads}

In this section we describe the eigenmodes of the leads, when there is no coupling to the molecule, where we closely follow Ezawa.~\cite{S_Ezawa1971,S_Patton2001} The equation of motion for an elastic medium is given by~\cite{S_Landau1986}
\begin{equation}\label{ewitts}
\rho\frac{\pd^2 u_i}{\pd t^2}=\frac{\pd \sigma_{ij}}{\pd x_j},
\end{equation}
where $\rho$ is the mass density of the medium, $u_i=u_i(\br,t)$ is a displacement vector component in the $i$ direction, $x_j\in\{x,y,z\}$ denotes Cartesian coordinate, and $\sigma_{ij}$ is a stress tensor given by
\begin{equation}\label{stote}
\frac{\sigma_{ij}}{\rho}=(c_l^2-2c_t^2)\frac{\pd u_k}{\pd x_k}\delta_{ij}+c_t^2\left(\frac{\pd u_i}{\pd x_j}+\frac{\pd u_j}{\pd x_i}\right)
\end{equation}
for an isotropic medium. Here $c_l$ is the velocity of the longitudinal wave and $c_t$ is the velocity of the transverse wave. Note that the Einstein summation convention over all repeated indices is implied. Equation (\ref{ewitts}) with stress tensor (\ref{stote}) can be conveniently rewritten as
\begin{equation}\label{ewefu}
\begin{aligned}
\frac{\pd^2 \bu}{\pd t^2}=
c_t^2\nabla^2\bu+(c_l^2-c_t^2)\nabla(\nabla\cdot\bu),
\end{aligned}
\end{equation}
where $\bu=(u_{x},u_{y},u_{z})$. Using Helmholtz decomposition we can represent the vector $\bu$ as a sum of two components $\bu=\bu_l+\bu_t$,
which satisfy the conditions $\nabla\times\bu_l=0$, $\nabla\cdot\bu_t=0$,
i.e.,  $\bu_l$ is an irrotational vector field and $\bu_t$ is a solenoidal vector field,
and satisfy the wave equation $\left(\frac{\pd^2}{\pd t^2}-c_{l/t}^2\nabla^2\right)\bu_{l/t}=0$.
The displacement vector also can be written in terms of the scalar potential $\phi$ and the vector potential $\bpsi$, i.e.,
$\bu_l=\nabla\phi$,
$\bu_t=\nabla\times\bpsi$, with $\nabla\cdot\bpsi=0$.
The potentials also satisfy the wave equations
\begin{subequations}\label{eomftppp}
\begin{align}
&\left(\frac{\pd^2}{\pd t^2}-c_t^2\nabla^2\right)\phi=0,\\
&\left(\frac{\pd^2}{\pd t^2}-c_l^2\nabla^2\right)\bpsi=0, \quad \text{with} \quad \nabla\cdot\bpsi=0.
\end{align}
\end{subequations}

We want to find the eigenmodes of a half space $\zz\leq0$ filled with an isotropic elastic medium, which has a stress-free surface. For every point of a stress-free surface we have to satisfy the boundary condition
\begin{equation}\label{ewefubc}
\sigma_{ij}n_j=0,
\end{equation}
where $\bn$ is the outward normal at each point of the surface. For a half space $\zz\leq0$ we have
$\bn=\{0,1,0\}$.
So we need to solve Eq. (\ref{ewefu}) with the boundary condition (\ref{ewefubc}). We start by choosing the following ansatz for the displacement vector
\begin{equation}\label{afdv}
\bu(\br,t)=\frac{1}{2\pi}\bff(\zz)\e^{\i\left(k_\xx\xx+k_\yy\yy-\omega t\right)}, \quad \bff=\{f_x,f_y,f_z\},
\quad [\bu(\br,t)]=[\bff(z)]=L,
\end{equation}
where $[A]$  denotes the dimension of the quantity $A$, and dimensions $L$, $T$, and $M$ stands for length, time, and mass dimension, respectively.
For definiteness we assume $\omega\geq0$. Inserting (\ref{afdv}) into Eq. (\ref{ewefu}) and the boundary condition (\ref{ewefubc}) we obtain
\begin{equation}\label{eomwual}
\left[\m{L}_0+\m{L}_1\left(-\i\frac{\pdd}{\pdd \zz}\right)
+\m{L}_2\left(-\i\frac{\pdd}{\pdd \zz}\right)^2\right]\bff(\zz)
=\omega^2 \bff(\zz),
\end{equation}
\begin{equation}\label{bcwual}
\left[\m{N}_0+\m{N}_1\left(-\i\frac{\pdd}{\pdd \zz}\right)\right]\bff(\zz)=0, \quad \text{for } \zz=0,
\end{equation}
where
\begin{subequations}
\begin{align}
&\m{L}_0=
\begin{pmatrix}
c_l^2k_{\xx}^2+c_t^2k_{\yy}^2 & (c_l^2-c_t^2)k_{\xx}k_{\yy} & 0 \\
(c_l^2-c_t^2)k_{\xx}k_{\yy}   & c_l^2k_{\yy}^2+c_t^2k_{\xx}^2 & 0 \\
0 & 0 & c_t^2(k_{\xx}^2+k_{\yy}^2)
\end{pmatrix},\\
&\m{L}_1=(c_l^2-c_t^2)
\begin{pmatrix}
0 & 0 & k_{\xx} \\
0 & 0 & k_{\yy} \\
k_{\xx} & k_{\yy} & 0
\end{pmatrix},\\
&\m{L}_2=\m{N}_1=
\begin{pmatrix}
c_t^2 & 0 & 0\\
0 & c_t^2 & 0\\
0 & 0 & c_l^2
\end{pmatrix},\\
&\m{N}_0=
\begin{pmatrix}
0 & 0 & c_t^2k_{\xx} \\
0 & 0 & c_t^2k_{\yy} \\
(c_l^2-2c_t^2)k_{\xx} & (c_l^2-2c_t^2)k_{\yy} & 0
\end{pmatrix}.
\end{align}
\end{subequations}

We anticipate that the eigenmodes, which we label as $\nu$, will be classified using the following quantities (``quantum numbers'')
\begin{equation}
\begin{aligned}
\nu=(\bk,k_{\zz},m), \quad \bk=\{k_{\xx},k_{\yy}\}.
\end{aligned}
\end{equation}
where $\bk$ is the wavevector in the $\xx\yy$ plane, $k_{\zz}$ is the wavevector perpendicular to the surface, and $m\in\{H,\pm,0,R\}$ labels the type of the mode, where there will be four types as discussed in forthcoming sections. The solutions will be normalized in the following way:
\begin{equation}\label{normcd}
\langle \bu^{(\nu')}, \bu^{(\nu)}\rangle=[\chi_{\nu}]\int \dif{\br}\bu^{\dag(\nu')}(\br)\bu^{(\nu)}(\br)=\delta_{\nu'\nu}[\chi_{\nu}],
\quad [\langle\bu^{(\nu')}, \bu^{(\nu)}\rangle]=L^5,
\end{equation}
where $\delta_{\nu'\nu}$ denotes Dirac delta function for continuous variables and Kroenecker delta for discrete variables.
The quantity $[\chi_{\nu}]$ denotes a particular dimension depending on whether $\nu$ contains discrete or continuous labels. Note that $[\bu({\br},t)]\neq [\bu^{(\nu)}(\br)]$.
The normalization condition (\ref{normcd}) can be written more explicitly as
\begin{equation}
\begin{aligned}
\langle \bu^{(\bk',k_\zz',m')}, \bu^{(\bk,k_\zz,m)}\rangle
&=[\chi_{m}]\delta_{\bk',\bk}\int_{-\infty}^{0}\dif{\zz}\bff^{\dag(\bk,k_\zz',m')}(\zz)\bff^{(\bk,k_\zz,m)}(\zz)
=\delta_{\bk',\bk}\delta_{k_\zz',k_\zz}\delta_{m',m}[\chi_{m}],
\end{aligned}
\end{equation}
and we get
\begin{equation}\label{ncwoz}
(\bff^{(\bk,k_\zz',m')},\bff^{(\bk,k_\zz,m)})=\int_{-\infty}^{0}\dif{\zz}\bff^{\dag(\bk,k_\zz',m')}(\zz)\bff^{(\bk,k_\zz,m)}(\zz)=\delta_{k_\zz',k_\zz}\delta_{m',m}.
\end{equation}
Here
\begin{equation}\label{unufnu}
\bu^{(\nu)}(\br)=\bu^{(\bk,k_\zz,m)}(\br)=\frac{1}{2\pi}\bff^{(\bk,k_\zz,m)}(\zz)\e^{\i\left(k_\xx\xx+k_\yy\yy\right)}, \quad \bk=\{k_\xx,k_\yy\},
\end{equation}
and $\langle A,B \rangle$ denotes inner product in the whole space, and $( A,B )$ in the $\zz$ direction. We have the following dimensions of $\bu^{(\bk,k_\zz,m)}$, $\bff^{(\bk,k_\zz,m)}$, and $[\chi_{m}]$:
\begin{equation}
\begin{aligned}{}
&[\bu^{(\bk,k_\zz,m)}]=[\bff^{(\bk,k_\zz,m)}]=1, \quad [\chi_{m}]=L^2 \quad \text{for} \quad m=H,\pm,0, \\
&[\bu^{(\bk,k_{R},R)}]=[\bff^{(\bk,k_{R},R)}]=L^{-1/2}, \quad [\chi_{R}]=L^3, \quad \text{for} \quad m=R,
\end{aligned}
\end{equation}
where the first line is the dimensions for the modes with continuous $k_\zz$ and the second line for the modes with discrete $k_\zz$.
For some of the derivations we will use the potentials $\phi$ and $\bpsi$. In analogy with (\ref{afdv}) we use the following ansatz
\begin{subequations}
\begin{align}
&\phi(\br,t)=\frac{1}{2\pi}f_{\phi}(\zz)\e^{\i(k_\xx\xx+k_\yy\yy-\omega t)},\\
&\bpsi(\br,t)=\frac{1}{2\pi}\bff_{\bpsi}(\zz)\e^{\i(k_\xx\xx+k_\yy\yy-\omega t)},
\end{align}
\end{subequations}
and we find that $\bff$ is expressed in terms of $f_{\phi}$ and $\bff_{\bpsi}$ as
\begin{subequations}\label{dvioppp}
\begin{equation}
f_\xx=\i k_\xx f_{\phi}+\i k_\yy f_{\psi_\zz}-\frac{\pdd}{\pdd{\zz}}f_{\psi_\yy},
\end{equation}
\begin{equation}
f_\yy=\i k_\yy f_{\phi}+\frac{\pdd}{\pdd{\zz}}f_{\psi_\xx}-\i k_\xx f_{\psi_\zz},
\end{equation}
\begin{equation}
f_\zz=\frac{\pdd}{\pdd{\zz}}f_{\phi}+\i k_\xx f_{\psi_\yy}-\i k_\yy f_{\psi_\xx}.
\end{equation}
\end{subequations}
Inserting the potentials into the wave equations (\ref{eomftppp}) we obtain
\begin{subequations}\label{ceomfppp}
\begin{align}
c_l^2\left[k_\xx^2+k_\yy^2-\frac{\pdd^2}{\pdd \zz^2}\right]f_{\phi}&=\omega^2 f_{\phi},\\
c_t^2\left[k_\xx^2+k_\yy^2-\frac{\pdd^2}{\pdd \zz^2}\right]\bff_{\bpsi}&=\omega^2 \bff_{\bpsi},\\
\i k_\xx f_{\psi_\xx}+\i k_\yy f_{\psi_\yy}+\frac{\pdd}{\pdd \zz}f_{\psi_\zz}&=0,
\end{align}
\end{subequations}
and from the boundary condition (\ref{bcwual}) we get
\begin{equation}\label{cbcfppp}
\left[\m{K}_0+\m{K}_1\left(-\i\frac{\pdd}{\pdd \zz}\right)+\m{K}_2\left(-\i\frac{\pdd}{\pdd \zz}\right)^2\right]\bF=0, \quad \text{for } \zz=0,
\quad \text{with} \quad \bF=\{f_{\phi},f_{\psi_x},f_{\psi_y},f_{\psi_z}\},
\end{equation}
where
\begin{subequations}
\begin{align}
\m{K}_0=&\begin{pmatrix}
0 & -c_t^2k_{\xx}k_{\yy} & c_t^2k_{\xx}^2 & 0 \\
0 & -c_t^2k_{\yy}^2 & c_t^2k_{\xx}k_{\yy} & 0 \\
(c_l^2-2c_t^2)(k_{\xx}^2+k_{\yy}^2) & 0 & 0 & 0
\end{pmatrix},\\
\m{K}_1=&\begin{pmatrix}
2c_t^2k_{\xx} & 0 & 0 & c_t^2k_{\yy} \\
2c_t^2k_{\yy} & 0 & 0  & -c_t^2k_{\xx} \\
0 & -2c_t^2k_{\yy} & 2c_t^2k_{\xx} & 0
\end{pmatrix},\\
\m{K}_2=&\begin{pmatrix}
0 & 0 & -c_t^2 & 0 \\
0 & c_t^2 & 0 & 0 \\
c_l^2 & 0 & 0 & 0
\end{pmatrix}.
\end{align}
\end{subequations}
We will perform calculations and will find the modes, where $\xx$ and $\yy$ axes are rotated such that the vector $\bk$ gets transformed into
$\bk=\{k_\xx,k_\yy\} \rightarrow \{\kappa,0\}$, with $\kappa=\sqrt{k_\xx^2+k_\yy^2}$.

\subsection{$SH$-mode, $m=H$}

For this mode we set $f_{\xx}=f_{\zz}=0$, and then the wave is polarized both to the direction of $\bk=\{\kappa,0\}$ and the ${\zz}$-axis, i.e., we get a shear wave with horizontal polarization ($SH$-mode), which we label as $m=H$.
For this polarization the equation of motion (\ref{eomwual}) and the boundary condition (\ref{bcwual}) are
\begin{equation}
\label{sodefshm}\left(c_t^2\frac{\dif^2}{\dif{\zz}^2}+\omega^2-c_t^2\kappa^2\right)f_{\yy}=0, \quad
\frac{\dif f_{\yy}}{\dif{\zz}}\bigg\vert_{{\zz}=0}=0.
\end{equation}
The solution to (\ref{sodefshm}), which is normalized according to the condition (\ref{ncwoz}) is
\begin{equation}
\boxed{
f_{\xx}^{(\kappa,k_{\beta},H)}=f_{\zz}^{(\kappa,k_{\beta},H)}=0, \quad f_{\yy}^{(\kappa,k_{\beta},H)}=\sqrt{\frac{2}{\pi}}\cos(k_{\beta} {\zz}),
\quad m=H,}
\end{equation}
and it has the eigenfrequency $\omega=c_t\sqrt{k_{\beta}^2+\kappa^2}$.

\subsection{Mixed $P-SV$ mode, $m=\pm$}
We proceed with the calculation using the potentials $\phi$ and $\bpsi$, and the corresponding equations of motion (\ref{ceomfppp}). In the coordinate system where $\bk=\{\kappa,0\}$ we get that $f_{\phi}$ with $f_{\psi_{\yy}}$ gets decoupled from $f_{\psi_{\xx}}$ with $f_{\psi_{\zz}}$, i.e., the equation of motion (\ref{ceomfppp}) and the boundary condition (\ref{cbcfppp}) simplifies to
\begin{subequations}\label{psoshm}
\begin{align}
c_t^2\left[\kappa^2-\frac{\pdd^2}{\pdd {\zz}^2}\right]f_{\psi_{{\xx},{\zz}}}&=\omega^2 f_{\psi_{{\xx},{\zz}}},\\
\i \kappa f_{\psi_{\xx}}+\frac{\pdd}{\pdd {\zz}}f_{\psi_{\zz}}&=0,\\
\i \kappa \frac{\pdd}{\pdd{\zz}}f_{\psi_{\zz}}-\frac{\pdd^2}{\pdd{\zz}^2}f_{\psi_{\xx}}&=0, \quad \text{at } {\zz}=0.
\end{align}
\end{subequations}
and
\begin{subequations}\label{sfom}
\begin{align}
c_l^2\left[\kappa^2-\frac{\pdd^2}{\pdd {\zz}^2}\right]f_{\phi}&=\omega^2 f_{\phi},\\
c_t^2\left[\kappa^2-\frac{\pdd^2}{\pdd {\zz}^2}\right]f_{\psi_{\yy}}&=\omega^2 f_{\psi_{\yy}},\\
\kappa^2 f_{\psi_{\yy}}-2\i\kappa\frac{\pdd}{\pdd{\zz}}f_{\phi}+\frac{\pdd^2}{\pdd{\zz}^2}f_{\psi_{\yy}}&=0,\quad \text{at } {\zz}=0,\\
(c_l^2-2c_t^2)\kappa^2 f_{\phi}-2c_t^2\i\kappa\frac{\pdd}{\pdd{\zz}}f_{\psi_{\yy}}-c_l^2\frac{\pdd^2}{\pdd{\zz}^2}f_{\phi}&=0,
\quad \text{at } {\zz}=0.
\end{align}
\end{subequations}
The solution of Eqs. (\ref{psoshm}) corresponds to the $SH$-mode when $f_{\phi}=f_{\psi_{\yy}}=0\ \rightarrow \ f_{\xx}=f_{\zz}=0$. For the mixed $P-SV$ mode (as well as for the modes in the next sections) we set $f_{\yy}=0 \ \rightarrow \ f_{\psi_{\xx}}=f_{\psi_{\zz}}=0$, and use the following ansatz for $f_{\phi}$ and $f_{\psi_{\yy}}$
\begin{subequations}\label{sfweppy}
\begin{align}
f_{\phi}&=A\e^{\i k_{\alpha} {\zz}}+B\e^{-\i k_{\alpha} {\zz}},\\
f_{\psi_{\yy}}&=C\e^{\i k_{\beta} {\zz}}+D\e^{-\i k_{\beta} {\zz}},
\end{align}
\end{subequations}
Inserting these solutions into Eqs. (\ref{sfom}a,b) we get
$\omega_{\alpha}=c_{l}\sqrt{k_{\alpha}^2+\kappa^2}$,
$\omega_{\beta}=c_{t}\sqrt{k_{\beta}^2+\kappa^2}$,
and the boundary condition (\ref{sfom}c,d) gives
\begin{subequations}\label{eqfabcd}
\begin{align}
-2\kappa k_{\alpha}[A-B]+(k_{\beta}^2-\kappa^2)[C+D]=0,\\
(k_{\beta}^2-\kappa^2)[A+B]+2\kappa k_{\beta}[C-D]=0.
\end{align}
\end{subequations}
After requiring that $\omega_{\alpha}=\omega_{\beta}$ we obtain
\begin{equation}
k_{\alpha}=\frac{c_t}{c_l}\sqrt{k_{\beta}^2+\kappa^2\left[1-\frac{c_l^2}{c_t^2}\right]}.
\end{equation}
There are two independent solutions for the above system of equations (\ref{eqfabcd}). We note that in this case the wave coming from $\phi$ is called a $P$-wave (pressure wave) and the one from $\bpsi$ a $SV$-wave (shear wave with vertical polarization). Also for a mixed $P-SV$ wave we have
\begin{align}\label{rovfmpsvm}
k_{\alpha}^2, \ k_{\beta}^2>0, \quad k_{\alpha}, \ k_{\beta}>0, \quad \rightarrow \quad k_{\beta}>\kappa\sqrt{\frac{c_l^2}{c_t^2}-1}.
\end{align}

As a first solution we pick a $P$-wave incident on the surface $A=1$, $C=0$, which from Eqs. (\ref{eqfabcd}) gives
\begin{subequations}
\begin{align}
f_{\phi}^{(\kappa,c,P)}&=\e^{\i k_{\alpha} {\zz}}-a\e^{-\i k_{\alpha} {\zz}},
\quad a=\frac{(k_{\beta}^2-\kappa^2)^2-4\kappa^2k_\alpha k_\beta}{(k_{\beta}^2-\kappa^2)^2+4\kappa^2k_\alpha k_\beta},\\
f_{\psi_{\yy}}^{(\kappa,c,P)}&=\sqrt{\frac{k_\alpha}{k_\beta}}b\e^{-\i k_{\beta} {\zz}},
\quad b=\frac{4\kappa\sqrt{k_\alpha k_\beta}(k_{\beta}^2-\kappa^2)}{(k_{\beta}^2-\kappa^2)^2+4\kappa^2k_\alpha k_\beta}.
\end{align}
\end{subequations}
Using the relations (\ref{dvioppp}) and the normalization condition (\ref{ncwoz}) we find the corresponding displacement vector $\bff({\zz})$ for a $P$-wave
\begin{equation}\label{pvdwff}
\begin{aligned}
f_{\xx}^{(\kappa,c,P)}&=\i\sqrt{\frac{k_{\beta}}{2\pi k_{\alpha}(1+k_{\beta}^2/\kappa^2)}}\left[
\e^{\i k_{\alpha} {\zz}}-a\e^{-\i k_{\alpha} {\zz}}
+\sqrt{\frac{k_{\alpha}k_{\beta}}{\kappa^2}}b\e^{-\i k_{\beta} {\zz}}\right],\\
f_{\yy}^{(\kappa,c,P)}&=0,\\
f_{\zz}^{(\kappa,c,P)}&=\i\sqrt{\frac{k_{\beta}}{2\pi k_{\alpha}(1+k_{\beta}^2/\kappa^2)}}\left[
\frac{k_{\alpha}}{\kappa}\left(\e^{\i k_{\alpha} {\zz}}+a\e^{-\i k_{\alpha} {\zz}}\right)
+\sqrt{\frac{k_{\alpha}}{k_{\beta}}}b\e^{-\i k_{\beta} {\zz}}\right], \quad m=P.
\end{aligned}
\end{equation}

When a $SV$-wave is the incident wave we have $A=0$, $C=1$,
and Eqs. (\ref{eqfabcd}) yield
\begin{subequations}
\begin{align}
f_{\phi}^{(\kappa,c,SV)}&=-\sqrt{\frac{k_{\beta}}{k_{\alpha}}}b\e^{-\i k_{\alpha} {\zz}},\\
f_{\psi_{\yy}}^{(\kappa,c,SV)}&=\e^{\i k_{\beta} {\zz}}-a\e^{-\i k_{\beta} {\zz}},
\end{align}
\end{subequations}
and
\begin{equation}\label{svvdwff}
\begin{aligned}
f_{\xx}^{(\kappa,c,SV)}&=\i\frac{1}{\sqrt{2\pi(1+k_{\beta}^2/\kappa^2)}}\left[
-\sqrt{\frac{k_{\beta}}{k_{\alpha}}}b\e^{-\i k_{\alpha} {\zz}}
-\frac{k_\beta}{\kappa}\left(\e^{\i k_{\beta} {\zz}}+a\e^{-\i k_{\beta} {\zz}}\right)\right],\\
f_{\yy}^{(\kappa,c,SV)}&=0,\\
f_{\zz}^{(\kappa,c,SV)}&=\i\frac{1}{\sqrt{2\pi(1+k_{\beta}^2/\kappa^2)}}\left[
\sqrt{\frac{k_{\alpha}k_{\beta}}{\kappa^2}}b\e^{-\i k_{\alpha} {\zz}}
+\e^{\i k_{\beta} {\zz}}-a\e^{-\i k_{\beta} {\zz}}\right], \quad m=SV.
\end{aligned}
\end{equation}

Having the $P$-wave and the $SV$-wave we will construct a so called mixed $P-SV$ wave in the following way
\begin{equation}
f_{\phi,\psi_{\yy}}^{(\kappa,c,\pm)}=\mp\i\sqrt{\frac{\kappa}{k_{\alpha}}}f_{\phi,\psi_{\yy}}^{(\kappa,c,P)}+\sqrt{\frac{\kappa}{k_{\beta}}}f_{\phi,\psi_{\yy}}^{(\kappa,c,SV)},
\end{equation}
which gives the following normalized displacement vector $\bff^{\pm}$
\begin{equation}
\bff^{(\kappa,c,\pm)}=\frac{1}{\sqrt{2}}\left[\mp\i\bff^{(\kappa,c,P)}+\bff^{(\kappa,c,SV)}\right],
\end{equation}
or more explicitly
\begin{equation}
\boxed{
\begin{aligned}
f_{\xx}^{(\kappa,c,\pm)}&=\frac{1}{\sqrt{4\pi(1+k_{\beta}^2/\kappa^2)}}
\left[\pm\sqrt{\frac{k_{\beta}}{k_\alpha}}\left(\e^{\i k_{\alpha} {\zz}}-\zeta_{\pm}\e^{-\i k_{\alpha} {\zz}}\right)
-\i\frac{k_{\beta}}{\kappa}\left(\e^{\i k_{\beta} {\zz}}+\zeta_{\pm}\e^{-\i k_{\beta} {\zz}}\right)\right],\\
f_{\yy}^{(\kappa,c,\pm)}&=0,\\
f_{\zz}^{(\kappa,c,\pm)}&=\frac{1}{\sqrt{4\pi (1+k_{\beta}^2/\kappa^2)}}
\left[\pm\sqrt{\frac{k_{\alpha}k_{\beta}}{\kappa^2}}\left(\e^{\i k_{\alpha} {\zz}}+\zeta_{\pm}\e^{-\i k_{\alpha} {\zz}}\right)
+\i\left(\e^{\i k_{\beta} {\zz}}-\zeta_{\pm}\e^{-\i k_{\beta} {\zz}}\right)\right],
\quad m=\pm,
\end{aligned}}
\end{equation}
where
\begin{equation}
\zeta_{\pm}=a\pm\i b, \quad \abs{\zeta_{\pm}}^2=a^2+b^2=1.
\end{equation}

\subsection{Mode with total reflection, $m=0$}

In this section we examine the modes which have
\begin{equation}\label{rovfmtrm}
k_{\alpha}^2<0, \ k_{\beta}^2>0 \quad \rightarrow \quad 0<k_{\beta}<\kappa\sqrt{\frac{c_l^2}{c_t^2}-1},
\end{equation}
and define
\begin{equation}
k_{\alpha}=\i k_{\gamma}=\i\frac{c_t}{c_l}\sqrt{\kappa^2\left[\frac{c_l^2}{c_t^2}-1\right]-k_{\beta}^2}, \quad 0<k_{\gamma}<\kappa\sqrt{1-\frac{c_t^2}{c_l^2}}.
\end{equation}
From the solution (\ref{sfweppy}) we see that we need to set $A=0$, because otherwise we would get exponentially increasing solution as ${\zz}\rightarrow-\infty$, which is unphysical. So we start with incident $SV$-wave $A=0$, $C=1$, and from Eqs. (\ref{eqfabcd}) we find the following potentials
\begin{subequations}
\begin{align}
f_{\phi}^{(\kappa,c,0)}&=-b\e^{k_{\gamma} {\zz}}, \quad b=\frac{4\kappa k_{\beta}(k_{\beta}^2-\kappa^2)}{(k_{\beta}^2-\kappa^2)^2+4\i \kappa^2k_{\gamma}k_{\beta}},\\
f_{\psi_{\yy}}^{(\kappa,c,0)}&=\e^{\i k_{\beta} {\zz}}-a\e^{-\i k_{\beta} {\zz}}, \quad a=\frac{(k_{\beta}^2-\kappa^2)^2-4\i \kappa^2 k_{\gamma}k_{\beta}}{(k_{\beta}^2-\kappa^2)^2+4\i \kappa^2k_{\gamma}k_{\beta}},
\end{align}
\end{subequations}
and such a the displacement vector for a mode with total reflection
\begin{equation}\label{svvdwff2}
\boxed{
\begin{aligned}
f_{\xx}^{(\kappa,c,0)}&=-\i \frac{1}{\sqrt{2\pi(1+k_{\beta}^2/\kappa^2)}}\left[
b\e^{k_{\gamma} {\zz}}
+\frac{k_{\beta}}{\kappa}\left(\e^{\i k_{\beta} {\zz}}+a\e^{-\i k_{\beta} {\zz}}\right)\right],\\
f_{\yy}^{(\kappa,c,0)}&=0,\\
f_{\zz}^{(\kappa,c,0)}&=\frac{1}{\sqrt{2\pi(1+k_{\beta}^2/\kappa^2)}}\left[
-\frac{k_{\gamma}}{\kappa} b\e^{k_{\gamma} {\zz}}
+\i\left(\e^{\i k_{\beta} {\zz}}-a\e^{-\i k_{\beta} {\zz}}\right)\right],\quad m=0.
\end{aligned}}
\end{equation}

\subsection{Rayleigh mode, $m=R$}
For the so-called Rayleigh mode we have
\begin{equation}\label{rovfmrm}
k_{\alpha}^2<0, \ k_{\beta}^2<0,
\end{equation}
and we define
%
\begin{align}
k_{\beta}=\i k_{\eta},\quad k_{\alpha}=\i k_{\gamma}=\i\frac{c_t}{c_l}\sqrt{\kappa^2\left[\frac{c_l^2}{c_t^2}-1\right]+k_{\eta}^2}.
\end{align}
%
We need to set $A=C=0$ in order not to have exponentially increasing terms in (\ref{sfweppy}), and then the set of equations (\ref{eqfabcd}) become
\begin{subequations}\label{eqfabcdrm}
\begin{align}
2\i\kappa k_{\gamma} B-(\kappa^2+k_{\eta}^2)D=0,\\
(\kappa^2+k_{\eta}^2)B+2\i\kappa k_\eta D=0.
\end{align}
\end{subequations}
In order to have non-trivial solutions the determinant of this set of equations has to be equal to zero, i.e.,
\begin{equation}\label{rlgcnd}
(\kappa^2+k_{\eta}^2)^2-4\kappa^2 k_\gamma k_\eta=0,
\end{equation}
which requires the wavevector $k_{\eta}$ to be a root $k_R$ of
\begin{equation}\label{rlgcnd2}
\xi^4+4\xi^3+2(3-8\nu)\xi^2-4(3-4\nu)\xi+1=0, \quad \text{with} \quad \xi=\frac{k_R^2}{\kappa^2}, \ \nu=\left(\frac{c_t}{c_l}\right)^2=\frac{1-2\sigma}{2(1-\sigma)},
\end{equation}
known as the Rayleigh condition.
Note that $\xi$ has to be in the interval $\xi\in[0,1]$.

We set the following coefficients for the Rayleigh mode $A=0$, $B=1$, $C=0$, the following potentials
\begin{subequations}
\begin{align}
f_{\phi}^{(\kappa,c,R)}&=\e^{k_{\gamma}{\zz}},\\
f_{\psi_{\yy}}^{(\kappa,c,R)}&=\frac{2\i\kappa k_{\gamma}}{\kappa^2+k_{\eta}^2}\e^{k_{\eta}{\zz}},
\end{align}
\end{subequations}
and the corresponding displacement vector $\bff({\zz})$
\begin{equation}\label{rmdwff}
\boxed{
\begin{aligned}
f_{\xx}^{(\kappa,c,R)}&=\i \sqrt{\frac{\kappa}{K(\sigma)}}\left[
\e^{k_{\gamma}{\zz}}
-\frac{2k_{\gamma}k_{\eta}}{\kappa^2+k_{\eta}^2}\e^{k_{\eta}{\zz}}\right],\\
f_{\yy}^{(\kappa,c,R)}&=0,\\
f_{\zz}^{(\kappa,c,R)}&= \sqrt{\frac{\kappa}{K(\sigma)}}\left[
\frac{k_{\gamma}}{\kappa}\e^{k_{\gamma}{\zz}}
-\frac{2\kappa k_{\gamma}}{\kappa^2+k_{\eta}^2}\e^{k_{\eta}{\zz}}\right],\quad m=R.
\end{aligned}}
\end{equation}

\subsection{Quantization of eigenmodes}

To get the the displacement vector in the full half-space $xyz$ we need to construct the eigenfunctions $f_i^{(\kappa,k_{\beta},m)}$ for an arbitrary direction of the wavevector $\bk=\{k_{\xx},k_{\yy}\}$. This we obtain by rotating the coordinate plane ${\xx}'{\yy}'$, where we have the wavevector $\bk'=\{\kappa,0\}$, to ${\xx}{\yy}$, where the wavevector becomes $\bk=\{k_{\xx},k_{\yy}\}$ with $\kappa=\sqrt{k_{\xx}^2+k_{\yy}^2}$. This is achieved by making the transformation
\begin{equation}
f_i^{(\bk,k_{\beta},m)}({\zz})=R_{ij}f_{j}^{(\kappa,k_{\beta},m)}({\zz}),
\quad u_i^{(\bk,k_{\beta},m)}(\br)=\frac{1}{2\pi}f_i^{(\bk,k_{\beta},m)}({\zz})\e^{\i(k_{\xx} {\xx}+k_{\yy} {\yy})},
\end{equation}
where the rotation matrix $R_{ij}$ is
\begin{equation}
R_{ij}=\begin{pmatrix}
\cos\theta & 0 & \sin\theta\\
0 & 1  & 0\\
-\sin\theta & 0 & \cos\theta
\end{pmatrix},
\end{equation}
and $\theta$ is the angle of rotation around the ${\zz}$ axis with
\begin{equation}
k_{\xx}=\kappa\cos\theta, \quad k_{\yy}=\kappa\sin\theta.
\end{equation}
By having $f_i^{(\bk,k_{\beta},m)}({\zz})$ we can obtain $u_i^{(\bk,k_{\beta},m)}({\zz})$ from Eq. (\ref{unufnu}), which satisfies the following completeness relation
\begin{equation}\label{tcsoffixyz}
\begin{aligned}
\sum_{\bk,m,k_{\beta}} u^{(\bk,k_{\beta},m)}_i(\br)u^{*(\bk,k_{\beta},m)}_j(\br')=\delta_{ij}\delta(\br-\br').
\end{aligned}
\end{equation}
Then the complete set of eigensfunctions (\ref{tcsoffixyz}) can be used to expand the phonon field $\hat{\bu}(\br,t)$ as
\begin{equation}\label{tdfov}
\hat{\bu}(\br,t)=\sum_{\nu}\sqrt{\frac{\hbar}{2\rho \omega_\nu}}
\left[\bu^{(\nu)}(\br)\e^{-\i\omega_\nu t}\aan_{\nu}
+\bu^{*(\nu)}(\br)\e^{\i\omega_\nu t}\ad_{\nu}\right],\quad \text{with}\quad \omega_{\bk,k_{\beta}}=c_l\sqrt{k_{\beta}^2+\abs{\bk}^2}>0,
\end{equation}
in terms of the operators $a_{\nu}$ and $a_{\nu}^{\dag}$, which satisfy the following commutation relation
\begin{equation}
[\aan_{\bk,k_{\beta},m},\ad_{\bk',k_{\beta}',m'}]=\delta_{m,m'}\delta_{k_{\beta},k_{\beta}'}\delta_{\bk,\bk'}.
\end{equation}
The canonical momentum for the phonon field (\ref{tdfov}) is
\begin{equation}\label{tcmfu}
\hat{\bpi}(\br,t)=\rho\frac{\pd \hat{\bu}(\br,t)}{\pd t}=-\i\sum_{\nu}\sqrt{\frac{\hbar\rho\omega_\nu}{2}}
\left[\bu^{(\nu)}(\br)\e^{-\i\omega_\nu t}\aan_{\nu}
-\bu^{*(\nu)}(\br)\e^{\i\omega_\nu t}\ad_{\nu}\right],
\end{equation}
and the following commutation relation is satisfied
\begin{equation}
[\hat{u}_i(\br,t),\hat{\pi}_j(\br',t)]=\i\hbar\delta_{ij}\delta(\br-\br').
\end{equation}

\section{\label{sec:hcpc}Heat current and phonon conductance}

We define the heat current due to phonons as the rate of change of the energy [as described by Hamiltonian (\ref{hamP})] in one particular lead $\alpha$~\cite{S_Segal2003,S_Mingo2006,S_Yamamoto2006,S_Wang2007}:
\begin{equation}\label{htcur}
Q_{\alpha,t}=-\avg{\pd_t H_{\alpha}(t)}=-\frac{\i}{\hbar}\avg{\left[H,H_{\alpha}\right](t)}.
\end{equation}
Here $A(t)=\e^{\i H t}A\e^{-\i H t}$ denotes the Heisenberg evolution of an operator. The commutator in Eq. (\ref{htcur}) yields
\begin{equation}
\begin{aligned}
Q_{\alpha,t}&=\frac{1}{\rho_\alpha}\sum_{\lbn \lbi \br}V_{\lbn,\lbi \alpha}(\br)\avg{\pi_{i\alpha}(\br,t)r_{\lbn}(t)}
=\frac{\i\hbar}{\rho_\alpha}\sum_{n\lbi \br}V_{\lbn,\lbi \alpha}(\br)G^{r\pi,<}_{\lbn,\lbi \alpha\br,tt},
\end{aligned}
\end{equation}
where $G^{r\pi,<}_{\lbn\lbbeta\rmbr{\br},tt'}=-\i/\hbar \avg{\pi_{\lbbeta}(\rmbr{\br,}t')r_{\lbn}(t)}$ is the lesser Green's function. The same time function $G^{r\pi,<}_{\lbn\lbbeta\rmbr{\br},tt}$ can be expressed in terms of $G^{ru,<}_{\lbn\lbbeta\rmbr{\br},tt'}=-\i/\hbar \avg{u_{\lbbeta}(\rmbr{\br,}t')r_{\lbn}(t)}$ as
\begin{equation}
G^{r\pi,<}_{\lbn\lbbeta\rmbr{\br},tt}=\lim_{t'\rightarrow t}\frac{\pd}{\pd t'}\rho_{\lbbeta}G^{ru,<}_{\lbn\lbbeta\rmbr{\br},tt'}.
\end{equation}
In the steady state the current becomes time independent and can be expressed in the following way
\begin{equation}\label{thcexpr}
\begin{aligned}
Q_{\alpha}&=Q_{\alpha,t=0}=\frac{\i\hbar}{\rho_\alpha}\int_{-\infty}^{+\infty}\frac{\dif{\omega}}{2\pi}
\sum_{\lbn\lbi\br}V_{\lbn,\lbi \alpha}(\br)G^{r\pi,<}_{\lbn,\lbi \alpha\br,\omega}
=-\hbar\int_{-\infty}^{+\infty}\frac{\dif\omega}{2\pi}\omega\sum_{\lbn\lbi\br}V_{\lbn,\lbi \alpha}(\br)G^{ru,<}_{\lbn, \lbi \alpha\br,\omega},
\end{aligned}
\end{equation}
where we have the Fourier transformation $G^{ru,<}_{\lbn\lbbeta\rmbr{\br},\omega}=\int_{-\infty}^{+\infty}\dif(t-t')\e^{\i\omega(t-t')}G^{ru,<}_{\lbn\lbbeta\rmbr{\br},tt'}$.
We calculate the above Green's function entering the current using the Keldysh technique~\cite{S_Keldysh1965} and closely follow Wang \textit{et al.}~[\onlinecite{S_Wang2007}].

We consider the following contour-ordered Green's functions
\begin{subequations}
\begin{align}
&S_{\lbbeta\rmbr{\br}\lbbeta'\rmbr{\br'},\tau\tau'}^{c}=-\frac{\i}{\hbar}\avgs{T_{c}u_{\lbbeta}(\rmbr{\br,}\tau)u_{\lbbeta'}(\rmbr{\br',}\tau')},\\
&G_{\lbbeta\rmbr{\br}\lbn,\tau\tau'}^{ur, c}=-\frac{\i}{\hbar}\avgs{T_{c}u_{\lbbeta}(\rmbr{\br,}\tau)r_{\lbn}(\tau')},\\
&G_{\lbn\lbbeta\rmbr{\br},\tau\tau'}^{ru, c}=-\frac{\i}{\hbar}\avgs{T_{c}r_{\lbn}(\tau)u_{\beta}(\rmbr{\br,}\tau')}, \\
&D_{\lbn\lbn',\tau\tau'}^{c}=-\frac{\i}{\hbar}\avgs{T_{c}r_{\lbn}(\tau)r_{\lbn'}(\tau')},
\end{align}
\end{subequations}
which satisfy the equations
\begin{subequations}\label{emgefgf}
\begin{align}
&G^{ur,c}_{\lbbeta\rmbr{\br}\lbn',\tau\tau'}=
\sum_{{\lbbeta_1\rmbr{\br_1}\lbn_2},\tau_1}S^{0,c}_{\lbbeta\rmbr{\br}\lbbeta_1\rmbr{\br_1},\tau\tau_1}V_{\lbbeta_1\lbn_2}\rmbr{(\br_1)}D^{c}_{\lbn_2\lbn',\tau_1\tau'}
+\sum_{\lbbeta_1\rmbr{\br_1},\tau_1}S^{0,c}_{\lbbeta\rmbr{\br}\lbbeta_1\rmbr{\br_1},\tau\tau_1}V_{\lbbeta_1}\rmbr{(\br_1)}G^{ur,c}_{\lbbeta_1\rmbr{\br_1}\lbn',\tau_1\tau'},\\
&G^{ru,c}_{\lbn\lbbeta'\rmbr{\br'},\tau\tau'}=
\sum_{{\lbn_1\lbbeta_2\rmbr{\br_2}},\tau_1}D^{c}_{\lbn\lbn_1,\tau\tau_1}V_{\lbn_1\lbbeta_2}\rmbr{(\br_2)}S^{0,c}_{\lbbeta_2\rmbr{\br_2}\lbbeta'\rmbr{\br'},\tau_1\tau'}
+\sum_{\lbbeta_1\rmbr{\br_1},\tau_1}G^{ru,c}_{\lbn\lbbeta_1\rmbr{\br_1},\tau\tau_1}V_{\lbbeta_1}\rmbr{(\br_1)}S^{0,c}_{\lbbeta_1\rmbr{\br_1}\lbbeta'\rmbr{\br'},\tau_1\tau'},\\
&D^{c}_{\lbn\lbn',\tau\tau'}=D^{0,c}_{\lbn\lbn',\tau\tau'}
+\sum_{\lbn_1\neq\lbn_2,\tau_1}D^{0,c}_{\lbn\lbn_1,\tau\tau_1}V_{\lbn_1\lbn_2}D^{c}_{\lbn_2\lbn',\tau_1\tau'}
+\sum_{\lbn_1\lbbeta_2\rmbr{\br_2},\tau_1}D^{0,c}_{\lbn\lbn_1,\tau\tau_1}V_{\lbn_1\lbbeta_2}\rmbr{(\br_2)}G^{ur,c}_{\lbbeta_2\rmbr{\br_2}\lbn',\tau_1\tau'},
\end{align}
\end{subequations}
on the Keldysh contour, where $\sum_{\tau_1}\rightarrow\int_{c_{\mrK}}\dif \tau_1\ldots$ denotes integration along the contour. The Green's functions $D^{0}$ and $S^{0}$ are calculated in \sectionname~\ref{sec:nisd} and correspond to a Hamiltonian consisting of (\ref{ham0}) and the terms $V_{nn}$ in (\ref{hamV}). Using the Larkin-Ovchinnikov representation for contour-ordered Green's functions~\cite{S_Larkin1975,S_Rammer1986}
\begin{equation}\label{LarOvRep}
G^c\quad\rightarrow\quad G=\begin{pmatrix}
G^R & G^K \\
0 & G^A
\end{pmatrix},
\end{equation}
we can Fourier transform Eqs. (\ref{emgefgf}) with respect to the time difference $t-t'$ to give
\begin{subequations}\label{emgefgfFT}
\begin{align}
&G^{ur}_{\lbbeta\rmbr{\br}\lbn',\omega}=
\sum_{{\lbbeta_1\rmbr{\br_1}\lbn_2}}S^{0}_{\lbbeta\rmbr{\br}\lbbeta_1\rmbr{\br_1},\omega}V_{\lbbeta_1\lbn_2}\rmbr{(\br_1)}D_{\lbn_2\lbn',\omega}
+\sum_{\lbbeta_1\rmbr{\br_1}}S^{0}_{\lbbeta\rmbr{\br}\lbbeta_1\rmbr{\br_1},\omega}V_{\lbbeta_1}\rmbr{(\br_1)}G^{ur}_{\lbbeta_1\rmbr{\br_1}\lbn',\omega},\\
&G^{ru}_{\lbn\lbbeta'\rmbr{\br'},\omega}=
\sum_{{\lbn_1\lbbeta_2\rmbr{\br_2}}}D_{\lbn\lbn_1,\omega}V_{\lbn_1\lbbeta_2}\rmbr{(\br_2)}S^{0}_{\lbbeta_2\rmbr{\br_2}\lbbeta'\rmbr{\br'},\omega}
+\sum_{\lbbeta_1\rmbr{\br_1}}G^{ru}_{\lbn\lbbeta_1\rmbr{\br_1},\omega}V_{\lbbeta_1}\rmbr{(\br_1)}S^{0}_{\lbbeta_1\rmbr{\br_1}\lbbeta'\rmbr{\br'},\omega},\\
&D_{\lbn\lbn',\omega}=D^{0}_{\lbn\lbn',\omega}
+\sum_{\lbn_1\neq\lbn_2}D^{0}_{\lbn\lbn_1,\omega}V_{\lbn_1\lbn_2}D_{\lbn_2\lbn',\omega}
+\sum_{\lbn_1\lbbeta_2\rmbr{\br_2}}D^{0}_{\lbn\lbn_1,\omega}V_{\lbn_1\lbbeta_2}\rmbr{(\br_2)}G^{ur}_{\lbbeta_2\rmbr{\br_2}\lbn',\omega}.
\end{align}
\end{subequations}
In Eq. (\ref{LarOvRep}) $G^{R}/G^{A}/G^{K}$ denote respectively retarded/advanced/Keldysh Green's functions, which for bosonic operators $a$ and $b$ are defined as
\begin{subequations}
\begin{align}
&G^{R}_{ab}(t,t')=-\frac{\i}{\hbar}\theta(t-t')\avgs{[a(t),b(t')]},\\
&G^{A}_{ab}(t,t')=\frac{\i}{\hbar}\theta(t'-t)\avgs{[a(t),b(t')]},\\
&G^{K}_{ab}(t,t')=-\frac{\i}{\hbar}\avgs{\{a(t),b(t')\}}.
\end{align}
\end{subequations}
By neglecting the subscripts in Eqs. (\ref{emgefgf}) and (\ref{emgefgfFT}), they can be written in a more compact form
\begin{subequations}\label{afoeom1}
\begin{align}
&G^{ur}=S^{0}VD+S^{0}VG^{ur},\\
&G^{ru}=DVS^{0}+G^{ru}VS^{0},\\
&D=D^{0}+D^0VD+D^{0}VG^{ur},
\end{align}
\end{subequations}
which can be expressed as
\begin{subequations}\label{dtefgf}
\begin{align}
&G^{ur}=\tilde{S}^{0}VD,\\
&G^{ru}=DV\tilde{S}^{0},\\
&D=\tilde{D}^{0}+\tilde{D}^{0}VG^{ur},
\end{align}
\end{subequations}
where we have introduced frequency shifted lead Green's function $\tilde{S}^0=[1-S^{0}V]^{-1}S^{0}$
and molecule Green's function in the normal mode basis $\tilde{D}^{0}=[1-D^{0}V]^{-1}D^{0}$. After inserting Eq. (\ref{dtefgf}a) into Eq. (\ref{dtefgf}c) we obtain the Dyson equation for the molecule
\begin{equation}
D=\tilde{D}^{0}+\tilde{D}^{0}\Sigma D, \quad \Sigma=V\tilde{S}^0V,
\end{equation}
with the self-energy $\Sigma$. In the calculation we will need separately the self-energy due to the left and right leads, which explicitly reads
\begin{equation}
\Sigma_{\lbn\lbn',\alpha}=\sum_{\lbi_1\br_1\lbi_2\br_2}
V_{\lbn,\lbi_1\alpha}(\br_1)
\tilde{S}^0_{\lbi_1\alpha\br_1, \lbi_2\alpha\br_2,\omega}
V_{i_2\alpha,\lbn'}(\br_2), \quad \alpha=L,R,
\end{equation}
where we have suppressed the frequency subscript for the self-energy.

Now we will express the heat current in terms of the molecule Green's function $D$. Using Eq. (\ref{dtefgf}b) and the Langreth rule $G^{ru,K}=D^{R}V\tilde{S}^{0,K}+D^{K}V\tilde{S}^{0,A}$, which can be seen from the Larkin-Ovchinnikov representation (\ref{LarOvRep}), we obtain
\begin{equation}\label{htcur2}
Q_{\alpha}=-\hbar\int_{-\infty}^{+\infty}\frac{\dif\omega}{2\pi}\frac{\omega}{2}
\Tr[D^{R}\Sigma_{\alpha}^{K}+D^{K}\Sigma_{\alpha}^{A}].
\end{equation}
We note that we have the relation $2G^{ru,<}_{\lbn\beta\rmbr{\br},tt}=G^{ru,K}_{\lbn\beta\rmbr{\br},tt}$ for the same time lesser Green's function. Because the heat current is real $Q_{\alpha}=Q_{\alpha}^{*}$ and conserved $Q_{L}=-Q_{R}$  Eq. (\ref{htcur2}) can be cast into the more symmetric form
\begin{equation}\label{htcur3}
\begin{aligned}
Q&=\frac{1}{4}(Q_{L}+Q_{L}^{*}-Q_{R}-Q_{R}^{*})\\
&=\frac{\hbar}{4}\int_{0}^{+\infty}\frac{\dif\omega}{2\pi}\omega
\Tr[(D^{R}-D^{A})(\Sigma_{\mrR}^{K}-\Sigma_{\mrL}^{K})+\i D^{K} (\Gamma_{\mrR}-\Gamma_{\mrL})],
\end{aligned}
\end{equation}
where we have introduced
\begin{equation}
\Gamma_{\alpha}=\i(\Sigma_{\alpha}^{R}-\Sigma_{\alpha}^{A}),
\end{equation}
and used the properties
$\left[D^{R}_{\omega}\right]^{\dag}=D^{A}_{\omega}$,
$\left[\Sigma_{\alpha,\omega}^{A}\right]^{\dag}=\Sigma_{\alpha,\omega}^{R}$,
$\left[D^{K}_{\omega}\right]^{\dag}=-D^{K}_{\omega}$,
$\left[\Sigma_{\alpha,\omega}^{K}\right]^{\dag}=-\Sigma_{\alpha,\omega}^{K}$.
For a quadratic model, like the one we consider in this paper [see Eq. (\ref{ham})], the heat current (\ref{htcur3}) can be rewritten in terms of the transmission $\m{T}_{\mrp}(\omega)$, which is a temperature independent function, as given by a Caroli type formula
\begin{equation}\label{tflb}
\m{T}_{\mrp}(\omega)=\Tr[\Gamma_{\mrL}(\omega)D^{R}(\omega)\Gamma_{\mrR}(\omega)D^{A}(\omega)].
\end{equation}

\subsection{Single mass model}

\begin{figure}[ht]
\begin{center}
\includegraphics[width=0.7\textwidth]{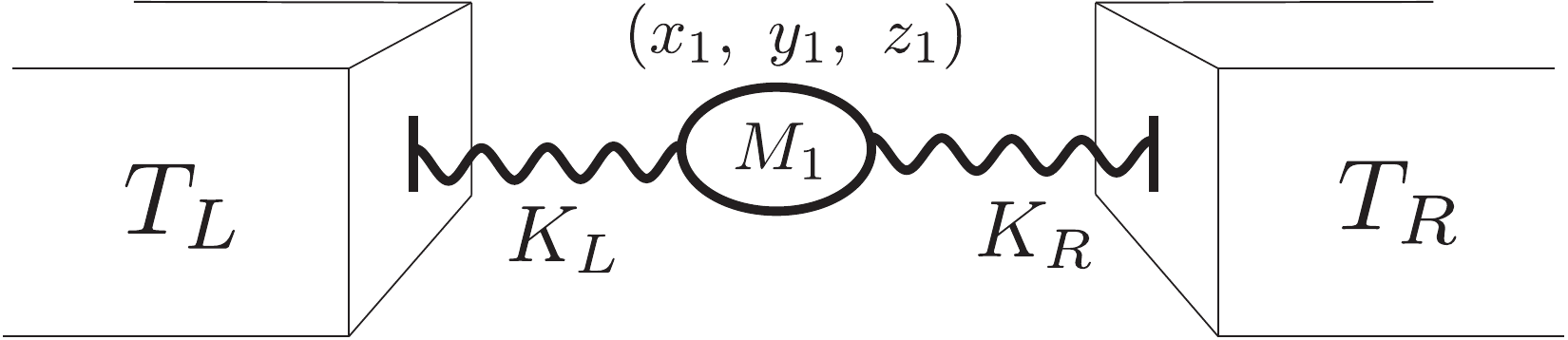}
\caption{\label{Sfig1} Model with a single mass in the junction.}
\end{center}
\end{figure}

When there is a single mass $M_1$ (see \figurename~\ref{Sfig1}) in the junction we have only coupling to the leads
\begin{equation}
K_{\lbi1,\lbi L}(\br)\equiv K_{\lbi L}\delta(\br-\br_{L}), \quad
K_{\lbi1,\lbi R}(\br)\equiv K_{\lbi R}\delta(\br-\br_{R}), \quad
\end{equation}
which gives the following mass weighted coordinate couplings $V$ to a single point
\begin{equation}
V_{\lbi1, \lbi1}=\frac{K_{\lbi L}+K_{\lbi R}}{M_1}\equiv \omega_{i}^2, \quad
V_{\lbi \alpha}=\frac{K_{\lbi \alpha}}{M_{\alpha}}, \quad
V_{\lbi 1, \lbi\alpha}=-\frac{K_{i\alpha}}{\sqrt{M_1 M_{\alpha}}}.
\end{equation}
We then obtain the following phonon transmission function
\begin{equation}\label{tosmitj}
\m{T}_{\mathrm{one}}^{\lbi}(\omega)=\left(\frac{K_{\lbi L}K_{\lbi R}}{M_1}\right)^2
\frac{2\ti_{\lbi L}\ti_{\lbi R}}{\left[\omega^2-\omega_{\lbi}^2-\frac{K_{\lbi L}^2}{M_1}\tr_{\lbi L}-\frac{K_{\lbi R}^2}{M_1}\tr_{\lbi R}\right]^2+\left[\frac{K_{\lbi L}^2}{M}\ti_{\lbi L}+\frac{K_{\lbi R}^2}{M}\ti_{\lbi R}\right]^2},
\end{equation}
where $\tr_{\lbi \alpha}$ and $\ti_{\lbi \alpha}$ describe real and imaginary parts of frequency shifted lead Green's function
\begin{equation}
\begin{aligned}
&\tilde{S}^{0,R}_{i\alpha\br_{\alpha},i\alpha\br_{\alpha},\omega}=M_{\alpha}(\tr_{\lbi \alpha}+\i \ti_{\lbi \alpha}), \\
&\tr_{\lbi \alpha}=\frac{\rr_{\lbi \alpha}-K_{\lbi \alpha}(\rr_{\lbi \alpha}^2+\ii_{\lbi \alpha}^2)}{(1-K_{\lbi \alpha}\rr_{\lbi \alpha})^2+(K_{\lbi \alpha}\ii_{\lbi \alpha})^2},
\quad
\ti_{\lbi \alpha}=\frac{\ii_{\lbi \alpha}}{(1-K_{\lbi \alpha}\rr_{\lbi \alpha})^2+(K_{\lbi \alpha}\ii_{\lbi \alpha})^2},
\end{aligned}
\end{equation}
with $\rr_{\lbi \alpha}$ and $\ii_{\lbi \alpha}$ being the real and imaginary parts of the non-interacting lead Green's function
\begin{equation}\label{mwcs0}
S^{0,R}_{i\alpha\br_{\alpha},i\alpha\br_{\alpha},\omega}=M_{\alpha}(\rr_{\lbi \alpha}+\i \ii_{\lbi \alpha}).
\end{equation}
We use the transmission (\ref{tosmitj}) to obtain the black dotted lines in \figurename~\rFigOne{c-f} of the main text.

\subsection{Two masses model}

When there are two masses $M_1$ and $M_2$ (see \figurename~\rFigOne{a}) in the junction we have the spring constants (\ref{twmsc}) which give the following mass weighted coordinate couplings $V$ to a single point
\begin{equation}
\begin{aligned}
&V_{\lbi1, \lbi1}=\frac{K_{\lbi}+K_{\lbi L}}{M_1}\equiv \omega_{\lbi 1}^2, \quad
V_{\lbi2, \lbi2}=\frac{K_{\lbi}+K_{\lbi R}}{M_2}\equiv \omega_{\lbi 2}^2, \quad
V_{\lbi \alpha}=\frac{K_{\lbi \alpha}}{M_{\alpha}}, \\
&V_{\lbi1, \lbi2}=-\frac{K_{\lbi}}{\sqrt{M_1M_2}}, \quad
V_{\lbi 1, \lbi L}=-\frac{K_{\lbi L}}{\sqrt{M_1 M_{L}}}, \quad
V_{\lbi 2, \lbi R}=-\frac{K_{\lbi R}}{\sqrt{M_2 M_{R}}}.
\end{aligned}
\end{equation}
This gives the phonon transmission function
\begin{equation}
\m{T}^{\lbi}_{\mathrm{two}}(\omega)=\left(\frac{K_{\lbi L}K_{\lbi}K_{\lbi R}}{M_1M_2}\right)^2\frac{2\ti_{\lbi L}\ti_{\lbi R}}{\m{R}^2+\m{I}^2},
\end{equation}
where all relevant functions entering the above transmission are expressed as
\begin{subequations}
\begin{align}
&\label{lmR}\begin{aligned}
\m{R}&=(\omega^2-\omega_{\lbi+}^2)(\omega^2-\omega_{\lbi-}^2)+\frac{K_{\lbi L}^2}{M_1}\frac{K_{\lbi R}^2}{M_2}(\tr_{\lbi L}\tr_{\lbi R}-\ti_{\lbi L}\ti_{\lbi R})\\
&-\left(\omega^2-\frac{\omega_{\lbi+}^2+\omega_{\lbi-}^2}{2}\right)\left(\frac{K_{\lbi L}^2}{M_1}\tr_{\lbi L}+\frac{K_{\lbi R}^2}{M_2}\tr_{\lbi R}\right)-\delta_{\lbi}\left(\frac{K_{\lbi L}^2}{M_1}\tr_{\lbi L}-\frac{K_{\lbi R}^2}{M_2}\tr_{\lbi R}\right),
 \end{aligned}\\
&\label{lmI}\begin{aligned}
\m{I}&=\left(\omega^2-\frac{\omega_{\lbi+}^2+\omega_{\lbi-}^2}{2}\right)\left(\frac{K_{\lbi L}^2}{M_1}\ti_{\lbi L}+\frac{K_{\lbi R}^2}{M_2}\ti_{\lbi R}\right)+\delta_{\lbi}\left(\frac{K_{\lbi L}^2}{M_1}\ti_{\lbi L}-\frac{K_{\lbi R}^2}{M_2}\ti_{\lbi R}\right)\\
&-\frac{K_{\lbi L}^2}{M_1}\frac{K_{\lbi R}^2}{M_2}(\ti_{\lbi L}\tr_{\lbi R}+\ti_{\lbi R}\tr_{\lbi L}),
\end{aligned}
\end{align}
\begin{equation}
\omega_{\lbi\pm}^2=\frac{\omega_{\lbi1}^2+\omega_{\lbi2}^2}{2}\pm\Delta_{\lbi},
\quad \Delta_{\lbi}=\sqrt{\delta^2_{\lbi}+V_{\lbi1,\lbi2}^2},
\quad \delta_{\lbi}=\frac{\omega_{\lbi1}^2-\omega_{\lbi2}^2}{2}.
\end{equation}
\end{subequations}

\section{\label{sec:nisd}The non-interacting particle $D^{0}$ and lead $S^{0}$ Green's functions}
In this section we present the non-interacting particle Green's function $D^{0}$ corresponding to the Hamiltonian $H_{0}=\sum_{\lbi\lbm}(p_{\lbi\lbm}^2+V_{\lbi\lbm,\lbi\lbm}r_{\lbi\lbm}^2)/2$ and the lead Green's functions $S^{0}$ corresponding to the Hamiltonian $H_{0}=\sum_{\alpha\nu}\hbar\omega_{\alpha\nu}\ad_{\alpha\nu}\aan_{\alpha\nu}$. We note that we use the mass weighted coordinates (\ref{mwcoord}). By using the equation of motion we obtain for the Fourier transformed retarded Green's functions
\begin{subequations}
\begin{align}
&D^{0,R}_{\lbi \lbm,\lbi'\lbm',\omega}=\frac{\delta_{\lbi\lbi'}\delta_{\lbm\lbm'}}{(\omega+\i\eta)^2-\omega_{\lbi\lbm}^2},\quad \omega_{\lbi\lbm}^2=\frac{K_{\lbi \lbm}}{M_\lbm},\\
&S^{0,R}_{\lbi\alpha\br\nu,\lbi'\alpha'\br'\nu',\omega}
=\frac{M_{\alpha}}{\rho_{\alpha}}\frac{c_{ii',\alpha,\nu,\br\br'}\delta_{\alpha\alpha'}\delta_{\nu\nu'}}{(\omega+\i\eta)^2-\omega_{\alpha\nu}^2},
\quad \omega_{\alpha\nu}=c_{\alpha t}\sqrt{k_{\beta}^2+\kappa^2},
\end{align}
\end{subequations}
where
\begin{equation}
c_{\lbi\lbi',\alpha,\nu,\br\br'}=\Real\left[u_{\lbi\alpha}^{(\nu)}(\br)u_{\lbi'\alpha}^{*(\nu)}(\br')\right]
=\frac{1}{2}\left[u_{\lbi\alpha}^{(\nu)}(\br)u_{\lbi'\alpha}^{*(\nu)}(\br')+u_{\lbi\alpha}^{*(\nu)}(\br)u_{\lbi'\alpha}^{(\nu)}(\br')\right],
\end{equation}
and $S^{0,R}_{\lbi\alpha\br\nu,\lbi'\alpha'\br'\nu',\omega}$ corresponds to
\begin{equation}
S^{0,R}_{\lbi\alpha\br\nu,\lbi'\alpha'\br'\nu',tt'}=
-\i\theta(t-t')\avgs{[u_{\lbi\alpha\nu}(\br,t),u_{\lbi'\alpha'\nu'}(\br',t')]},
\end{equation}
with
$u_{\lbi\alpha\nu}(\br,t)=\sqrt{M_{\alpha}\hbar/2\rho \omega_\nu}[u^{(\nu)}_{\lbi\alpha}(\br)\e^{-\i\omega_\nu t}\aan_{\nu}+u^{*(\nu)}_{\lbi\alpha}(\br)\e^{\i\omega_\nu t}\ad_{\nu}]$ being a mass weighted displacement vector for a particular mode $\nu$.
We are interested in the $\nu\nu'$ summed lead Green's function $S^{0,R}_{\lbi\alpha\br,\lbi'\alpha'\br',\omega}$, when the molecules are attached only to the surface of the leads $\br_{\alpha}=(x,y_{\alpha},z)$, $\br_{\alpha}'=(x',y_{\alpha},z')$. So we need the coefficient
\begin{equation}
\begin{aligned}
c_{\alpha,\lbi\lbi',\nu,\br_{\alpha}\br_{\alpha}'}
=\frac{1}{2(2\pi)^2}
\Big[&f_{\lbi\alpha}^{(\nu)}({\zz}_{\alpha})f_{\lbi'\alpha}^{*(\nu)}({\zz}_{\alpha})\e^{\i k_{\xx}({\xx}-{\xx}')}\e^{\i k_{\yy}({\yy}-{\yy}')}\\
+&f_{\lbi\alpha}^{*(\nu)}({\zz}_{\alpha})f_{\lbi'\alpha}^{(\nu)}({\zz}_{\alpha})\e^{-\i k_{\xx}({\xx}-{\xx}')}\e^{-\i k_{\yy}({\yy}-{\yy}')}\Big],
\end{aligned}
\end{equation}
and then the required $\nu\nu'$ summed lead Green's function is
\begin{align}\label{misgf}
\begin{aligned}
&S^{0,R/A}_{\lbi\alpha\br_{\alpha},\lbi'\alpha'\br_{\alpha}',\omega}=
\frac{M_{\alpha}\delta_{\alpha\alpha'}}{2(2\pi)^2\rho_{\alpha}}\sum_{m=H,\pm,0,R}\int_{0}^{\kappa_D}\dif{\kappa}\kappa\int_{k_1}^{k_2}\dif{k}_{\beta}\int_{0}^{2\pi}\dif{\theta} \frac{1}{(\omega\pm\i\eta)^2-c_t^2(k_{\beta}^2+\kappa^2)}\\
&\quad\times\left[f_{\lbi\alpha}^{(\nu)}({\zz}_{\alpha})f_{\lbi'\alpha}^{*(\nu)}({\zz}_{\alpha})\e^{\i k_{\xx}({\xx}-{\xx}')}\e^{\i k_{\yy}({\yy}-{\yy}')}
+f_{\lbi\alpha}^{*(\nu)}({\zz}_{\alpha})f_{\lbi'\alpha}^{(\nu)}({\zz}_{\alpha})\e^{-\i k_{\xx}({\xx}-{\xx}')}\e^{-\i k_{\yy}({\yy}-{\yy}')}\right],
\end{aligned}
\end{align}
where the different modes $m=H,\pm,0,R$ and the relevant integration intervals $k_1,k_2$ for these modes are described in \sectionname~\ref{sec:emotl}. We have the following expressions for the functions $f_{\lbi\alpha,{\zz}_{\alpha}}^{(\kappa,\theta,k_{\beta},m)}$, which enter (\ref{misgf})

\footnotesize
\begin{equation}\label{masp}
\begin{aligned}
H: \quad &f_{{\xx}}=-\sin\theta\sqrt{\frac{2}{\pi}}, \quad k_1=0, \ k_2=+\infty, \\
&f_{{\yy}}=\cos\theta\sqrt{\frac{2}{\pi}}.  \\
\pm: \quad &f_{{\xx}}=\frac{\cos\theta}{\sqrt{4\pi(1+k_{\beta}^2/\kappa^2)}}
\left[\pm\sqrt{\frac{k_{\beta}}{k_\alpha}}\left(1-\zeta_{\pm}\right)
-\i\frac{k_{\beta}}{\kappa}\left(1+\zeta_{\pm}\right)\right],
\quad \zeta_{\pm}=a\pm\i b, \quad k_{\alpha}=\frac{c_t}{c_l}\sqrt{k_{\beta}^2+\kappa^2\left[1-\frac{c_l^2}{c_t^2}\right]},\\
&f_{{\yy}}=\frac{\sin\theta}{\sqrt{4\pi(1+k_{\beta}^2/\kappa^2)}}
\left[\pm\sqrt{\frac{k_{\beta}}{k_\alpha}}\left(1-\zeta_{\pm}\right)
-\i\frac{k_{\beta}}{\kappa}\left(1+\zeta_{\pm}\right)\right],
\quad a=\frac{(k_{\beta}^2-\kappa^2)^2-4\kappa^2k_\alpha k_\beta}{(k_{\beta}^2-\kappa^2)^2+4\kappa^2k_\alpha k_\beta}, \\
&\phantom{..............................................................................................}
b=\frac{4\kappa\sqrt{k_\alpha k_\beta}(k_{\beta}^2-\kappa^2)}{(k_{\beta}^2-\kappa^2)^2+4\kappa^2k_\alpha k_\beta}, \\
&f_{{\zz}}=\frac{1}{\sqrt{4\pi (1+k_{\beta}^2/\kappa^2)}}
\left[\pm\sqrt{\frac{k_{\alpha}k_{\beta}}{\kappa^2}}\left(1+\zeta_{\pm}\right)
+\i\left(1-\zeta_{\pm}\right)\right], \quad k_1=\kappa\sqrt{\frac{c_l^2}{c_t^2}-1}, \ k_2=+\infty. \\
0: \quad &f_{{\xx}}=-\i \frac{\cos\theta}{\sqrt{2\pi(1+k_{\beta}^2/\kappa^2)}}\left[
b+\frac{k_{\beta}}{\kappa}\left(1+a\right)\right], \quad k_{\gamma}=\frac{c_t}{c_l}\sqrt{\kappa^2\left[\frac{c_l^2}{c_t^2}-1\right]-k_{\beta}^2}, \quad k_{\alpha}=\i k_{\gamma}, \\
&f_{{\yy}}=-\i\frac{\sin\theta}{\sqrt{2\pi(1+k_{\beta}^2/\kappa^2)}}\left[
b+\frac{k_{\beta}}{\kappa}\left(1+a\right)\right], \quad a=\frac{(k_{\beta}^2-\kappa^2)^2-4\i \kappa^2 k_{\gamma}k_{\beta}}{(k_{\beta}^2-\kappa^2)^2+4\i \kappa^2k_{\gamma}k_{\beta}}, \\
&\phantom{........................................................................}
b=\frac{4\kappa k_{\beta}(k_{\beta}^2-\kappa^2)}{(k_{\beta}^2-\kappa^2)^2+4\i \kappa^2k_{\gamma}k_{\beta}}.\\
&f_{{\zz}}=\frac{1}{\sqrt{2\pi(1+k_{\beta}^2/\kappa^2)}}\left[
-\frac{k_{\gamma}}{\kappa} b
+\i\left(1-a\right)\right], \quad k_1=0, \ k_2=\kappa\sqrt{\frac{c_l^2}{c_t^2}-1}. \\
R: \quad &f_{{\xx}}=\i\cos\theta \sqrt{\frac{\kappa}{K(\sigma)}}\left[
1-\frac{2k_{\gamma}k_{\eta}}{\kappa^2+k_{\eta}^2}\right], \quad k_{\beta}=\i k_{\eta}, \\
&f_{{\yy}}=\i\sin\theta \sqrt{\frac{\kappa}{K(\sigma)}}\left[
1-\frac{2k_{\gamma}k_{\eta}}{\kappa^2+k_{\eta}^2}\right], \quad K(\sigma)=\frac{(k_{\gamma}-k_{\eta})(\kappa^2 k_{\gamma}-\kappa^2 k_{\eta}+2k_{\gamma}k_{\eta}^2)}{2\kappa k_{\gamma}k_{\eta}^2},\\
&f_{{\zz}}=\sqrt{\frac{\kappa}{K(\sigma)}}\left[
\frac{k_{\gamma}}{\kappa}-\frac{2\kappa k_{\gamma}}{\kappa^2+k_{\eta}^2}\right],
\quad \text{there is no integral over $k_{\eta}$, because $k_{\eta}=k_R$}, \\
\end{aligned}
\end{equation}
\normalsize
where we have supressed the labels $\alpha$, ${\zz}_{\alpha}$, and ${(\kappa,\theta,k_{\beta},m)}$ for simplicity. We can obtain the advanced $S^{0,A}$ and Keldysh $S^{0,K}$ Green's functions by noting that in equilibrium for a bosonic Green's function $G_{\omega}$ we have the relations $G^{A}_{\omega}=[G^{R}_{\omega}]^{\dag}$ and $G^{K}_{\omega}=\left[2n(\omega)+1\right](G^R_{\omega}-G^A_{\omega})$.

If the molecule couples to a single point of the lead $\br_{\alpha}=(x_{\alpha},y_{\alpha},z_{\alpha})$ then from Eq. (\ref{misgf}) and Eq. (\ref{mwcs0}) we obtain for the imaginary part for different modes $m$ of $S^{0,R}_{i\alpha\br_{\alpha},i\alpha\br_{\alpha},\omega}$:
\begin{fleqn}
\begin{subequations}
\begin{equation}
\ii^{H}_{\parallel}=-\frac{\omega}{4\pi \rho c_t^3}\int_{0}^{1}\frac{x\dif{x}}{\sqrt{1-x^2}}\approx-\frac{\omega}{4\pi \rho c_t^3},
\end{equation}
\begin{equation}
\ii^{\pm}_{\parallel}=-\frac{\omega}{8\pi \rho c_t^3}\int_{0}^{1/c_r}
\frac{c_rx\sqrt{1-x^2}\dif{x}}{c_r(1-2x^2)^2+4x^2\sqrt{1-x^2}\sqrt{1-(c_rx)^2}}\approx-0.042\frac{\omega}{4\pi \rho c_t^3},
\end{equation}
\begin{equation}
\ii^{(0)}_{\parallel}=-\frac{\omega}{4\pi \rho c_t^3}\int_{1/c_r}^{1}
\frac{c_r^2x\sqrt{1-x^2}(1-2x^2)^2\dif{x}}{c_r^2\left[1-8x^2(1-x^2)(1-2x^2)\right]-16x^4(1-x^2)}\approx-0.168\frac{\omega}{4\pi \rho c_t^3},
\end{equation}
\begin{equation}
\ii^{R}_{\parallel}=-\frac{\omega}{4\pi \rho c_t^3}
\frac{\pi\xi\sqrt{(1-\xi)(c_r^2-1+\xi)}[c_r(1+\xi)-2\sqrt{\xi(c_r^2-1+\xi)}]^2}{c_r(1-\xi^2)^2(\sqrt{c_r^2-1+\xi}-c_r\sqrt{\xi})[(1+2\xi)\sqrt{c_r^2-1+\xi}-c_r\sqrt{\xi}]}
\approx-0.151\frac{\omega}{4\pi \rho c_t^3},
\end{equation}
\begin{equation}
\ii^{\pm}_{\perp}=-\frac{\omega}{4\pi \rho c_t^3}\int_{0}^{1/c_r}
\frac{x\sqrt{1-(c_rx)^2}\dif{x}}{c_r(1-2x^2)^2+4x^2\sqrt{1-x^2}\sqrt{1-(c_rx)^2}}\approx-0.021\frac{\omega}{4\pi \rho c_t^3},
\end{equation}
\begin{equation}
\ii^{(0)}_{\perp}=-\frac{2\omega}{\pi \rho c_t^3}\int_{1/c_r}^{1}
\frac{x^3\sqrt{1-x^2}[(c_rx)^2-1]\dif{x}}{c_r^2\left[1-8x^2(1-x^2)(1-2x^2)\right]-16x^4(1-x^2)}\approx-0.540\frac{\omega}{4\pi \rho c_t^3},
\end{equation}
\begin{equation}
\ii^{R}_{\perp}=-\frac{\omega}{2\pi \rho c_t^3}
\frac{\pi\xi(1-\xi)^{5/2}(c_r^2-1+\xi)^{3/2}}{c_r(1-\xi^2)^2(\sqrt{c_r^2-1+\xi}-c_r\sqrt{\xi})[(1+2\xi)\sqrt{c_r^2-1+\xi}-c_r\sqrt{\xi}]}
\approx-0.863\frac{\omega}{4\pi \rho c_t^3}.
\end{equation}
\end{subequations}
\end{fleqn}

\noindent For the gold leads we have
\begin{equation}
c_r=\frac{c_l}{c_t}=\sqrt{\frac{2(1-\sigma)}{1-2\sigma}}\approx 2.693,
\quad \text{where $\sigma=0.42$ is Poisson ratio},
\end{equation}
and $0<\xi<1$ is determined from Eq. (\ref{rlgcnd2})
\begin{equation}
\xi^4+4\xi^3+2\left(3-\frac{8}{c_r^2}\right)\xi^2-4\left(3-\frac{4}{c_r^2}\right)\xi+1=0, \quad \xi\approx0.107.
\end{equation}
Then the total values of the imaginary parts are
\begin{equation}\label{iipii}
\ii_{\parallel}\approx -1.404\frac{\omega}{4\pi \rho c_t^3}\equiv -A_{\parallel}\omega, \quad \ii_{\perp}\approx -1.445\frac{\omega}{4\pi \rho c_t^3}\equiv -A_{\perp}\omega,
\end{equation}
Note that $\ii_{\parallel}$ corresponds to the $\ii_{\xx\xx}$, $\ii_{\yy\yy}$ components and $\ii_{\perp}$ corresponds to $\ii_{\zz\zz}$.

We can calculate the real part using the Kramers-Kr\"{o}nig relations (also known as Hilbert transform). For the function $G(\omega)$, which is analytic in the upper complex half-plane (the retarded one), and which decays as $\omega^{-a}$ with $a>0$, we have
\begin{equation}\label{krkrrel}
\begin{aligned}
&\Realp{G(\omega)}=\frac{1}{\pi}\m{P}\int_{-\infty}^{+\infty}\dif{\omega'}\frac{\Imagp{G(\omega')}}{\omega'-\omega},\\
&\Imagp{G(\omega)}=-\frac{1}{\pi}\m{P}\int_{-\infty}^{+\infty}\dif{\omega'}\frac{\Realp{G(\omega')}}{\omega'-\omega}.
\end{aligned}
\end{equation}
In our case the retarded function $S^{0,R}_{i\alpha\br_{\alpha},i\alpha\br_{\alpha},\omega}$ satisfies all the above mentioned conditions if we cut off the frequency $\omega$ in the imaginary parts of (\ref{iipii}) by the Debye frequency $\omega_{\mathrm{D}}$. In this case after applying the transformation Eq. (\ref{krkrrel}) to Eq. (\ref{iipii}) we obtain
\begin{equation}\label{rrdoc}
\rr_{\parallel/\perp}=-\frac{A_{\parallel/\perp}}{\pi}\left(2\omega_{\mathrm{D}}+\omega\ln\absB{\frac{\omega_D-\omega}{\omega_D+\omega}}\right).
\end{equation}

\section{DFT calculations}

\subsection{Spring constants}

\begin{figure}[ht]
\begin{center}
\includegraphics[width=0.9\textwidth]{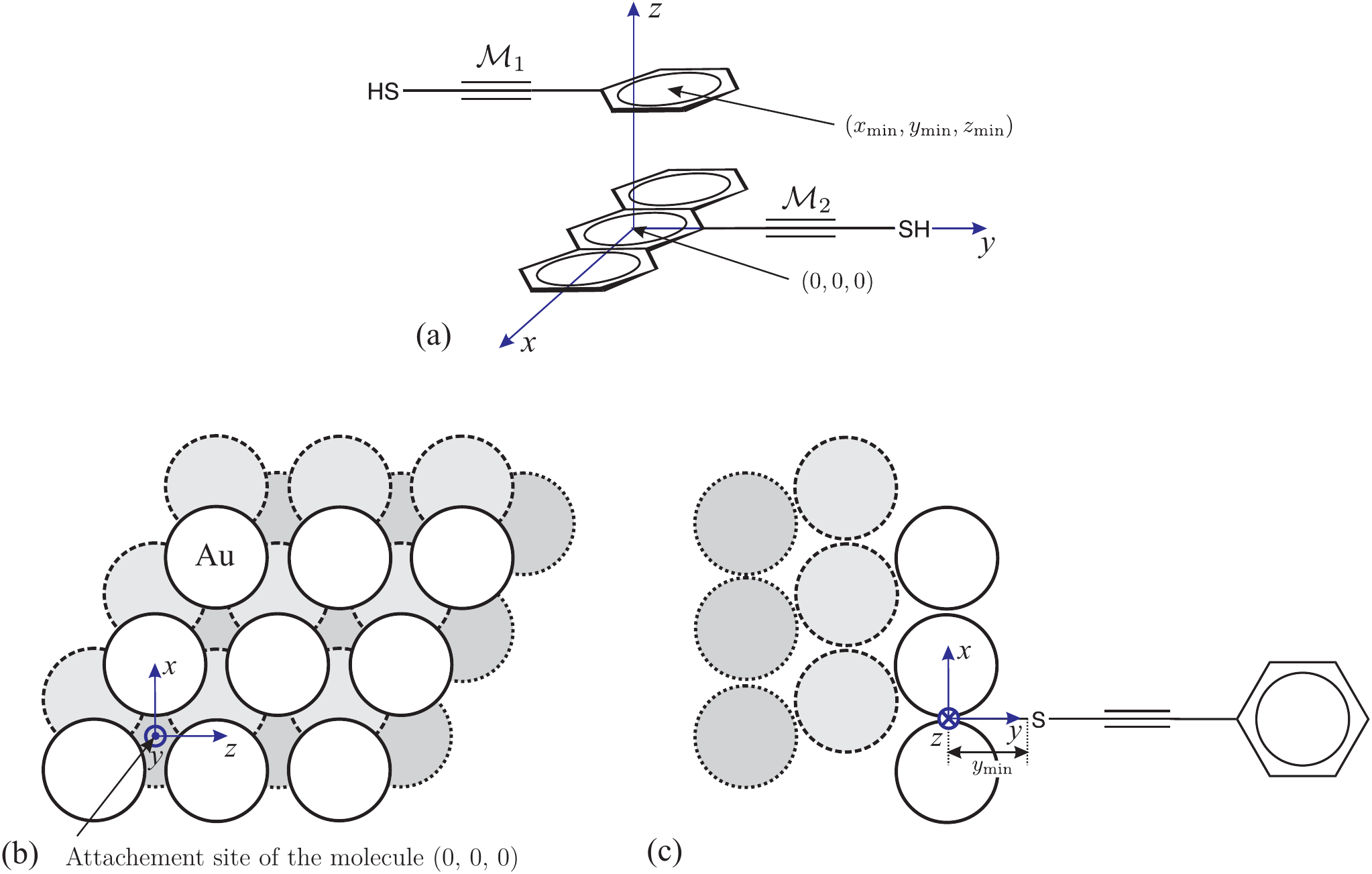}
\caption{\label{Sfig9} (a) The coordinate system (example given for the $\m{M}_{1}\m{M}_{2}$ stacking combination) used for determining the relative positions between the stacked molecules. The molecule $\m{M}_{2}$ is kept at the position $(0,0,0)$ and the molecule $\m{M}_1$ is moved. (b), (c) The configuration of the molecule attached to the hollow site of the (111) gold surface (example given for molecule $\m{M}_1$). Periodic boundary conditions are applied along $x$ and $z$ directions and three layers (c) are considered in the $y$ direction.}
\end{center}
\end{figure}

The effective spring constants for $\pi$-stacked molecules described by the model shown in \figurename~\rFigOne{a} were obtained by fitting the energy landscape around the energy minimum position to Hook's law $E \sim\frac{1}{2}K_{i}(\Delta{x_{i}})^2$, with $K_{i}$ being the spring constant and $\Delta{x_{i}}$ the displacement from the energy minimum position in the direction $x_{i}\in\{x,y,z\}$.~\cite{S_Markussen2013} The energy landscape was obtained using the density functional theory (DFT) as described in the main text.

\begin{figure}
\begin{center}
\includegraphics[width=0.76\textwidth]{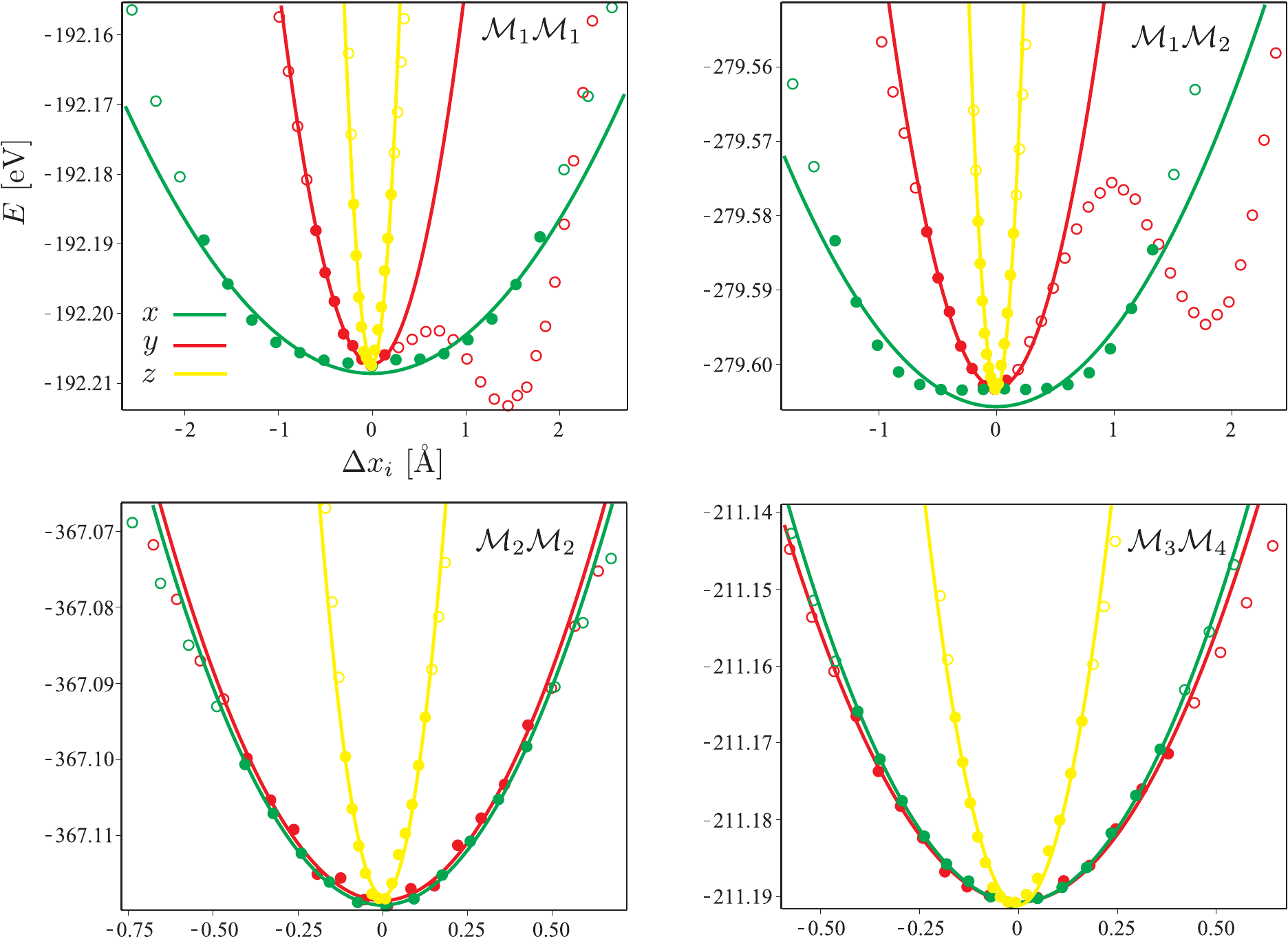}
\caption{\label{Sfig10} The energy landscape $E$ in different directions $\Delta{x}_{i}$, $i=x,y,z$ from the minimum position for isolated $\pi$-stacked molecules.}
\end{center}
\end{figure}

The middle spring constant $K_{i}$ was calculated for the two isolated $\pi$-stacked molecules (not attached to the leads). An example of the coordinate system which was used is shown in \figurename~\ref{Sfig9}a for the $\m{M}_{1}\m{M}_{2}$ stacking combination, where the molecule $\m{M}_{2}$ is kept at the position $(0,0,0)$ and the molecule $\m{M}_1$ is moved. The resulting energy landscape around the minimum position for different combinations of the stacking is shown in \figurename~\ref{Sfig10}, where the points denoted by the solid circles denote the points which were fitted (these points correspond to an energy interval from the minimum energy $E_{0}$ up to $E_{0}+k_{\mathrm{B}}T$, where $T=300 \ \mathrm{K}$ is room temperature). We see that most of the landscapes are fitted well to Hook's law, except for the landscape in the $y$ direction for the stacking combinations $\m{M}_1\m{M}_1$ and $\m{M}_1\m{M}_2$.
The landscape looks anharmonic, because an additional minima appears within the energy interval of interest due to interactions between the binding groups of one molecule with the $\pi$-system of the other.
For the stacking combination $\m{M}_1\m{M}_1$ we assume that the molecules are arranged in the local minimum position $\Delta{y}=0$ and then coupled to the leads. The coupling to the leads converts this local minimum to the global one, because the spring constants between the lead and the molecule in the $y$ direction are much larger than between the molecules and it becomes energetically more favorable to sit in the position $\Delta{y}=0$ for the chosen lead separation. The minimum positions obtained from the DFT calculation and the ones obtained after the fitting are summarized in \tablename~\ref{Stbl1}. Additionally, we note that the energy was calculated within the accuracy of $\Delta{E}=0.5 \ \mathrm{meV}$, which is the reason we are not able to specify the minimum position in the $x$ direction for stacking combinations $\m{M}_1\m{M}_1$ and $\m{M}_1\m{M}_2$ more accurately.
Also for the stacking combination $\m{M}_{2}\m{M}_2$ there are two symmetrically positioned minima of the same energy along the $x$ direction. The two minima have the same energy because one configuration of the molecules at one minimum can be mapped to the configuration at the other minimum by $180^{\circ}$ rotation around the $x$ axis of the coordinate system given in \figurename~\ref{Sfig9}a. In our calculations we consider the position given by the negative $x_{\mathrm{min}}=-1.26\text{ \AA}$, while the other minimum appears at $+1.26\text{ \AA}$. Additionally, the values of the middle spring constant $K_i$ and the resulting heat conductances $\kappatwo^{i}$, $\kappaone^{i}$ and their comparison are summarized in \tablename~\ref{tbl1}.

\begin{figure}
\begin{center}
\includegraphics[width=0.76\textwidth]{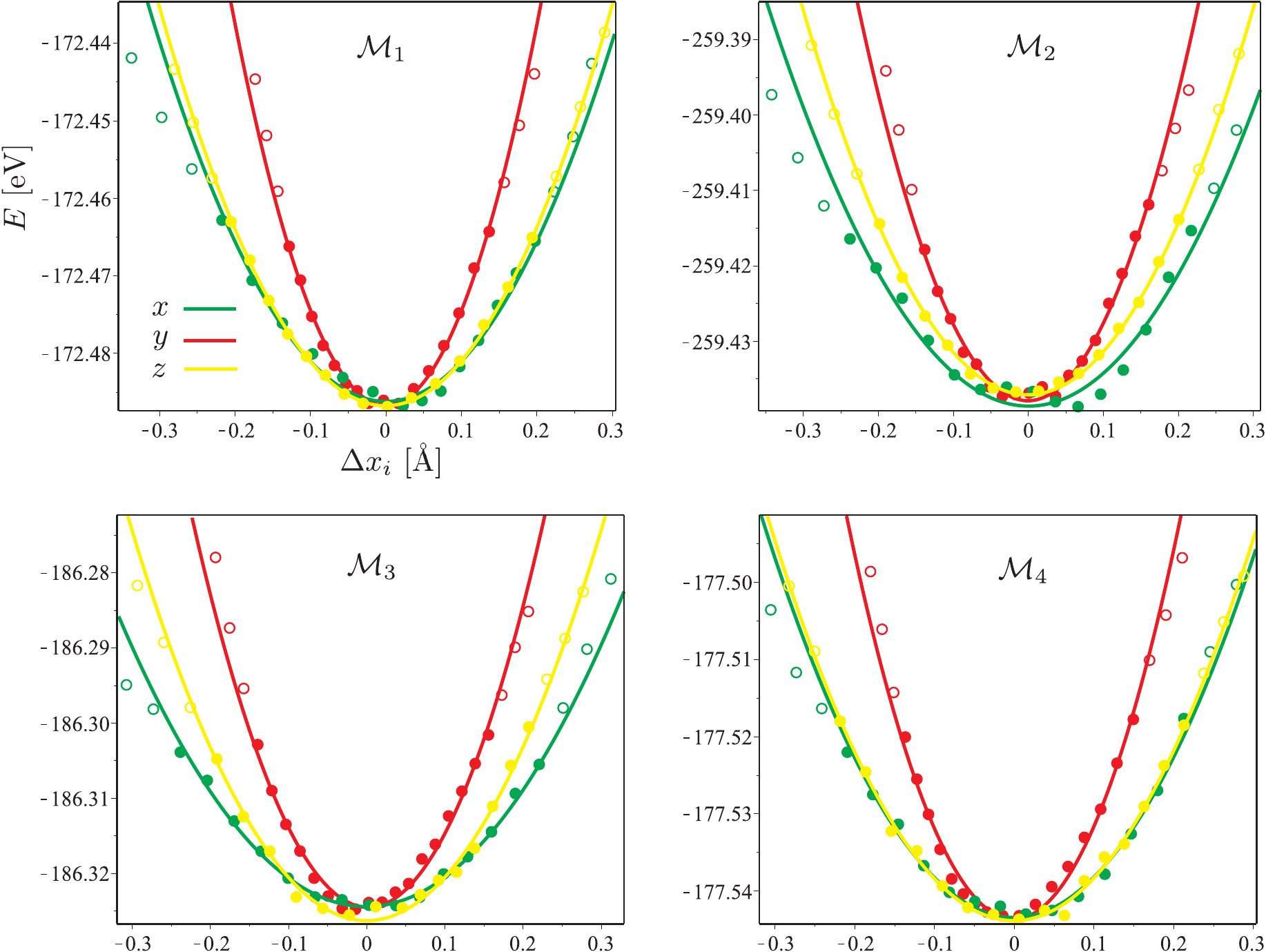}
\caption{\label{Sfig11} The energy landscape $E$ in different directions $\Delta{x}_{i}$, $i=x,y,z$ from the minimum position for different molecules attached to the hollow site of (111) surface of the gold lead.}
\end{center}
\end{figure}

The spring constants to the leads were calculated for every molecule $\m{M}_{i}$ separately. The configuration of the attached molecule on the (111) gold surface hollow site is depicted in \figurename~\ref{Sfig9}a,b. Periodic boundary conditions are applied along the $x$ and $z$ directions and in the $y$ direction three layers of gold are considered. The energy landscape for different molecules coupled to the lead is shown in \figurename~\ref{Sfig11} and the minimum position is given also in \tablename~\ref{Stbl1}.
The molecule-lead couplings, $K_{i\alpha}$, are summarized in the second part of \tablename~\ref{tbl1}. For different molecules, they are similar because the coupling strength is mainly determined by the bonding of the sulphur to the gold.
We note that within the elasticity theory the leads are isotropical in the $xz$ plane and the difference between $K_{x\alpha}$ and $K_{z\alpha}$ arise because the gold (111) surface structure and the molecules break the rotational symmetry $C_{\infty}$ in the $xz$ plane.

\begin{table}
\begin{center}
\begin{tabular}{l|cccc}
  \hline\hline
         & $\m{M}_1\m{M}_1$ & $\m{M}_1\m{M}_2$ & $\m{M}_2\m{M}_2$ & $\m{M}_3\m{M}_4$ \\
  \hline
  $x_{\mathrm{min}}$ [\AA]  & $\sim$0.00 (-0.08) & $\sim$0.00 (-0.08) & $\mp$1.26 ($\mp$1.28) & -0.56 (-0.56) \\
  $y_{\mathrm{min}}$ [\AA]  & 1.44 (1.44) & 0.76 (0.77) & 0.99 (0.97) & 1.03 (1.01) \\
  $z_{\mathrm{min}}$ [\AA]  & 3.49 (3.49) & 3.41 (3.41) & 3.41 (3.40) & 3.27 (3.27) \\
  \hline\hline
         & $\m{M}_1$ & $\m{M}_2$ & $\m{M}_3$ & $\m{M}_4$ \\
  \hline
  $x_{\mathrm{min}}$ [\AA]  & 0.05 (0.03) & 0.06 (-0.01) & -0.01 (-0.04) & -0.03 (-0.05) \\
  $y_{\mathrm{min}}$ [\AA]  & 2.18 (2.16) & 2.16 (2.20) & 2.19 (2.20) & 2.16 (2.17) \\
  $z_{\mathrm{min}}$ [\AA]  & -0.07 (-0.07) & -0.08 (-0.06) & -0.08 (-0.06) & -0.06 (-0.07) \\
  \hline\hline
\end{tabular}
\end{center}
\caption{\label{Stbl1} The minimum positions of the isolated molecules in different stacking configurations $\m{M}_i\m{M}_j$ and of the molecules $\m{M}_i$ coupled to the (111) surface of the gold lead. The first number gives the minimum position obtained from DFT and the second number in parenthesis gives the minimum position obtained after fitting the energy landscape to Hook's law. The coordinates are explained in \figurename~\ref{Sfig9}.}
\end{table}

\begin{table}
\begin{center}
\begin{tabular}{l|cccc}
  \hline\hline
         & $\m{M}_1\m{M}_1$ & $\m{M}_1\m{M}_2$ & $\m{M}_2\m{M}_2$ & $\m{M}_3\m{M}_4$ \\ 
  \hline
  $K_x/K_{0}$   & 0.014 & 0.026 & 0.290 & 0.387 \\ 
  $K_y/K_{0}$   & 0.138 & 0.156 & 0.306 & 0.356 \\ 
  $K_z/K_{0}$   & 1.469 & 2.407 & 3.760 & 2.380 \\ 
  \hline
  $\kappatwo^{x}/\kappa_{0}$  & \cbl{0.002} & \cbl{0.001} & \crd{0.078} & \crd{0.111} \\  
  $\kappatwo^{y}/\kappa_{0}$  & \cbl{0.040} & \cbl{0.027} & \crd{0.107} & \crd{0.143} \\  
  $\kappatwo^{z}/\kappa_{0}$  & \crd{0.217} & \crd{0.145} & \crd{0.132} & \crd{0.182} \\  
  \hline
  $\kappaone^{x}/\kappa_{0}$  & \cbl{0.095} & \cbl{0.062} & \crd{0.045} & \crd{0.064} \\  
  $\kappaone^{y}/\kappa_{0}$  & \cbl{0.177} & \cbl{0.126} & \crd{0.095} & \crd{0.139} \\  
  $\kappaone^{z}/\kappa_{0}$  & \crd{0.105} & \crd{0.076} & \crd{0.059} & \crd{0.084} \\  
  \hline
  $\kappatwo/\kappa_{0}$  & \cbl{0.259} & \cbl{0.173} & \crd{0.317} & \crd{0.436} \\  
  $\kappaone/\kappa_{0}$  & \cbl{0.377} & \cbl{0.264} & \crd{0.199} & \crd{0.287} \\
  $\kappatwo/\kappaone$   & \cbl{1/1.456} & \cbl{1/1.527} & \crd{1.593} & \crd{1.519} \\
  \hline\hline
         & $\m{M}_1$ & $\m{M}_2$ & $\m{M}_3$ & $\m{M}_4$ \\ 
  \hline
  $K_{x\alpha}/K_{0}$   & 1.303 & 1.109 & 0.974 & 1.309 \\ 
  $K_{y\alpha}/K_{0}$   & 3.109 & 2.577 & 2.583 & 3.001 \\ 
  $K_{z\alpha}/K_{0}$   & 1.446 & 1.436 & 1.459 & 1.386 \\ 
  \hline\hline
\end{tabular}
\end{center}
\caption{\label{tbl1} The coupling to the leads $K_{i\alpha}$ and the middle spring constant $K_i$ obtained from DFT, the resulting phonon conductances $\kappatwo^{i}$, $\kappaone^{i}$ in three different directions, and comparison of total phonon conductance $\kappatwo$ between two masses and a single mass phonon conductance $\kappaone$. The blue values of conductance denote situation when $\kappatwo<\kappaone$ and the red ones \textit{vice versa}.}
\end{table}

\subsection{Electronic transmission $T_{\mre}$}

\begin{figure}
\begin{center}
\includegraphics[width=0.76\textwidth]{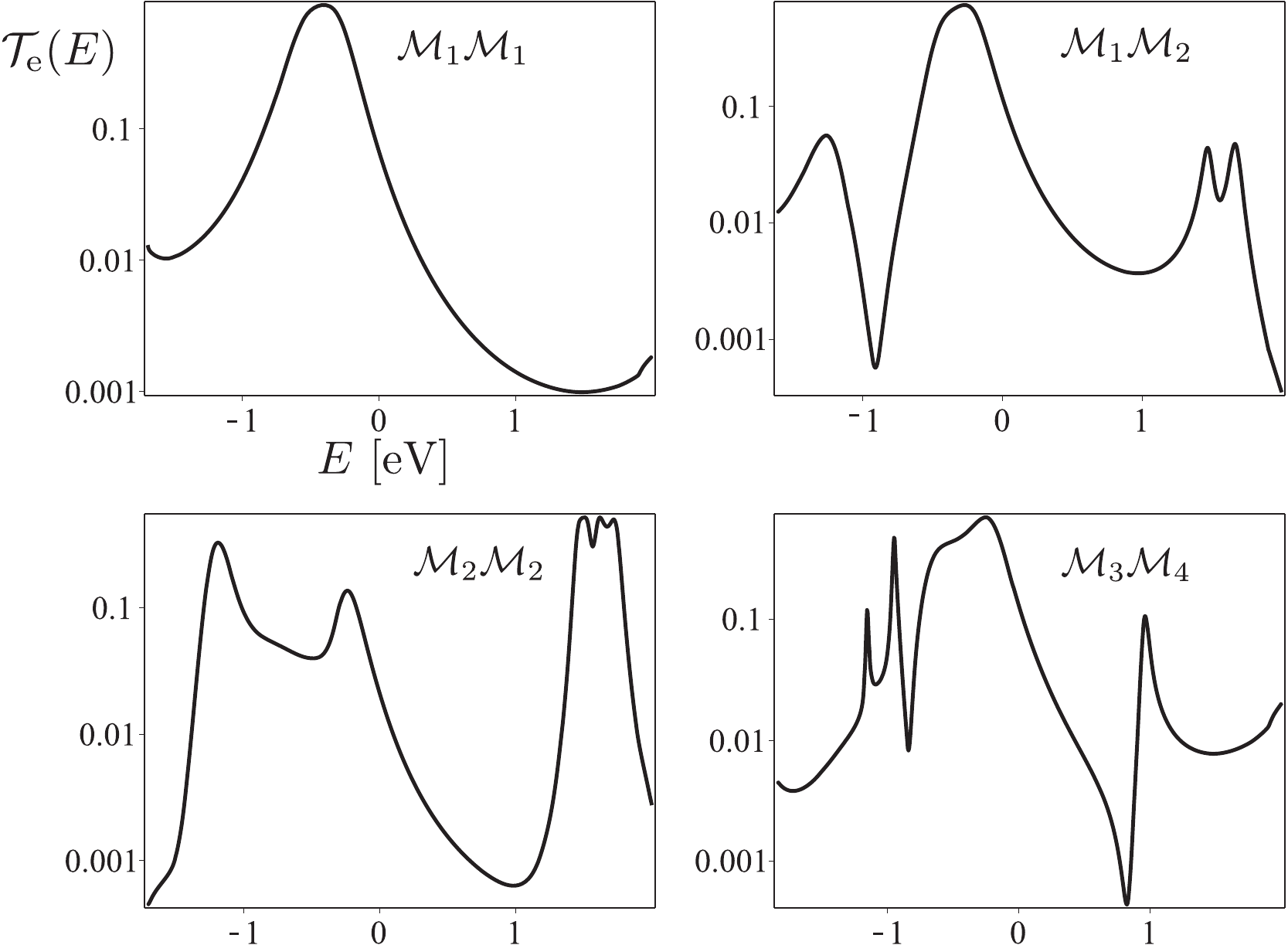}
\caption{\label{Sfig12} The electronic transmission $\m{T}_{\mre}(E)$ obtained from DFT at the minimum position for different stackings.}
\end{center}
\end{figure}

The current through a molecule is calculated with a Green's function implementation of the Landauer-B{\"{u}}ttiker approach:
\begin{equation}
I(V) = \frac{2e}{h}\int^{\infty}_{-\infty}\dif{E}(f_{\mrL}-f_{\mrR})\m{T}_{\mre}(E),
\end{equation}
where $f=1/\{\exp[\beta_{\alpha}(E-\mu_{\alpha})]+1\}$ is the Fermi function for the electrode $\alpha=\mrL,\mrR$ with the inverse temperature $\beta_{\alpha}=1/(k_{\mathrm{B}}T)$, the chemical potential $\mu_{\alpha}$, and $\m{T}_{\mre}(E)$ is the energy dependent transmission. Following the standard DFT-Landauer approach, we calculate the zero-bias transmission function
\begin{equation}
\m{T}_{\mre}(E)=\Tr[\Gamma_{\mrL}G^{R}\Gamma_{\mrR}G^{A}],
\end{equation}		
where $\Gamma_{\mrL}$ and $\Gamma_{\mrR}$ are half the imaginary parts of the left and right electrode self-energies, respectively, and $G^{R}$ and $G^{A}$ are the electron retarded and advanced Green's functions of the scattering region. The electronic transmissions are calculated with the GPAW DFT code using an atomic orbital basis set corresponding to double-zeta plus polarization and the Perdew-Burke-Ernzerhof (PBE) exchange-correlation functional. The Monckhorst-Pack $k$-point sampling is $4\times4\times1$. The molecules are chemisorbed (the terminal hydrogen atoms are removed) onto an FCC (111) hollow site with an optimal distance to the Au surface obtained from the adsorption energy curve. The calculated electronic transmission $\m{T}_{\mre}(E)$ at the minimum position for different stacking combinations is shown in \figurename~\ref{Sfig12}.

 When the inelastic scattering is neglected, the Seebeck coefficient $S$, the electronic conductance $\sigma_{\mre}$, and the electronic heat conductance $\kappa_{\mre}$ can be expressed through the electron transmission $\m{T}_{\mre}(E)$ as~\cite{S_Sivan1986}
\begin{equation}
\sigma_{\mre}=e^2 L_{0}, \quad
S=\frac{1}{eT}\frac{L_{1}}{L_{0}}, \quad
\kappa_{\mre}=\frac{1}{T}\left(L_{2}-\frac{L_1^2}{L_{0}}\right),
\end{equation}
where
\begin{equation}\label{Lmf}
L_{m}=\frac{2}{h}\int_{-\infty}^{+\infty}\dif{E}(E-\mu)^{m}\left(-\frac{\pd f(E,\mu,T)}{\pd E}\right)\m{T}_{\mre}(E).
\end{equation}

\subsection{Figure of merit $ZT$}

\begin{figure}[ht!]
\begin{center}
\includegraphics[width=0.76\columnwidth]{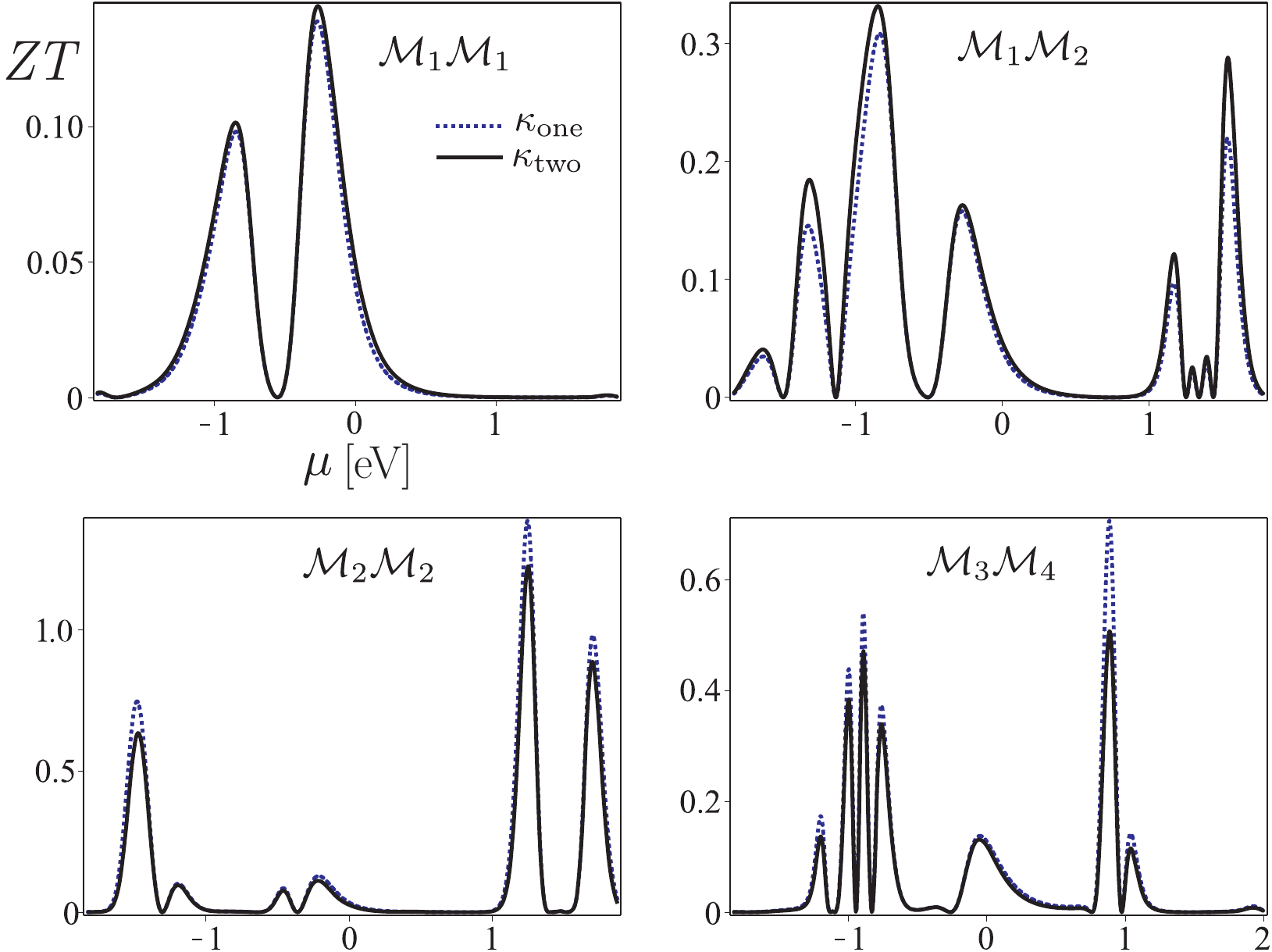}
\caption{\label{Sfig8} The figure of merit $ZT$ as a function of the chemical potential $\mu$ at $T=300 \ \mathrm{K}$ for all considered stackings. The solid black lines show $ZT$ when system has a weak link in the middle (two masses model) and dashed blue lines show $ZT$ when the center of mass vibrational degrees of freedom are described by single mass in the junction. }
\end{center}
\end{figure}

The resulting figure of merit $ZT$ at room temperature $T=300 \ \mathrm{K}$ for all considered configurations of stacking is depicted in \figurename~\ref{Sfig8}. As mentioned in the main text for the stacking $\m{M}_1\m{M}_2$, which had the largest reduction in the phonon conductance, the optimized values of $ZT$ as a function of chemical potential $\mu$, get the largest increase. For the stacking $\m{M}_1\m{M}_1$, at the optimized values of $ZT$, the main contribution to the thermal transport comes from the electrons, and the influence of the decreased phonon conductance is not substantial. The stacking combinations $\m{M}_2\m{M}_2$ and $\m{M}_3\m{M}_4$ have an increased thermal conductance.


\begin{thebibliography}{50}%
\makeatletter
\providecommand \@ifxundefined [1]{%
 \@ifx{#1\undefined}
}%
\providecommand \@ifnum [1]{%
 \ifnum #1\expandafter \@firstoftwo
 \else \expandafter \@secondoftwo
 \fi
}%
\providecommand \@ifx [1]{%
 \ifx #1\expandafter \@firstoftwo
 \else \expandafter \@secondoftwo
 \fi
}%
\providecommand \natexlab [1]{#1}%
\providecommand \enquote  [1]{``#1''}%
\providecommand \bibnamefont  [1]{#1}%
\providecommand \bibfnamefont [1]{#1}%
\providecommand \citenamefont [1]{#1}%
\providecommand \href@noop [0]{\@secondoftwo}%
\providecommand \href [0]{\begingroup \@sanitize@url \@href}%
\providecommand \@href[1]{\@@startlink{#1}\@@href}%
\providecommand \@@href[1]{\endgroup#1\@@endlink}%
\providecommand \@sanitize@url [0]{\catcode `\\12\catcode `\$12\catcode
  `\&12\catcode `\#12\catcode `\^12\catcode `\_12\catcode `\%12\relax}%
\providecommand \@@startlink[1]{}%
\providecommand \@@endlink[0]{}%
\providecommand \url  [0]{\begingroup\@sanitize@url \@url }%
\providecommand \@url [1]{\endgroup\@href {#1}{\urlprefix }}%
\providecommand \urlprefix  [0]{URL }%
\providecommand \Eprint [0]{\href }%
\providecommand \doibase [0]{http://dx.doi.org/}%
\providecommand \selectlanguage [0]{\@gobble}%
\providecommand \bibinfo  [0]{\@secondoftwo}%
\providecommand \bibfield  [0]{\@secondoftwo}%
\providecommand \translation [1]{[#1]}%
\providecommand \BibitemOpen [0]{}%
\providecommand \bibitemStop [0]{}%
\providecommand \bibitemNoStop [0]{.\EOS\space}%
\providecommand \EOS [0]{\spacefactor3000\relax}%
\providecommand \BibitemShut  [1]{\csname bibitem#1\endcsname}%
\let\auto@bib@innerbib\@empty
\bibitem [{\citenamefont {Giazotto}\ \emph {et~al.}(2006)\citenamefont
  {Giazotto}, \citenamefont {Heikkil\"{a}}, \citenamefont {Luukanen},
  \citenamefont {Savin},\ and\ \citenamefont {Pekola}}]{Giazotto2006}%
  \BibitemOpen
  \bibfield  {author} {\bibinfo {author} {\bibfnamefont {F.}~\bibnamefont
  {Giazotto}}, \bibinfo {author} {\bibfnamefont {T.}~\bibnamefont
  {Heikkil\"{a}}}, \bibinfo {author} {\bibfnamefont {A.}~\bibnamefont
  {Luukanen}}, \bibinfo {author} {\bibfnamefont {A.}~\bibnamefont {Savin}}, \
  and\ \bibinfo {author} {\bibfnamefont {J.}~\bibnamefont {Pekola}},\ }\href
  {\doibase 10.1103/RevModPhys.78.217} {\bibfield  {journal} {\bibinfo
  {journal} {Rev. Mod. Phys.}\ }\textbf {\bibinfo {volume} {78}},\ \bibinfo
  {pages} {217} (\bibinfo {year} {2006})}\BibitemShut {NoStop}%
\bibitem [{\citenamefont {Dubi}\ and\ \citenamefont
  {{Di Ventra}}(2011)}]{Dubi2011}%
  \BibitemOpen
  \bibfield  {author} {\bibinfo {author} {\bibfnamefont {Y.}~\bibnamefont
  {Dubi}}\ and\ \bibinfo {author} {\bibfnamefont {M.}~\bibnamefont
  {{Di Ventra}}},\ }\href {\doibase 10.1103/RevModPhys.83.131} {\bibfield
  {journal} {\bibinfo  {journal} {Rev. Mod. Phys.}\ }\textbf {\bibinfo {volume}
  {83}},\ \bibinfo {pages} {131} (\bibinfo {year} {2011})}\BibitemShut
  {NoStop}%
\bibitem [{\citenamefont {Rowe}(2005)}]{Rowe2005}%
  \BibitemOpen
  \bibfield  {author} {\bibinfo {author} {\bibfnamefont {D.~M.}\ \bibnamefont
  {Rowe}},\ }\href@noop {} {\emph {\bibinfo {title} {{Thermoelectrics Handbook:
  Macro to Nano}}}}\ (\bibinfo  {publisher} {Taylor \& Francis},\ \bibinfo
  {year} {2005})\BibitemShut {NoStop}%
\bibitem [{\citenamefont {Goldsmid}(1964)}]{Goldsmid1964}%
  \BibitemOpen
  \bibfield  {author} {\bibinfo {author} {\bibfnamefont {H.~J.}\ \bibnamefont
  {Goldsmid}},\ }\href@noop {} {\emph {\bibinfo {title} {{Thermoelectric
  Refrigeration}}}}\ (\bibinfo  {publisher} {Plenum Press},\ \bibinfo {address}
  {New York},\ \bibinfo {year} {1964})\BibitemShut {NoStop}%
\bibitem [{\citenamefont {Snyder}\ and\ \citenamefont
  {Toberer}(2008)}]{Snyder2008}%
  \BibitemOpen
  \bibfield  {author} {\bibinfo {author} {\bibfnamefont {G.~J.}\ \bibnamefont
  {Snyder}}\ and\ \bibinfo {author} {\bibfnamefont {E.~S.}\ \bibnamefont
  {Toberer}},\ }\href {\doibase 10.1038/nmat2090} {\bibfield  {journal}
  {\bibinfo  {journal} {Nat. Mater.}\ }\textbf {\bibinfo {volume} {7}},\
  \bibinfo {pages} {105} (\bibinfo {year} {2008})}\BibitemShut {NoStop}%
\bibitem [{\citenamefont {Segal}, \citenamefont {Nitzan},\ and\ \citenamefont
  {H\"{a}nggi}(2003)}]{Segal2003}%
  \BibitemOpen
  \bibfield  {author} {\bibinfo {author} {\bibfnamefont {D.}~\bibnamefont
  {Segal}}, \bibinfo {author} {\bibfnamefont {A.}~\bibnamefont {Nitzan}}, \
  and\ \bibinfo {author} {\bibfnamefont {P.}~\bibnamefont {H\"{a}nggi}},\
  }\href {\doibase 10.1063/1.1603211} {\bibfield  {journal} {\bibinfo
  {journal} {J. Chem. Phys.}\ }\textbf {\bibinfo {volume} {119}},\ \bibinfo
  {pages} {6840} (\bibinfo {year} {2003})}\BibitemShut {NoStop}%
\bibitem [{\citenamefont {Mingo}\ \emph {et~al.}(2003)\citenamefont {Mingo},
  \citenamefont {Yang}, \citenamefont {Li},\ and\ \citenamefont
  {Majumdar}}]{Mingo2003}%
  \BibitemOpen
  \bibfield  {author} {\bibinfo {author} {\bibfnamefont {N.}~\bibnamefont
  {Mingo}}, \bibinfo {author} {\bibfnamefont {L.}~\bibnamefont {Yang}},
  \bibinfo {author} {\bibfnamefont {D.}~\bibnamefont {Li}}, \ and\ \bibinfo
  {author} {\bibfnamefont {A.}~\bibnamefont {Majumdar}},\ }\href {\doibase
  10.1021/nl034721i} {\bibfield  {journal} {\bibinfo  {journal} {Nano Lett.}\
  }\textbf {\bibinfo {volume} {3}},\ \bibinfo {pages} {1713} (\bibinfo {year}
  {2003})}\BibitemShut {NoStop}%
\bibitem [{\citenamefont {Wang}\ and\ \citenamefont {Wang}(2007)}]{Wang2007a}%
  \BibitemOpen
  \bibfield  {author} {\bibinfo {author} {\bibfnamefont {J.}~\bibnamefont
  {Wang}}\ and\ \bibinfo {author} {\bibfnamefont {J.-S.}\ \bibnamefont
  {Wang}},\ }\href {\doibase 10.1063/1.2748342} {\bibfield  {journal} {\bibinfo
   {journal} {Appl. Phys. Lett.}\ }\textbf {\bibinfo {volume} {90}},\ \bibinfo
  {pages} {241908} (\bibinfo {year} {2007})}\BibitemShut {NoStop}%
\bibitem [{\citenamefont {Galperin}, \citenamefont {Nitzan},\ and\
  \citenamefont {Ratner}(2007)}]{Galperin2007}%
  \BibitemOpen
  \bibfield  {author} {\bibinfo {author} {\bibfnamefont {M.}~\bibnamefont
  {Galperin}}, \bibinfo {author} {\bibfnamefont {A.}~\bibnamefont {Nitzan}}, \
  and\ \bibinfo {author} {\bibfnamefont {M.}~\bibnamefont {Ratner}},\ }\href
  {\doibase 10.1103/PhysRevB.75.155312} {\bibfield  {journal} {\bibinfo
  {journal} {Phys. Rev. B}\ }\textbf {\bibinfo {volume} {75}},\ \bibinfo
  {pages} {155312} (\bibinfo {year} {2007})}\BibitemShut {NoStop}%
\bibitem [{\citenamefont {Markussen}, \citenamefont {Jauho},\ and\
  \citenamefont {Brandbyge}(2008)}]{Markussen2008}%
  \BibitemOpen
  \bibfield  {author} {\bibinfo {author} {\bibfnamefont {T.}~\bibnamefont
  {Markussen}}, \bibinfo {author} {\bibfnamefont {A.-P.}\ \bibnamefont
  {Jauho}}, \ and\ \bibinfo {author} {\bibfnamefont {M.}~\bibnamefont
  {Brandbyge}},\ }\href {\doibase 10.1021/nl8020889} {\bibfield  {journal}
  {\bibinfo  {journal} {Nano Lett.}\ }\textbf {\bibinfo {volume} {8}},\
  \bibinfo {pages} {3771} (\bibinfo {year} {2008})}\BibitemShut {NoStop}%
\bibitem [{\citenamefont {Boukai}\ \emph {et~al.}(2008)\citenamefont {Boukai},
  \citenamefont {Bunimovich}, \citenamefont {Tahir-Kheli}, \citenamefont {Yu},
  \citenamefont {{Goddard III}},\ and\ \citenamefont {Heath}}]{Boukai2008}%
  \BibitemOpen
  \bibfield  {author} {\bibinfo {author} {\bibfnamefont {A.~I.}\ \bibnamefont
  {Boukai}}, \bibinfo {author} {\bibfnamefont {Y.}~\bibnamefont {Bunimovich}},
  \bibinfo {author} {\bibfnamefont {J.}~\bibnamefont {Tahir-Kheli}}, \bibinfo
  {author} {\bibfnamefont {J.-K.}\ \bibnamefont {Yu}}, \bibinfo {author}
  {\bibfnamefont {W.~A.}\ \bibnamefont {{Goddard III}}}, \ and\ \bibinfo
  {author} {\bibfnamefont {J.~R.}\ \bibnamefont {Heath}},\ }\href {\doibase
  10.1038/nature06458} {\bibfield  {journal} {\bibinfo  {journal} {Nature}\
  }\textbf {\bibinfo {volume} {451}},\ \bibinfo {pages} {168} (\bibinfo {year}
  {2008})}\BibitemShut {NoStop}%
\bibitem [{\citenamefont {Hochbaum}\ \emph {et~al.}(2008)\citenamefont
  {Hochbaum}, \citenamefont {Chen}, \citenamefont {Delgado}, \citenamefont
  {Liang}, \citenamefont {Garnett}, \citenamefont {Najarian}, \citenamefont
  {Majumdar},\ and\ \citenamefont {Yang}}]{Hochbaum2008}%
  \BibitemOpen
  \bibfield  {author} {\bibinfo {author} {\bibfnamefont {A.~I.}\ \bibnamefont
  {Hochbaum}}, \bibinfo {author} {\bibfnamefont {R.}~\bibnamefont {Chen}},
  \bibinfo {author} {\bibfnamefont {R.~D.}\ \bibnamefont {Delgado}}, \bibinfo
  {author} {\bibfnamefont {W.}~\bibnamefont {Liang}}, \bibinfo {author}
  {\bibfnamefont {E.~C.}\ \bibnamefont {Garnett}}, \bibinfo {author}
  {\bibfnamefont {M.}~\bibnamefont {Najarian}}, \bibinfo {author}
  {\bibfnamefont {A.}~\bibnamefont {Majumdar}}, \ and\ \bibinfo {author}
  {\bibfnamefont {P.}~\bibnamefont {Yang}},\ }\href {\doibase
  10.1038/nature06381} {\bibfield  {journal} {\bibinfo  {journal} {Nature}\
  }\textbf {\bibinfo {volume} {451}},\ \bibinfo {pages} {163} (\bibinfo {year}
  {2008})}\BibitemShut {NoStop}%
\bibitem [{\citenamefont {Blanc}\ \emph {et~al.}(2013)\citenamefont {Blanc},
  \citenamefont {Rajabpour}, \citenamefont {Volz}, \citenamefont {Fournier},\
  and\ \citenamefont {Bourgeois}}]{Blanc2013}%
  \BibitemOpen
  \bibfield  {author} {\bibinfo {author} {\bibfnamefont {C.}~\bibnamefont
  {Blanc}}, \bibinfo {author} {\bibfnamefont {A.}~\bibnamefont {Rajabpour}},
  \bibinfo {author} {\bibfnamefont {S.}~\bibnamefont {Volz}}, \bibinfo {author}
  {\bibfnamefont {T.}~\bibnamefont {Fournier}}, \ and\ \bibinfo {author}
  {\bibfnamefont {O.}~\bibnamefont {Bourgeois}},\ }\href {\doibase
  10.1063/1.4816590} {\bibfield  {journal} {\bibinfo  {journal} {Appl. Phys.
  Lett.}\ }\textbf {\bibinfo {volume} {103}},\ \bibinfo {pages} {043109}
  (\bibinfo {year} {2013})},\ \Eprint {http://arxiv.org/abs/arXiv:1302.4422v2}
  {arXiv:1302.4422v2} \BibitemShut {NoStop}%
\bibitem [{\citenamefont {Hicks}\ and\ \citenamefont
  {Dresselhaus}(1993)}]{Hicks1993a}%
  \BibitemOpen
  \bibfield  {author} {\bibinfo {author} {\bibfnamefont {L.}~\bibnamefont
  {Hicks}}\ and\ \bibinfo {author} {\bibfnamefont {M.}~\bibnamefont
  {Dresselhaus}},\ }\href {\doibase 10.1103/PhysRevB.47.16631} {\bibfield
  {journal} {\bibinfo  {journal} {Phys. Rev. B}\ }\textbf {\bibinfo {volume}
  {47}},\ \bibinfo {pages} {16631} (\bibinfo {year} {1993})}\BibitemShut
  {NoStop}%
\bibitem [{\citenamefont {Majumdar}(2004)}]{Majumdar2004}%
  \BibitemOpen
  \bibfield  {author} {\bibinfo {author} {\bibfnamefont {A.}~\bibnamefont
  {Majumdar}},\ }\href {\doibase 10.1126/science.1093164} {\bibfield  {journal}
  {\bibinfo  {journal} {Science}\ }\textbf {\bibinfo {volume} {303}},\ \bibinfo
  {pages} {777} (\bibinfo {year} {2004})}\BibitemShut {NoStop}%
\bibitem [{\citenamefont {Dresselhaus}\ \emph {et~al.}(2007)\citenamefont
  {Dresselhaus}, \citenamefont {Chen}, \citenamefont {Tang}, \citenamefont
  {Yang}, \citenamefont {Lee}, \citenamefont {Wang}, \citenamefont {Ren},
  \citenamefont {Fleurial},\ and\ \citenamefont {Gogna}}]{Dresselhaus2007}%
  \BibitemOpen
  \bibfield  {author} {\bibinfo {author} {\bibfnamefont {M.}~\bibnamefont
  {Dresselhaus}}, \bibinfo {author} {\bibfnamefont {G.}~\bibnamefont {Chen}},
  \bibinfo {author} {\bibfnamefont {M.}~\bibnamefont {Tang}}, \bibinfo {author}
  {\bibfnamefont {R.}~\bibnamefont {Yang}}, \bibinfo {author} {\bibfnamefont
  {H.}~\bibnamefont {Lee}}, \bibinfo {author} {\bibfnamefont {D.}~\bibnamefont
  {Wang}}, \bibinfo {author} {\bibfnamefont {Z.}~\bibnamefont {Ren}}, \bibinfo
  {author} {\bibfnamefont {J.-P.}\ \bibnamefont {Fleurial}}, \ and\ \bibinfo
  {author} {\bibfnamefont {P.}~\bibnamefont {Gogna}},\ }\href {\doibase
  10.1002/adma.200600527} {\bibfield  {journal} {\bibinfo  {journal} {Adv.
  Mater.}\ }\textbf {\bibinfo {volume} {19}},\ \bibinfo {pages} {1043}
  (\bibinfo {year} {2007})}\BibitemShut {NoStop}%
\bibitem [{\citenamefont {Appleyard}\ \emph {et~al.}(2000)\citenamefont
  {Appleyard}, \citenamefont {Nicholls}, \citenamefont {Pepper}, \citenamefont
  {Tribe}, \citenamefont {Simmons},\ and\ \citenamefont
  {Ritchie}}]{Appleyard2000}%
  \BibitemOpen
  \bibfield  {author} {\bibinfo {author} {\bibfnamefont {N.}~\bibnamefont
  {Appleyard}}, \bibinfo {author} {\bibfnamefont {J.}~\bibnamefont {Nicholls}},
  \bibinfo {author} {\bibfnamefont {M.}~\bibnamefont {Pepper}}, \bibinfo
  {author} {\bibfnamefont {W.}~\bibnamefont {Tribe}}, \bibinfo {author}
  {\bibfnamefont {M.}~\bibnamefont {Simmons}}, \ and\ \bibinfo {author}
  {\bibfnamefont {D.}~\bibnamefont {Ritchie}},\ }\href {\doibase
  10.1103/PhysRevB.62.R16275} {\bibfield  {journal} {\bibinfo  {journal} {Phys.
  Rev. B}\ }\textbf {\bibinfo {volume} {62}},\ \bibinfo {pages} {R16275}
  (\bibinfo {year} {2000})}\BibitemShut {NoStop}%
\bibitem [{\citenamefont {Boese}\ and\ \citenamefont
  {Fazio}(2001)}]{Boese2001}%
  \BibitemOpen
  \bibfield  {author} {\bibinfo {author} {\bibfnamefont {D.}~\bibnamefont
  {Boese}}\ and\ \bibinfo {author} {\bibfnamefont {R.}~\bibnamefont {Fazio}},\
  }\href {\doibase 10.1209/epl/i2001-00559-8} {\bibfield  {journal} {\bibinfo
  {journal} {EPL-Europhys. Lett.}\ }\textbf {\bibinfo {volume} {56}},\ \bibinfo
  {pages} {576} (\bibinfo {year} {2001})}\BibitemShut {NoStop}%
\bibitem [{\citenamefont {Kubala}, \citenamefont {K\"{o}nig},\ and\
  \citenamefont {Pekola}(2008)}]{Kubala2008}%
  \BibitemOpen
  \bibfield  {author} {\bibinfo {author} {\bibfnamefont {B.}~\bibnamefont
  {Kubala}}, \bibinfo {author} {\bibfnamefont {J.}~\bibnamefont {K\"{o}nig}}, \
  and\ \bibinfo {author} {\bibfnamefont {J.}~\bibnamefont {Pekola}},\ }\href
  {\doibase 10.1103/PhysRevLett.100.066801} {\bibfield  {journal} {\bibinfo
  {journal} {Phys. Rev. Lett.}\ }\textbf {\bibinfo {volume} {100}},\ \bibinfo
  {pages} {066801} (\bibinfo {year} {2008})}\BibitemShut {NoStop}%
\bibitem [{\citenamefont {Reddy}\ \emph {et~al.}(2007)\citenamefont {Reddy},
  \citenamefont {Jang}, \citenamefont {Segalman},\ and\ \citenamefont
  {Majumdar}}]{Reddy2007}%
  \BibitemOpen
  \bibfield  {author} {\bibinfo {author} {\bibfnamefont {P.}~\bibnamefont
  {Reddy}}, \bibinfo {author} {\bibfnamefont {S.-Y.}\ \bibnamefont {Jang}},
  \bibinfo {author} {\bibfnamefont {R.~A.}\ \bibnamefont {Segalman}}, \ and\
  \bibinfo {author} {\bibfnamefont {A.}~\bibnamefont {Majumdar}},\ }\href
  {\doibase 10.1126/science.1137149} {\bibfield  {journal} {\bibinfo  {journal}
  {Science}\ }\textbf {\bibinfo {volume} {315}},\ \bibinfo {pages} {1568}
  (\bibinfo {year} {2007})}\BibitemShut {NoStop}%
\bibitem [{\citenamefont {Murphy}, \citenamefont {Mukerjee},\ and\
  \citenamefont {Moore}(2008)}]{Murphy2008}%
  \BibitemOpen
  \bibfield  {author} {\bibinfo {author} {\bibfnamefont {P.}~\bibnamefont
  {Murphy}}, \bibinfo {author} {\bibfnamefont {S.}~\bibnamefont {Mukerjee}}, \
  and\ \bibinfo {author} {\bibfnamefont {J.}~\bibnamefont {Moore}},\ }\href
  {\doibase 10.1103/PhysRevB.78.161406} {\bibfield  {journal} {\bibinfo
  {journal} {Phys. Rev. B}\ }\textbf {\bibinfo {volume} {78}},\ \bibinfo
  {pages} {161406} (\bibinfo {year} {2008})}\BibitemShut {NoStop}%
\bibitem [{\citenamefont {Finch}, \citenamefont {Garcia-Suarez},\ and\
  \citenamefont {Lambert}(2009)}]{Finch2009}%
  \BibitemOpen
  \bibfield  {author} {\bibinfo {author} {\bibfnamefont {C.}~\bibnamefont
  {Finch}}, \bibinfo {author} {\bibfnamefont {V.}~\bibnamefont
  {Garcia-Suarez}}, \ and\ \bibinfo {author} {\bibfnamefont {C.}~\bibnamefont
  {Lambert}},\ }\href {\doibase 10.1103/PhysRevB.79.033405} {\bibfield
  {journal} {\bibinfo  {journal} {Phys. Rev. B}\ }\textbf {\bibinfo {volume}
  {79}},\ \bibinfo {pages} {033405} (\bibinfo {year} {2009})}\BibitemShut
  {NoStop}%
\bibitem [{\citenamefont {Solomon}\ \emph {et~al.}(2008)\citenamefont
  {Solomon}, \citenamefont {Andrews}, \citenamefont {Hansen}, \citenamefont
  {Goldsmith}, \citenamefont {Wasielewski}, \citenamefont {{Van Duyne}},\ and\
  \citenamefont {Ratner}}]{Solomon2008}%
  \BibitemOpen
  \bibfield  {author} {\bibinfo {author} {\bibfnamefont {G.~C.}\ \bibnamefont
  {Solomon}}, \bibinfo {author} {\bibfnamefont {D.~Q.}\ \bibnamefont
  {Andrews}}, \bibinfo {author} {\bibfnamefont {T.}~\bibnamefont {Hansen}},
  \bibinfo {author} {\bibfnamefont {R.~H.}\ \bibnamefont {Goldsmith}}, \bibinfo
  {author} {\bibfnamefont {M.~R.}\ \bibnamefont {Wasielewski}}, \bibinfo
  {author} {\bibfnamefont {R.~P.}\ \bibnamefont {{Van Duyne}}}, \ and\ \bibinfo
  {author} {\bibfnamefont {M.~A.}\ \bibnamefont {Ratner}},\ }\href {\doibase
  10.1063/1.2958275} {\bibfield  {journal} {\bibinfo  {journal} {J. Chem.
  Phys.}\ }\textbf {\bibinfo {volume} {129}},\ \bibinfo {pages} {054701}
  (\bibinfo {year} {2008})}\BibitemShut {NoStop}%
\bibitem [{\citenamefont {Karlstr\"{o}m}\ \emph {et~al.}(2011)\citenamefont
  {Karlstr\"{o}m}, \citenamefont {Linke}, \citenamefont {Karlstr\"{o}m},\ and\
  \citenamefont {Wacker}}]{Karlstrom2011}%
  \BibitemOpen
  \bibfield  {author} {\bibinfo {author} {\bibfnamefont {O.}~\bibnamefont
  {Karlstr\"{o}m}}, \bibinfo {author} {\bibfnamefont {H.}~\bibnamefont
  {Linke}}, \bibinfo {author} {\bibfnamefont {G.}~\bibnamefont
  {Karlstr\"{o}m}}, \ and\ \bibinfo {author} {\bibfnamefont {A.}~\bibnamefont
  {Wacker}},\ }\href {\doibase 10.1103/PhysRevB.84.113415} {\bibfield
  {journal} {\bibinfo  {journal} {Phys. Rev. B}\ }\textbf {\bibinfo {volume}
  {84}},\ \bibinfo {pages} {113415} (\bibinfo {year} {2011})}\BibitemShut
  {NoStop}%
\bibitem [{\citenamefont {Bergfield}\ and\ \citenamefont
  {Stafford}(2009)}]{Bergfield2009}%
  \BibitemOpen
  \bibfield  {author} {\bibinfo {author} {\bibfnamefont {J.~P.}\ \bibnamefont
  {Bergfield}}\ and\ \bibinfo {author} {\bibfnamefont {C.~A.}\ \bibnamefont
  {Stafford}},\ }\href {\doibase 10.1021/nl901554s} {\bibfield  {journal}
  {\bibinfo  {journal} {Nano Lett.}\ }\textbf {\bibinfo {volume} {9}},\
  \bibinfo {pages} {3072} (\bibinfo {year} {2009})}\BibitemShut {NoStop}%
\bibitem [{\citenamefont {Seldenthuis}\ \emph {et~al.}(2008)\citenamefont
  {Seldenthuis}, \citenamefont {van~der Zant}, \citenamefont {Ratner},\ and\
  \citenamefont {Thijssen}}]{Seldenthuis2008}%
  \BibitemOpen
  \bibfield  {author} {\bibinfo {author} {\bibfnamefont {J.~S.}\ \bibnamefont
  {Seldenthuis}}, \bibinfo {author} {\bibfnamefont {H.~S.~J.}\ \bibnamefont
  {van~der Zant}}, \bibinfo {author} {\bibfnamefont {M.~A.}\ \bibnamefont
  {Ratner}}, \ and\ \bibinfo {author} {\bibfnamefont {J.~M.}\ \bibnamefont
  {Thijssen}},\ }\href {\doibase 10.1021/nn800170h} {\bibfield  {journal}
  {\bibinfo  {journal} {ACS Nano}\ }\textbf {\bibinfo {volume} {2}},\ \bibinfo
  {pages} {1445} (\bibinfo {year} {2008})}\BibitemShut {NoStop}%
\bibitem [{\citenamefont {Giese}(2002)}]{Giese2002}%
  \BibitemOpen
  \bibfield  {author} {\bibinfo {author} {\bibfnamefont {B.}~\bibnamefont
  {Giese}},\ }\href {\doibase 10.1146/annurev.biochem.71.083101.134037}
  {\bibfield  {journal} {\bibinfo  {journal} {Annu. Rev. Biochem.}\ }\textbf
  {\bibinfo {volume} {71}},\ \bibinfo {pages} {51} (\bibinfo {year}
  {2002})}\BibitemShut {NoStop}%
\bibitem [{\citenamefont {Watson}\ \emph {et~al.}(2004)\citenamefont {Watson},
  \citenamefont {J\"{a}ckel}, \citenamefont {Severin}, \citenamefont {Rabe},\
  and\ \citenamefont {M\"{u}llen}}]{Watson2004}%
  \BibitemOpen
  \bibfield  {author} {\bibinfo {author} {\bibfnamefont {M.~D.}\ \bibnamefont
  {Watson}}, \bibinfo {author} {\bibfnamefont {F.}~\bibnamefont {J\"{a}ckel}},
  \bibinfo {author} {\bibfnamefont {N.}~\bibnamefont {Severin}}, \bibinfo
  {author} {\bibfnamefont {J.~P.}\ \bibnamefont {Rabe}}, \ and\ \bibinfo
  {author} {\bibfnamefont {K.}~\bibnamefont {M\"{u}llen}},\ }\href {\doibase
  10.1021/ja037520p} {\bibfield  {journal} {\bibinfo  {journal} {J. Am. Chem.
  Soc.}\ }\textbf {\bibinfo {volume} {126}},\ \bibinfo {pages} {1402} (\bibinfo
  {year} {2004})}\BibitemShut {NoStop}%
\bibitem [{\citenamefont {Wu}\ \emph {et~al.}(2008)\citenamefont {Wu},
  \citenamefont {Gonz\'{a}lez}, \citenamefont {Huber}, \citenamefont {Grunder},
  \citenamefont {Mayor}, \citenamefont {Sch\"{o}nenberger},\ and\ \citenamefont
  {Calame}}]{Wu2008}%
  \BibitemOpen
  \bibfield  {author} {\bibinfo {author} {\bibfnamefont {S.}~\bibnamefont
  {Wu}}, \bibinfo {author} {\bibfnamefont {M.~T.}\ \bibnamefont
  {Gonz\'{a}lez}}, \bibinfo {author} {\bibfnamefont {R.}~\bibnamefont {Huber}},
  \bibinfo {author} {\bibfnamefont {S.}~\bibnamefont {Grunder}}, \bibinfo
  {author} {\bibfnamefont {M.}~\bibnamefont {Mayor}}, \bibinfo {author}
  {\bibfnamefont {C.}~\bibnamefont {Sch\"{o}nenberger}}, \ and\ \bibinfo
  {author} {\bibfnamefont {M.}~\bibnamefont {Calame}},\ }\href {\doibase
  10.1038/nnano.2008.237} {\bibfield  {journal} {\bibinfo  {journal} {Nat.
  Nanotechnol.}\ }\textbf {\bibinfo {volume} {3}},\ \bibinfo {pages} {569}
  (\bibinfo {year} {2008})}\BibitemShut {NoStop}%
\bibitem [{\citenamefont {Mart\'{\i}n}\ \emph {et~al.}(2010)\citenamefont
  {Mart\'{\i}n}, \citenamefont {Grace}, \citenamefont {Bryce}, \citenamefont
  {Wang}, \citenamefont {Jitchati}, \citenamefont {Batsanov}, \citenamefont
  {Higgins}, \citenamefont {Lambert},\ and\ \citenamefont
  {Nichols}}]{Martin2010}%
  \BibitemOpen
  \bibfield  {author} {\bibinfo {author} {\bibfnamefont {S.}~\bibnamefont
  {Mart\'{\i}n}}, \bibinfo {author} {\bibfnamefont {I.}~\bibnamefont {Grace}},
  \bibinfo {author} {\bibfnamefont {M.~R.}\ \bibnamefont {Bryce}}, \bibinfo
  {author} {\bibfnamefont {C.}~\bibnamefont {Wang}}, \bibinfo {author}
  {\bibfnamefont {R.}~\bibnamefont {Jitchati}}, \bibinfo {author}
  {\bibfnamefont {A.~S.}\ \bibnamefont {Batsanov}}, \bibinfo {author}
  {\bibfnamefont {S.~J.}\ \bibnamefont {Higgins}}, \bibinfo {author}
  {\bibfnamefont {C.~J.}\ \bibnamefont {Lambert}}, \ and\ \bibinfo {author}
  {\bibfnamefont {R.~J.}\ \bibnamefont {Nichols}},\ }\href {\doibase
  10.1021/ja103327f} {\bibfield  {journal} {\bibinfo  {journal} {J. Am. Chem.
  Soc.}\ }\textbf {\bibinfo {volume} {132}},\ \bibinfo {pages} {9157} (\bibinfo
  {year} {2010})}\BibitemShut {NoStop}%
\bibitem [{\citenamefont {Schneebeli}\ \emph {et~al.}(2011)\citenamefont
  {Schneebeli}, \citenamefont {Kamenetska}, \citenamefont {Cheng},
  \citenamefont {Skouta}, \citenamefont {Friesner}, \citenamefont
  {Venkataraman},\ and\ \citenamefont {Breslow}}]{Schneebeli2011}%
  \BibitemOpen
  \bibfield  {author} {\bibinfo {author} {\bibfnamefont {S.~T.}\ \bibnamefont
  {Schneebeli}}, \bibinfo {author} {\bibfnamefont {M.}~\bibnamefont
  {Kamenetska}}, \bibinfo {author} {\bibfnamefont {Z.}~\bibnamefont {Cheng}},
  \bibinfo {author} {\bibfnamefont {R.}~\bibnamefont {Skouta}}, \bibinfo
  {author} {\bibfnamefont {R.~A.}\ \bibnamefont {Friesner}}, \bibinfo {author}
  {\bibfnamefont {L.}~\bibnamefont {Venkataraman}}, \ and\ \bibinfo {author}
  {\bibfnamefont {R.}~\bibnamefont {Breslow}},\ }\href {\doibase
  10.1021/ja111320n} {\bibfield  {journal} {\bibinfo  {journal} {J. Am. Chem.
  Soc.}\ }\textbf {\bibinfo {volume} {133}},\ \bibinfo {pages} {2136} (\bibinfo
  {year} {2011})}\BibitemShut {NoStop}%
\bibitem [{\citenamefont {Wang}\ \emph {et~al.}(2007)\citenamefont {Wang},
  \citenamefont {Carter}, \citenamefont {Lagutchev}, \citenamefont {Koh},
  \citenamefont {Seong}, \citenamefont {Cahill},\ and\ \citenamefont
  {Dlott}}]{Wang2007b}%
  \BibitemOpen
  \bibfield  {author} {\bibinfo {author} {\bibfnamefont {Z.}~\bibnamefont
  {Wang}}, \bibinfo {author} {\bibfnamefont {J.~A.}\ \bibnamefont {Carter}},
  \bibinfo {author} {\bibfnamefont {A.}~\bibnamefont {Lagutchev}}, \bibinfo
  {author} {\bibfnamefont {Y.~K.}\ \bibnamefont {Koh}}, \bibinfo {author}
  {\bibfnamefont {N.-H.}\ \bibnamefont {Seong}}, \bibinfo {author}
  {\bibfnamefont {D.~G.}\ \bibnamefont {Cahill}}, \ and\ \bibinfo {author}
  {\bibfnamefont {D.~D.}\ \bibnamefont {Dlott}},\ }\href {\doibase
  10.1126/science.1145220} {\bibfield  {journal} {\bibinfo  {journal}
  {Science}\ }\textbf {\bibinfo {volume} {317}},\ \bibinfo {pages} {787}
  (\bibinfo {year} {2007})}\BibitemShut {NoStop}%
\bibitem [{\citenamefont {Losego}\ \emph {et~al.}(2012)\citenamefont {Losego},
  \citenamefont {Grady}, \citenamefont {Sottos}, \citenamefont {Cahill},\ and\
  \citenamefont {Braun}}]{Losego2012}%
  \BibitemOpen
  \bibfield  {author} {\bibinfo {author} {\bibfnamefont {M.~D.}\ \bibnamefont
  {Losego}}, \bibinfo {author} {\bibfnamefont {M.~E.}\ \bibnamefont {Grady}},
  \bibinfo {author} {\bibfnamefont {N.~R.}\ \bibnamefont {Sottos}}, \bibinfo
  {author} {\bibfnamefont {D.~G.}\ \bibnamefont {Cahill}}, \ and\ \bibinfo
  {author} {\bibfnamefont {P.~V.}\ \bibnamefont {Braun}},\ }\href {\doibase
  10.1038/nmat3303} {\bibfield  {journal} {\bibinfo  {journal} {Nat. Mater.}\
  }\textbf {\bibinfo {volume} {11}},\ \bibinfo {pages} {502} (\bibinfo {year}
  {2012})}\BibitemShut {NoStop}%
\bibitem [{\citenamefont {O'Brien}\ \emph {et~al.}(2013)\citenamefont
  {O'Brien}, \citenamefont {Shenogin}, \citenamefont {Liu}, \citenamefont
  {Chow}, \citenamefont {Laurencin}, \citenamefont {Mutin}, \citenamefont
  {Yamaguchi}, \citenamefont {Keblinski},\ and\ \citenamefont
  {Ramanath}}]{O'Brien2013}%
  \BibitemOpen
  \bibfield  {author} {\bibinfo {author} {\bibfnamefont {P.~J.}\ \bibnamefont
  {O'Brien}}, \bibinfo {author} {\bibfnamefont {S.}~\bibnamefont {Shenogin}},
  \bibinfo {author} {\bibfnamefont {J.}~\bibnamefont {Liu}}, \bibinfo {author}
  {\bibfnamefont {P.~K.}\ \bibnamefont {Chow}}, \bibinfo {author}
  {\bibfnamefont {D.}~\bibnamefont {Laurencin}}, \bibinfo {author}
  {\bibfnamefont {P.~H.}\ \bibnamefont {Mutin}}, \bibinfo {author}
  {\bibfnamefont {M.}~\bibnamefont {Yamaguchi}}, \bibinfo {author}
  {\bibfnamefont {P.}~\bibnamefont {Keblinski}}, \ and\ \bibinfo {author}
  {\bibfnamefont {G.}~\bibnamefont {Ramanath}},\ }\href {\doibase
  10.1038/nmat3465} {\bibfield  {journal} {\bibinfo  {journal} {Nat. Mater.}\
  }\textbf {\bibinfo {volume} {12}},\ \bibinfo {pages} {118} (\bibinfo {year}
  {2013})}\BibitemShut {NoStop}%
\bibitem [{\citenamefont {Meier}\ \emph {et~al.}(2014)\citenamefont {Meier},
  \citenamefont {Menges}, \citenamefont {Nirmalraj}, \citenamefont
  {H\"{o}lscher}, \citenamefont {Riel},\ and\ \citenamefont
  {Gotsmann}}]{Meier2014}%
  \BibitemOpen
  \bibfield  {author} {\bibinfo {author} {\bibfnamefont {T.}~\bibnamefont
  {Meier}}, \bibinfo {author} {\bibfnamefont {F.}~\bibnamefont {Menges}},
  \bibinfo {author} {\bibfnamefont {P.}~\bibnamefont {Nirmalraj}}, \bibinfo
  {author} {\bibfnamefont {H.}~\bibnamefont {H\"{o}lscher}}, \bibinfo {author}
  {\bibfnamefont {H.}~\bibnamefont {Riel}}, \ and\ \bibinfo {author}
  {\bibfnamefont {B.}~\bibnamefont {Gotsmann}},\ }\href {\doibase
  10.1103/PhysRevLett.113.060801} {\bibfield  {journal} {\bibinfo  {journal}
  {Phys. Rev. Lett.}\ }\textbf {\bibinfo {volume} {113}},\ \bibinfo {pages}
  {060801} (\bibinfo {year} {2014})}\BibitemShut {NoStop}%
\bibitem [{\citenamefont {Landau}\ \emph {et~al.}(1986)\citenamefont {Landau},
  \citenamefont {Lifshitz}, \citenamefont {Kosevich},\ and\ \citenamefont
  {Pitaevskii}}]{Landau1986}%
  \BibitemOpen
  \bibfield  {author} {\bibinfo {author} {\bibfnamefont {L.~D.}\ \bibnamefont
  {Landau}}, \bibinfo {author} {\bibfnamefont {E.~M.}\ \bibnamefont
  {Lifshitz}}, \bibinfo {author} {\bibfnamefont {A.~M.}\ \bibnamefont
  {Kosevich}}, \ and\ \bibinfo {author} {\bibfnamefont {L.~P.}\ \bibnamefont
  {Pitaevskii}},\ }\href@noop {} {\emph {\bibinfo {title} {{Theory of
  Elasticity}}}}\ (\bibinfo  {publisher} {Pergamon Press},\ \bibinfo {address}
  {Oxford},\ \bibinfo {year} {1986})\BibitemShut {NoStop}%
\bibitem [{\citenamefont {Ezawa}(1971)}]{Ezawa1971}%
  \BibitemOpen
  \bibfield  {author} {\bibinfo {author} {\bibfnamefont {H.}~\bibnamefont
  {Ezawa}},\ }\href {\doibase 10.1016/0003-4916(71)90149-7} {\bibfield
  {journal} {\bibinfo  {journal} {Ann. Phys.}\ }\textbf {\bibinfo {volume}
  {67}},\ \bibinfo {pages} {438} (\bibinfo {year} {1971})}\BibitemShut
  {NoStop}%
\bibitem [{\citenamefont {Patton}\ and\ \citenamefont
  {Geller}(2001)}]{Patton2001}%
  \BibitemOpen
  \bibfield  {author} {\bibinfo {author} {\bibfnamefont {K.}~\bibnamefont
  {Patton}}\ and\ \bibinfo {author} {\bibfnamefont {M.}~\bibnamefont
  {Geller}},\ }\href {\doibase 10.1103/PhysRevB.64.155320} {\bibfield
  {journal} {\bibinfo  {journal} {Phys. Rev. B}\ }\textbf {\bibinfo {volume}
  {64}},\ \bibinfo {pages} {155320} (\bibinfo {year} {2001})}\BibitemShut
  {NoStop}%
\bibitem [{\citenamefont {B\"{u}ttiker}\ \emph {et~al.}(1985)\citenamefont
  {B\"{u}ttiker}, \citenamefont {Imry}, \citenamefont {Landauer},\ and\
  \citenamefont {Pinhas}}]{Buttiker1985}%
  \BibitemOpen
  \bibfield  {author} {\bibinfo {author} {\bibfnamefont {M.}~\bibnamefont
  {B\"{u}ttiker}}, \bibinfo {author} {\bibfnamefont {Y.}~\bibnamefont {Imry}},
  \bibinfo {author} {\bibfnamefont {R.}~\bibnamefont {Landauer}}, \ and\
  \bibinfo {author} {\bibfnamefont {S.}~\bibnamefont {Pinhas}},\ }\href
  {\doibase 10.1103/PhysRevB.31.6207} {\bibfield  {journal} {\bibinfo
  {journal} {Phys. Rev. B}\ }\textbf {\bibinfo {volume} {31}},\ \bibinfo
  {pages} {6207} (\bibinfo {year} {1985})}\BibitemShut {NoStop}%
\bibitem [{\citenamefont {Caroli}\ \emph {et~al.}(1971)\citenamefont {Caroli},
  \citenamefont {Combescot}, \citenamefont {Nozieres},\ and\ \citenamefont
  {Saint-James}}]{Caroli1971}%
  \BibitemOpen
  \bibfield  {author} {\bibinfo {author} {\bibfnamefont {C.}~\bibnamefont
  {Caroli}}, \bibinfo {author} {\bibfnamefont {R.}~\bibnamefont {Combescot}},
  \bibinfo {author} {\bibfnamefont {P.}~\bibnamefont {Nozieres}}, \ and\
  \bibinfo {author} {\bibfnamefont {D.}~\bibnamefont {Saint-James}},\ }\href
  {\doibase 10.1088/0022-3719/4/8/018} {\bibfield  {journal} {\bibinfo
  {journal} {J. Phys. C Solid State Phys.}\ }\textbf {\bibinfo {volume} {4}},\
  \bibinfo {pages} {916} (\bibinfo {year} {1971})}\BibitemShut {NoStop}%
\bibitem [{Sup()}]{SupplementalMaterial}%
  \BibitemOpen
  \href@noop {} {\bibinfo  {journal} {See supplemental material}\ }\BibitemShut {NoStop}%
\bibitem [{\citenamefont {Samsonov}(1968)}]{Samsonov1968}%
  \BibitemOpen
\bibfield  {journal} {  }\bibfield  {author} {\bibinfo {author} {\bibfnamefont
  {G.~V.}\ \bibnamefont {Samsonov}},\ }\href@noop {} {\emph {\bibinfo {title}
  {{Handbook of the Physicochemical Properties of the Elements}}}}\ (\bibinfo
  {publisher} {IFI/Plenum},\ \bibinfo {address} {New York-Washington},\
  \bibinfo {year} {1968})\BibitemShut {NoStop}%
\bibitem [{\citenamefont {Kittel}(2004)}]{Kittel2004}%
  \BibitemOpen
  \bibfield  {author} {\bibinfo {author} {\bibfnamefont {C.}~\bibnamefont
  {Kittel}},\ }\href@noop {} {\emph {\bibinfo {title} {{Introduction to Solid
  State Physics}}}}\ (\bibinfo  {publisher} {John Wiley \& Sons, Inc.},\
  \bibinfo {address} {New York},\ \bibinfo {year} {2004})\BibitemShut {NoStop}%
\bibitem [{\citenamefont {Enkovaara}\ \emph {et~al.}(2010)\citenamefont
  {Enkovaara}, \citenamefont {Rostgaard}, \citenamefont {Mortensen},
  \citenamefont {Chen}, \citenamefont {Dułak}, \citenamefont {Ferrighi},
  \citenamefont {Gavnholt}, \citenamefont {Glinsvad}, \citenamefont {Haikola},
  \citenamefont {Hansen}, \citenamefont {Kristoffersen}, \citenamefont
  {Kuisma}, \citenamefont {Larsen}, \citenamefont {Lehtovaara}, \citenamefont
  {Ljungberg}, \citenamefont {Lopez-Acevedo}, \citenamefont {Moses},
  \citenamefont {Ojanen}, \citenamefont {Olsen}, \citenamefont {Petzold},
  \citenamefont {Romero}, \citenamefont {Stausholm-M{\o}ller}, \citenamefont
  {Strange}, \citenamefont {Tritsaris}, \citenamefont {Vanin}, \citenamefont
  {Walter}, \citenamefont {Hammer}, \citenamefont {H\"{a}kkinen}, \citenamefont
  {Madsen}, \citenamefont {Nieminen}, \citenamefont {N{\o}rskov}, \citenamefont
  {Puska}, \citenamefont {Rantala}, \citenamefont {Schi{\o}tz}, \citenamefont
  {Thygesen},\ and\ \citenamefont {Jacobsen}}]{Enkovaara2010}%
  \BibitemOpen
  \bibfield  {author} {\bibinfo {author} {\bibfnamefont {J.}~\bibnamefont
  {Enkovaara}}, \bibinfo {author} {\bibfnamefont {C.}~\bibnamefont
  {Rostgaard}}, \bibinfo {author} {\bibfnamefont {J.~J.}\ \bibnamefont
  {Mortensen}}, \bibinfo {author} {\bibfnamefont {J.}~\bibnamefont {Chen}},
  \bibinfo {author} {\bibfnamefont {M.}~\bibnamefont {Dułak}}, \bibinfo
  {author} {\bibfnamefont {L.}~\bibnamefont {Ferrighi}}, \bibinfo {author}
  {\bibfnamefont {J.}~\bibnamefont {Gavnholt}}, \bibinfo {author}
  {\bibfnamefont {C.}~\bibnamefont {Glinsvad}}, \bibinfo {author}
  {\bibfnamefont {V.}~\bibnamefont {Haikola}}, \bibinfo {author} {\bibfnamefont
  {H.~A.}\ \bibnamefont {Hansen}}, \bibinfo {author} {\bibfnamefont {H.~H.}\
  \bibnamefont {Kristoffersen}}, \bibinfo {author} {\bibfnamefont
  {M.}~\bibnamefont {Kuisma}}, \bibinfo {author} {\bibfnamefont {A.~H.}\
  \bibnamefont {Larsen}}, \bibinfo {author} {\bibfnamefont {L.}~\bibnamefont
  {Lehtovaara}}, \bibinfo {author} {\bibfnamefont {M.}~\bibnamefont
  {Ljungberg}}, \bibinfo {author} {\bibfnamefont {O.}~\bibnamefont
  {Lopez-Acevedo}}, \bibinfo {author} {\bibfnamefont {P.~G.}\ \bibnamefont
  {Moses}}, \bibinfo {author} {\bibfnamefont {J.}~\bibnamefont {Ojanen}},
  \bibinfo {author} {\bibfnamefont {T.}~\bibnamefont {Olsen}}, \bibinfo
  {author} {\bibfnamefont {V.}~\bibnamefont {Petzold}}, \bibinfo {author}
  {\bibfnamefont {N.~A.}\ \bibnamefont {Romero}}, \bibinfo {author}
  {\bibfnamefont {J.}~\bibnamefont {Stausholm-M{\o}ller}}, \bibinfo {author}
  {\bibfnamefont {M.}~\bibnamefont {Strange}}, \bibinfo {author} {\bibfnamefont
  {G.~A.}\ \bibnamefont {Tritsaris}}, \bibinfo {author} {\bibfnamefont
  {M.}~\bibnamefont {Vanin}}, \bibinfo {author} {\bibfnamefont
  {M.}~\bibnamefont {Walter}}, \bibinfo {author} {\bibfnamefont
  {B.}~\bibnamefont {Hammer}}, \bibinfo {author} {\bibfnamefont
  {H.}~\bibnamefont {H\"{a}kkinen}}, \bibinfo {author} {\bibfnamefont
  {G.~K.~H.}\ \bibnamefont {Madsen}}, \bibinfo {author} {\bibfnamefont {R.~M.}\
  \bibnamefont {Nieminen}}, \bibinfo {author} {\bibfnamefont {J.~K.}\
  \bibnamefont {N{\o}rskov}}, \bibinfo {author} {\bibfnamefont
  {M.}~\bibnamefont {Puska}}, \bibinfo {author} {\bibfnamefont {T.~T.}\
  \bibnamefont {Rantala}}, \bibinfo {author} {\bibfnamefont {J.}~\bibnamefont
  {Schi{\o}tz}}, \bibinfo {author} {\bibfnamefont {K.~S.}\ \bibnamefont
  {Thygesen}}, \ and\ \bibinfo {author} {\bibfnamefont {K.~W.}\ \bibnamefont
  {Jacobsen}},\ }\href {\doibase 10.1088/0953-8984/22/25/253202} {\bibfield
  {journal} {\bibinfo  {journal} {J. Phys.-Condens. Matter}\ }\textbf {\bibinfo
  {volume} {22}},\ \bibinfo {pages} {253202} (\bibinfo {year}
  {2010})}\BibitemShut {NoStop}%
\bibitem [{\citenamefont {Tkatchenko}\ and\ \citenamefont
  {Scheffler}(2009)}]{Tkatchenko2009}%
  \BibitemOpen
  \bibfield  {author} {\bibinfo {author} {\bibfnamefont {A.}~\bibnamefont
  {Tkatchenko}}\ and\ \bibinfo {author} {\bibfnamefont {M.}~\bibnamefont
  {Scheffler}},\ }\href {\doibase 10.1103/PhysRevLett.102.073005} {\bibfield
  {journal} {\bibinfo  {journal} {Phys. Rev. Lett.}\ }\textbf {\bibinfo
  {volume} {102}},\ \bibinfo {pages} {073005} (\bibinfo {year}
  {2009})}\BibitemShut {NoStop}%
\bibitem [{\citenamefont {Sivan}\ and\ \citenamefont {Imry}(1986)}]{Sivan1986}%
  \BibitemOpen
  \bibfield  {author} {\bibinfo {author} {\bibfnamefont {U.}~\bibnamefont
  {Sivan}}\ and\ \bibinfo {author} {\bibfnamefont {Y.}~\bibnamefont {Imry}},\
  }\href {\doibase 10.1103/PhysRevB.33.551} {\bibfield  {journal} {\bibinfo
  {journal} {Phys. Rev. B}\ }\textbf {\bibinfo {volume} {33}},\ \bibinfo
  {pages} {551} (\bibinfo {year} {1986})}\BibitemShut {NoStop}%
\bibitem [{\citenamefont {Mahan}\ and\ \citenamefont {Sofo}(1996)}]{Mahan1996}%
  \BibitemOpen
  \bibfield  {author} {\bibinfo {author} {\bibfnamefont {G.~D.}\ \bibnamefont
  {Mahan}}\ and\ \bibinfo {author} {\bibfnamefont {J.~O.}\ \bibnamefont
  {Sofo}},\ }\href
  {http://www.pubmedcentral.nih.gov/articlerender.fcgi?artid=38761\&tool=pmcentrez\&rendertype=abstract}
  {\bibfield  {journal} {\bibinfo  {journal} {P. Natl. Acad. Sci. USA}\
  }\textbf {\bibinfo {volume} {93}},\ \bibinfo {pages} {7436} (\bibinfo {year}
  {1996})}\BibitemShut {NoStop}%
\bibitem [{\citenamefont {Humphrey}\ and\ \citenamefont
  {Linke}(2005)}]{Humphrey2005}%
  \BibitemOpen
  \bibfield  {author} {\bibinfo {author} {\bibfnamefont {T.}~\bibnamefont
  {Humphrey}}\ and\ \bibinfo {author} {\bibfnamefont {H.}~\bibnamefont
  {Linke}},\ }\href {\doibase 10.1103/PhysRevLett.94.096601} {\bibfield
  {journal} {\bibinfo  {journal} {Phys. Rev. Lett.}\ }\textbf {\bibinfo
  {volume} {94}},\ \bibinfo {pages} {096601} (\bibinfo {year}
  {2005})}\BibitemShut {NoStop}%
\bibitem [{\citenamefont {Esposito}, \citenamefont {Lindenberg},\ and\
  \citenamefont {{Van den Broeck}}(2009)}]{Esposito2009}%
  \BibitemOpen
  \bibfield  {author} {\bibinfo {author} {\bibfnamefont {M.}~\bibnamefont
  {Esposito}}, \bibinfo {author} {\bibfnamefont {K.}~\bibnamefont
  {Lindenberg}}, \ and\ \bibinfo {author} {\bibfnamefont {C.}~\bibnamefont
  {{Van den Broeck}}},\ }\href {\doibase 10.1209/0295-5075/85/60010} {\bibfield
   {journal} {\bibinfo  {journal} {EPL-Europhys. Lett.}\ }\textbf {\bibinfo
  {volume} {85}},\ \bibinfo {pages} {60010} (\bibinfo {year}
  {2009})}\BibitemShut {NoStop}%
\bibitem [{\citenamefont {Leijnse}, \citenamefont {Wegewijs},\ and\
  \citenamefont {Flensberg}(2010)}]{Leijnse2010}%
  \BibitemOpen
  \bibfield  {author} {\bibinfo {author} {\bibfnamefont {M.}~\bibnamefont
  {Leijnse}}, \bibinfo {author} {\bibfnamefont {M.~R.}\ \bibnamefont
  {Wegewijs}}, \ and\ \bibinfo {author} {\bibfnamefont {K.}~\bibnamefont
  {Flensberg}},\ }\href {\doibase 10.1103/PhysRevB.82.045412} {\bibfield
  {journal} {\bibinfo  {journal} {Phys. Rev. B}\ }\textbf {\bibinfo {volume}
  {82}},\ \bibinfo {pages} {045412} (\bibinfo {year} {2010})},\ \Eprint
  {http://arxiv.org/abs/1004.4500} {1004.4500} \BibitemShut {NoStop}%
\end{thebibliography}

\begin{thebibliography}{12}%
\makeatletter
\providecommand \@ifxundefined [1]{%
 \@ifx{#1\undefined}
}%
\providecommand \@ifnum [1]{%
 \ifnum #1\expandafter \@firstoftwo
 \else \expandafter \@secondoftwo
 \fi
}%
\providecommand \@ifx [1]{%
 \ifx #1\expandafter \@firstoftwo
 \else \expandafter \@secondoftwo
 \fi
}%
\providecommand \natexlab [1]{#1}%
\providecommand \enquote  [1]{``#1''}%
\providecommand \bibnamefont  [1]{#1}%
\providecommand \bibfnamefont [1]{#1}%
\providecommand \citenamefont [1]{#1}%
\providecommand \href@noop [0]{\@secondoftwo}%
\providecommand \href [0]{\begingroup \@sanitize@url \@href}%
\providecommand \@href[1]{\@@startlink{#1}\@@href}%
\providecommand \@@href[1]{\endgroup#1\@@endlink}%
\providecommand \@sanitize@url [0]{\catcode `\\12\catcode `\$12\catcode
  `\&12\catcode `\#12\catcode `\^12\catcode `\_12\catcode `\%12\relax}%
\providecommand \@@startlink[1]{}%
\providecommand \@@endlink[0]{}%
\providecommand \url  [0]{\begingroup\@sanitize@url \@url }%
\providecommand \@url [1]{\endgroup\@href {#1}{\urlprefix }}%
\providecommand \urlprefix  [0]{URL }%
\providecommand \Eprint [0]{\href }%
\providecommand \doibase [0]{http://dx.doi.org/}%
\providecommand \selectlanguage [0]{\@gobble}%
\providecommand \bibinfo  [0]{\@secondoftwo}%
\providecommand \bibfield  [0]{\@secondoftwo}%
\providecommand \translation [1]{[#1]}%
\providecommand \BibitemOpen [0]{}%
\providecommand \bibitemStop [0]{}%
\providecommand \bibitemNoStop [0]{.\EOS\space}%
\providecommand \EOS [0]{\spacefactor3000\relax}%
\providecommand \BibitemShut  [1]{\csname bibitem#1\endcsname}%
\let\auto@bib@innerbib\@empty
\bibitem [{\citenamefont {Ezawa}(1971)}]{S_Ezawa1971}%
  \BibitemOpen
  \bibfield  {author} {\bibinfo {author} {\bibfnamefont {H.}~\bibnamefont
  {Ezawa}},\ }\href {\doibase 10.1016/0003-4916(71)90149-7} {\bibfield
  {journal} {\bibinfo  {journal} {Ann. Phys.}\ }\textbf {\bibinfo {volume}
  {67}},\ \bibinfo {pages} {438} (\bibinfo {year} {1971})}\BibitemShut
  {NoStop}%
\bibitem [{\citenamefont {Patton}\ and\ \citenamefont
  {Geller}(2001)}]{S_Patton2001}%
  \BibitemOpen
  \bibfield  {author} {\bibinfo {author} {\bibfnamefont {K.}~\bibnamefont
  {Patton}}\ and\ \bibinfo {author} {\bibfnamefont {M.}~\bibnamefont
  {Geller}},\ }\href {\doibase 10.1103/PhysRevB.64.155320} {\bibfield
  {journal} {\bibinfo  {journal} {Phys. Rev. B}\ }\textbf {\bibinfo {volume}
  {64}},\ \bibinfo {pages} {155320} (\bibinfo {year} {2001})}\BibitemShut
  {NoStop}%
\bibitem [{\citenamefont {Landau}\ \emph {et~al.}(1986)\citenamefont {Landau},
  \citenamefont {Lifshitz}, \citenamefont {Kosevich},\ and\ \citenamefont
  {Pitaevskii}}]{S_Landau1986}%
  \BibitemOpen
  \bibfield  {author} {\bibinfo {author} {\bibfnamefont {L.~D.}\ \bibnamefont
  {Landau}}, \bibinfo {author} {\bibfnamefont {E.~M.}\ \bibnamefont
  {Lifshitz}}, \bibinfo {author} {\bibfnamefont {A.~M.}\ \bibnamefont
  {Kosevich}}, \ and\ \bibinfo {author} {\bibfnamefont {L.~P.}\ \bibnamefont
  {Pitaevskii}},\ }\href@noop {} {\emph {\bibinfo {title} {{Theory of
  Elasticity}}}}\ (\bibinfo  {publisher} {Pergamon Press},\ \bibinfo {address}
  {Oxford},\ \bibinfo {year} {1986})\BibitemShut {NoStop}%
\bibitem [{\citenamefont {Segal}, \citenamefont {Nitzan},\ and\ \citenamefont
  {H\"{a}nggi}(2003)}]{S_Segal2003}%
  \BibitemOpen
  \bibfield  {author} {\bibinfo {author} {\bibfnamefont {D.}~\bibnamefont
  {Segal}}, \bibinfo {author} {\bibfnamefont {A.}~\bibnamefont {Nitzan}}, \
  and\ \bibinfo {author} {\bibfnamefont {P.}~\bibnamefont {H\"{a}nggi}},\
  }\href {\doibase 10.1063/1.1603211} {\bibfield  {journal} {\bibinfo
  {journal} {J. Chem. Phys.}\ }\textbf {\bibinfo {volume} {119}},\ \bibinfo
  {pages} {6840} (\bibinfo {year} {2003})}\BibitemShut {NoStop}%
\bibitem [{\citenamefont {Mingo}(2006)}]{S_Mingo2006}%
  \BibitemOpen
  \bibfield  {author} {\bibinfo {author} {\bibfnamefont {N.}~\bibnamefont
  {Mingo}},\ }\href {\doibase 10.1103/PhysRevB.74.125402} {\bibfield  {journal}
  {\bibinfo  {journal} {Phys. Rev. B}\ }\textbf {\bibinfo {volume} {74}},\
  \bibinfo {pages} {125402} (\bibinfo {year} {2006})}\BibitemShut {NoStop}%
\bibitem [{\citenamefont {Yamamoto}\ and\ \citenamefont
  {Watanabe}(2006)}]{S_Yamamoto2006}%
  \BibitemOpen
  \bibfield  {author} {\bibinfo {author} {\bibfnamefont {T.}~\bibnamefont
  {Yamamoto}}\ and\ \bibinfo {author} {\bibfnamefont {K.}~\bibnamefont
  {Watanabe}},\ }\href {\doibase 10.1103/PhysRevLett.96.255503} {\bibfield
  {journal} {\bibinfo  {journal} {Phys. Rev. Lett.}\ }\textbf {\bibinfo
  {volume} {96}},\ \bibinfo {pages} {255503} (\bibinfo {year}
  {2006})}\BibitemShut {NoStop}%
\bibitem [{\citenamefont {Wang}\ \emph {et~al.}(2007)\citenamefont {Wang},
  \citenamefont {Zeng}, \citenamefont {Wang},\ and\ \citenamefont
  {Gan}}]{S_Wang2007}%
  \BibitemOpen
  \bibfield  {author} {\bibinfo {author} {\bibfnamefont {J.-S.}\ \bibnamefont
  {Wang}}, \bibinfo {author} {\bibfnamefont {N.}~\bibnamefont {Zeng}}, \bibinfo
  {author} {\bibfnamefont {J.}~\bibnamefont {Wang}}, \ and\ \bibinfo {author}
  {\bibfnamefont {C.~K.}\ \bibnamefont {Gan}},\ }\href {\doibase
  10.1103/PhysRevE.75.061128} {\bibfield  {journal} {\bibinfo  {journal} {Phys.
  Rev. E}\ }\textbf {\bibinfo {volume} {75}},\ \bibinfo {pages} {061128}
  (\bibinfo {year} {2007})}\BibitemShut {NoStop}%
\bibitem [{\citenamefont {Keldysh}(1965)}]{S_Keldysh1965}%
  \BibitemOpen
  \bibfield  {author} {\bibinfo {author} {\bibfnamefont {L.~V.}\ \bibnamefont
  {Keldysh}},\ }\href {http://www.jetp.ac.ru/cgi-bin/dn/e\_020\_04\_1018.pdf}
  {\bibfield  {journal} {\bibinfo  {journal} {Sov. Phys. JETP-USSR}\ }\textbf
  {\bibinfo {volume} {20}},\ \bibinfo {pages} {1018} (\bibinfo {year}
  {1965})}\BibitemShut {NoStop}%
\bibitem [{\citenamefont {Larkin}\ and\ \citenamefont
  {Ovchinnikov}(1975)}]{S_Larkin1975}%
  \BibitemOpen
  \bibfield  {author} {\bibinfo {author} {\bibfnamefont {A.~I.}\ \bibnamefont
  {Larkin}}\ and\ \bibinfo {author} {\bibfnamefont {Y.~N.}\ \bibnamefont
  {Ovchinnikov}},\ }\href {http://jetp.ac.ru/cgi-bin/dn/e\_041\_05\_0960.pdf}
  {\bibfield  {journal} {\bibinfo  {journal} {Sov. Phys. JETP-USSR}\ }\textbf
  {\bibinfo {volume} {41}},\ \bibinfo {pages} {960} (\bibinfo {year}
  {1975})}\BibitemShut {NoStop}%
\bibitem [{\citenamefont {Rammer}\ and\ \citenamefont
  {Smith}(1986)}]{S_Rammer1986}%
  \BibitemOpen
  \bibfield  {author} {\bibinfo {author} {\bibfnamefont {J.}~\bibnamefont
  {Rammer}}\ and\ \bibinfo {author} {\bibfnamefont {H.}~\bibnamefont {Smith}},\
  }\href {\doibase 10.1103/RevModPhys.58.323} {\bibfield  {journal} {\bibinfo
  {journal} {Rev. Mod. Phys.}\ }\textbf {\bibinfo {volume} {58}},\ \bibinfo
  {pages} {323} (\bibinfo {year} {1986})}\BibitemShut {NoStop}%
\bibitem [{\citenamefont {Markussen}(2013)}]{S_Markussen2013}%
  \BibitemOpen
  \bibfield  {author} {\bibinfo {author} {\bibfnamefont {T.}~\bibnamefont
  {Markussen}},\ }\href {\doibase 10.1063/1.4849178} {\bibfield  {journal}
  {\bibinfo  {journal} {J. Chem. Phys.}\ }\textbf {\bibinfo {volume} {139}},\
  \bibinfo {pages} {244101} (\bibinfo {year} {2013})},\ \Eprint
  {http://arxiv.org/abs/1308.2543} {1308.2543} \BibitemShut {NoStop}%
\bibitem [{\citenamefont {Sivan}\ and\ \citenamefont {Imry}(1986)}]{S_Sivan1986}%
  \BibitemOpen
  \bibfield  {author} {\bibinfo {author} {\bibfnamefont {U.}~\bibnamefont
  {Sivan}}\ and\ \bibinfo {author} {\bibfnamefont {Y.}~\bibnamefont {Imry}},\
  }\href {\doibase 10.1103/PhysRevB.33.551} {\bibfield  {journal} {\bibinfo
  {journal} {Phys. Rev. B}\ }\textbf {\bibinfo {volume} {33}},\ \bibinfo
  {pages} {551} (\bibinfo {year} {1986})}\BibitemShut {NoStop}%
\end{thebibliography}
\end{document}